\documentclass[
reprint,
superscriptaddress,
 amsmath,amssymb,
 aps,
]{revtex4-2}

\usepackage{amsmath}
\usepackage{amssymb}
\usepackage{amsthm}
\usepackage{mathtools}
\usepackage{mathdots}
\usepackage{amsfonts}
\usepackage{bbm}

\usepackage{xr-hyper}

\makeatletter
\newcommand*{\addFileDependency}[1]{
  \typeout{(#1)}
  \@addtofilelist{#1}
  \IfFileExists{#1}{}{\typeout{No file #1.}}
}
\makeatother

\usepackage{bbold}
\usepackage{newpxtext}
\usepackage{epsfig,color,dsfont,upgreek,physics}
\usepackage{mathrsfs}
\usepackage{multirow}

\usepackage{graphicx}
\usepackage{dcolumn} 
\usepackage{bm} 
\usepackage{float} 
\usepackage[driverfallback=dvipdfm]{hyperref} 
\usepackage{longtable}
\usepackage{ulem}   
\usepackage{hyperref}

\allowdisplaybreaks

\newcommand{\bs}[1]{{\boldsymbol{#1}}}

\newcommand{\blue}[1]{{\textcolor{blue}{#1}}}

\renewcommand\[{\begin{equation}}
\renewcommand\]{\end{equation}}

\begin{document}

\title{Unconventional Self-Similar Hofstadter Superconductivity from Repulsive Interactions}

\author{Daniel Shaffer}
\affiliation
{
Department  of  Physics,  Emory  University,  400 Dowman Drive, Atlanta,  GA  30322,  USA
}

\author{Jian Wang}
\affiliation
{
Department  of  Physics,  Emory  University,  400 Dowman Drive, Atlanta,  GA  30322,  USA
}

\author{Luiz H. Santos}
 \email{luiz.santos@emory.edu}
\affiliation
{
Department  of  Physics,  Emory  University,  400 Dowman Drive, Atlanta,  GA  30322,  USA
}

\begin{abstract}

Fractal Hofstadter bands have become widely accessible with the advent of moir\'e superlattices, opening the door to studies of the effect of interactions in these systems.
In this work we employ a renormalization group (RG) analysis to demonstrate that the combination of repulsive interactions with the presence of a tunable manifold of Van Hove singularities provides a new mechanism for driving unconventional superconductivity in Hofstadter bands. 
Specifically, the number of Van Hove singularities at the Fermi energy can be controlled by varying the flux per unit cell and the electronic filling, leading to instabilities toward nodal superconductivity and chiral topological superconductivity with Chern number $\mathcal{C} = \pm 6$. 
The latter is characterized by a self-similar fixed trajectory of the RG flow and an emerging self-similarity symmetry of the order parameter. Our results establish Hofstadter quantum materials such as moir\'e heterostructures as promising platforms for realizing novel reentrant Hofstadter superconductors.

\end{abstract}

\date{\today}

\maketitle

\section{Introduction}

It has long been theoretically suggested that, contrary to the conventional view \cite{GinzburgLandau50,BCS57,Abrikosov57}, superconductivity (SC) can reemerge in Landau levels in the presence of strong magnetic fields, provided there are attractive interactions \cite{RasoltTesanovich92}. More recently it has been proposed that such reentrant
SC in reconstructed electron bands forming Landau levels can theoretically occur in magic angle twisted bilayer graphene (TBG) \cite{chaudhary2021Quantum}. TBG and other 2D moir\'e superlattices are particularly attractive for realizing reentrant SC as they can host SC at zero magnetic field at low density carrier regimes \cite{cao2018unconventional}, such that only relatively modest magnetic fields are required to achieve the quanutm limit of Landau levels.
However, several challenges have stood in the way of observing reentrant SC in experiment, among them the role of repulsive interactions that make quantum Hall states natural competitors of such reentrant SC in Landau levels.

In this work we propose that this issue can be circumvented in Hofstadter bands that{\textcolor{blue}{,}} unlike Landau levels{\textcolor{blue}{,}} have a finite bandwidth \(W\) \cite{Azbel64,Hofstadter76}, allowing a weak-coupling renormalization group (RG) treatment of repulsive electronic interactions. This is especially relevant for moir\'e systems since their nanometer scale unit cells enable the realization of Hofstadter bands in experimentally accessible magnetic fields at which the magnetic flux per super unit cell $\Phi = B A_{\textrm{uc}}$ is comparable to the flux quantum $\Phi_0 = 2\pi \hbar/e = 2\pi$ in natural units \cite{Dean13,Ponomarenko13,Hunt13,Forsythe18,Wang15,Spanton18,Saito21}. 
Beyond the rich phenomena in Hofstadter-Chern insulators \cite{wang_classification_2020,Herzog-Arbeitman_Hofstadter2020,Spanton2018,sharpe2019emergent,serlin2020intrinsic,Saito2021,xie2021fractional,nuckolls2020strongly,wu_chern_2021,das2021symmetry,choi2021correlation,park2021flavour,stepanov2021competing,pierce2021unconventional,yu2021correlated}, a recent classification \cite{ShafferWangSantos21} has shown that Hofstadter bands may support novel Hofstadter superconductors (HSC) characterized by spontaneous breaking of the magnetic translation group (MTG) symmetries \cite{Zak64_1, Zak64_2, Brown64}, leading to multi-component finite momentum Cooper pairing similar to pair-density wave states \cite{Agterberg20}. HSCs embody a new form of reentrant superconductivity in Hofstadter bands, in which the large flux per unit cell makes the magnetic length comparable to the lattice scale,  thus generalizing the Landau level reentrant SC state.

\begin{figure}[h]
\centering
\includegraphics[width=0.49\textwidth]{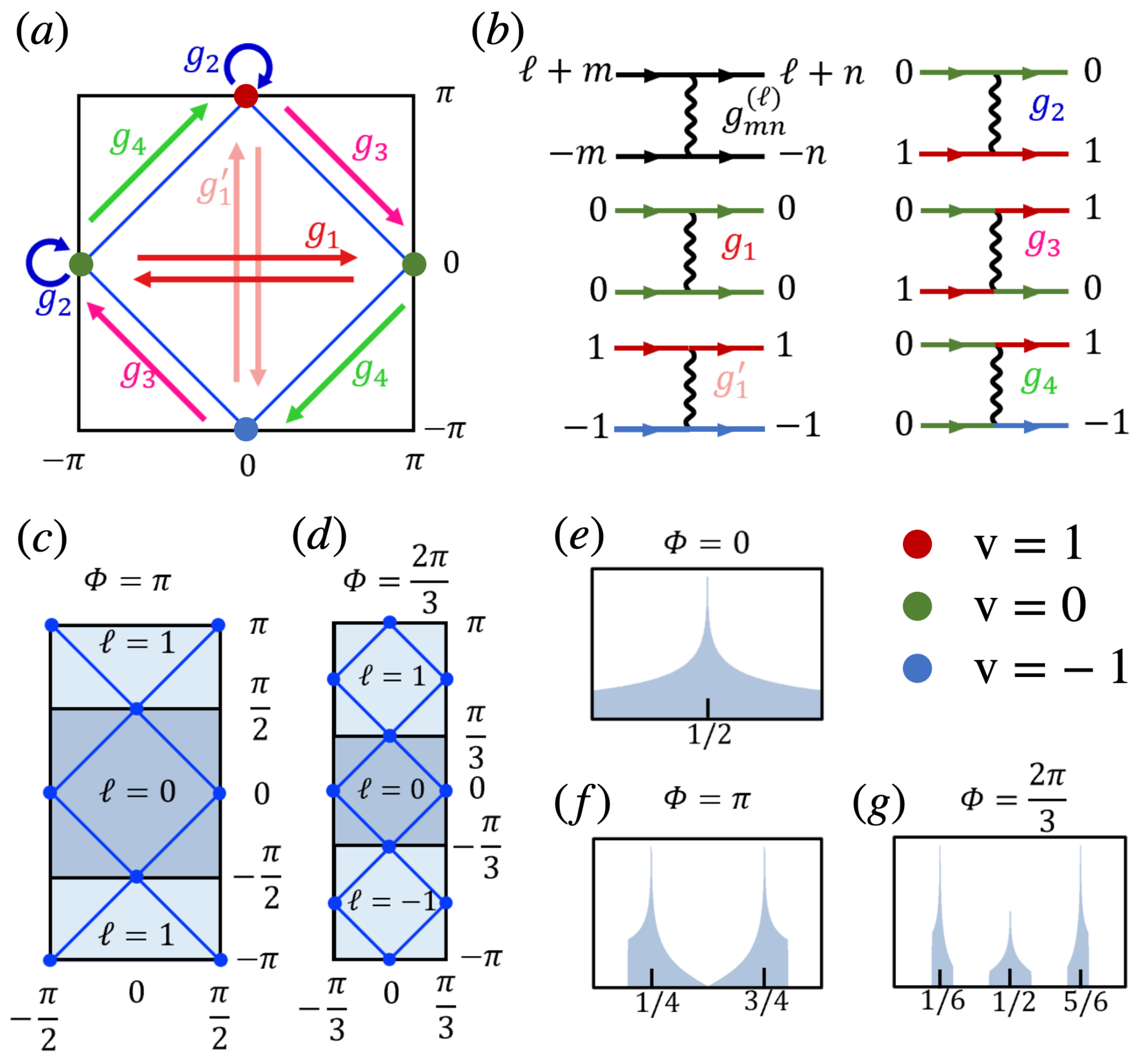}
\caption{Van Hove singularities and relevant interactions in the square Hofstadter model. VHSs are shown at (a) zero, (c) \(\pi\), and (d) \(2\pi/3\) flux, and (e-g) the corresponding peaks in the density of states at indicated fillings. Due to the MTG symmetry, the magnetic Brillouin zone (MBZ) splits into \(q\) (energy degenerate) reduced magnetic Brillouin zones (rMBZ) labeled with \(\ell=0,\dots,q-1\). In each band there are a total of \(2q\) VHSs occurring at momenta \(\mathbf{K}_{\ell,\mathrm{v}}=\left((1+\mathrm{v})\frac{\pi}{q},(\mathrm{v}+2 p \ell)\frac{\pi}{q}\right)\), such that there is a pair of VHSs in each rMBZ labelled with a VHS index \(\mathrm{v}=0,\pm1\), with the identification of VHS \(\ell,1\) and \(\ell+1,-1\).
Arrows in (a) and the Feynman diagrams in (b) show the types of interaction processes considered in the RG analysis: intra-VHS processes \(g_1\) and \(g_{1'}\) (red and light red); inter-VHS forward scattering \(g_2\) (blue); exchange \(g_3\) (magenta); and pair-hopping \(g_4\). The VHS index color-coded in (a) and (b) as red/green/blue for \(\mathrm{v}=1,0,-1\) respectively, and the black diagram shows the additional rMBZ indices \(\ell,m,n=0,\dots,q-1\) carried by the coupling constants \(g^{(\ell)}_{mn}\), \(\ell\) denoting the total momentum of the interacting pair.}
\label{fig:VHS}
\end{figure}

Though pairing in Hofstadter bands has been studied earlier using mean-field calculations with phenomenological attractive interactions \cite{Maska02, MoSudbo02, ZhaiOktel10, Iskin15a,JeonJain19, SohalFradkin20, SchirmerJain22}, no microscopic mechanism leading to this attraction has so far been proposed.
Here we show that HSCs can arise from repulsive interactions due to the competition of electronic orders near Van Hove singularities (VHS) that provide a logarithmic enhancement of the density of states (DOS) \cite{vanHove1953theoccurrence}.
Such a scenario of competing orders near VHSs underlies several proposed mechanisms of unconventional superconductivity through repulsive interactions, for example in cuprates \cite{schulz1987superconductivity,dzyaloshinskiui1987maximal,markiewicz1997survey},
doped graphene \cite{Nandkishore2012,WangFunctionalRG2012,KieselCompeting2012,GonzalezGraphene2008} and moir\'e graphene superlattices
\cite{isobe2018unconventional, Sherkunov-Betouras2018, Liu-Chiral2018, Kennes-Strongcorrelations-2018, You-Vishwanath-2019, Lin-Nandkishore-2020, hsu_topological_2020, classen_competing_2020, Chichinadze2020Nematicsuperconductivity}.
Furthermore, we go beyond mean-field by using an RG analysis \cite{shankar_renormalization-group_1994,polchinski1992effective,maiti_superconductivity_2013}, extending it to the new realm of Hofstadter electronic bands and uncovering a new pathway to realize reentrant superconductivity in moir\'e superlattices.
This approach allows us to treat the interplay of all logarithmically divergent instabilities on equal footing, and thus to additionally study the competition of superconductivity with charge/spin density wave (CDW/SDW) orders,
thus going beyond earlier mean-field studies of CDW and SDW in Hofstadter systems in \cite{MishraShankar16} and \cite{HongSalk99,HongSSLeeSalk00} respectively. The RG analysis also provides an alternative scenario to fractionalization in Hofstadter bands \cite{Kol1993, Wen91, MollerCooper09, Moller2015, ScaffidiSimon14, MotrukZaletelMong16,Lee2018Emergent,Sohal-2018,AndrewsMoller18, AndrewsSoluyanov20}.

Famously, the Hofstadter spectrum has a fractal nature characterized by a self-similar structure as a consequence of the MTG symmetries \cite{Hofstadter76,Thouless83}. We find that, remarkably, some aspects of this self-similarity are passed on to the RG flow and some of the resulting instabilities. First, we find that for all flux values, the RG flow has a particular fixed trajectory that is equivalent to multiple copies of the RG flow in the absence of the magnetic flux. We therefore refer to it as a self-similar fixed trajectory. As there are in principle many other fixed trajectories of the RG flow, it is not a given that the self-similar trajectory is reached for given set of interactions. Nevertheless we find that the self-similar trajectory is reached in our model in some cases. Second, in those cases the superconducting instability occurs by the same VHS mechanism as proposed in cuprates \cite{schulz1987superconductivity,dzyaloshinskiui1987maximal,markiewicz1997survey}, but the resulting order parameter also repeats in a self-similar fashion in the magnetic Brillouin zone. The self-similarity of the order parameter can be expressed as an emergent symmetry, which we call the self-similar symmetry, and which we show implies a highly non-local character of the order. These self-similarity properties illustrate how the MTG symmetries of the Hofstadter system can lead to novel phenomena via the VHS patch RG mechanism.

As a proof of principle, we work with the repulsive fermionic square lattice Hofstadter-Hubbard (HH) model with onsite interaction $U>0$ and flux $\Phi = 2\pi p/q$ that is a rational multiple of the flux quantum. Importantly, we focus on the regime $q \sim 1$ in which the Hofstadter bands have a bandwidth \(W\) comparable to that of the original band at zero field, which allows us to investigate electronic instabilities in a controlled weak coupling regime $U/W \ll 1$. While a hexagonal lattice would better approximate twisted bilayer graphene, which is the best studied superconducting moir\'e system, we establish our results on the square lattice since it still allows us to capture the essential correlation effects in Hofstadter bands while working with a simpler band structure, as shown in Sec.\ref{Sec:Model}.
Nevertheless, we stress that while the competition of electronic orders and their resulting instabilities can depend on the underlying lattice and interactions, the weak coupling RG framework developed here is of general applicability, and thus represents an important step towards the investigation of electronic instabilities in a wider class of two dimensional Hofstadter superlattices, including moir\'e graphene.

Additionally, the square HH model can more easily be realized in cold atom systems \cite{HofstatterDemlerLukin02, mueller2004artificial, Lewenstein07, Goldman10, gerbier2010gauge, Aidelsburger11, hauke2012non, Miyake13, AidelsburgerBloch13, celi2014synthetic}, though the focus in that field has been on bosonic \cite{Niemeyer99,Balents05,SorensenDemlerLukin05,HafeziDemlerLukin07,Oktel07,PowellDasSarma11} and time-reversal invariant fermionic \cite{OrthHofstetter13,Wang14,Peotta15,UmucalilarIskin17,Zeng19,Lin21} HH models (note that the latter coincides with the regular fermionic Hofstadter-Hubbard model at \(q=2\), i.e. at \(\pi\)-flux).
In addition, more recently single layer cuprates exhibiting critical temperatures close to their bulk values have been fabricated \cite{yu2019high}, opening an avenue for realizing twisted cuprate moir\'e systems with square lattices for which our model may be directly applicable. Such twisted heterostructures have recently been studied theoretically \cite{BilleKlemm01, can2021high, volkov2020magic, VolkovKimPixley21, SongVishwanath21}, with few-layer twisted interfaces already realized in experiment \cite{Zhu21,zhao2021emergent}. It remains to be seen whether Hofstadter physics can be realized in twisted cuprates, but, if it is, a reentrant HSC phase may be possible in this system.

The MTG symmetries play a key role in our analysis. In particular, they imply the presence of $2q$ VHSs per Hofstadter band, as shown in Fig.~\ref{fig:VHS}. The magnetic flux \(\Phi = 2\pi p/q\) thus acts as a knob controlling the number of VHSs in the system, which completely alters the RG flow and thus the possible instabilities of the system. This is well illustrated by the two distinct reentrant HSC phases that we find at \(\pi\)-flux (i.e. \(q=2\)) and at \(2\pi/3\)-flux (\(q=3\)). For the former case, we identify a nodal SC phase that respects all MTG symmetries as the winning RG instability at \(1/4\) and \(3/4\) lattice filling, even with perfect nesting in the competing SDW channel that is degenerate with the SC channel in the absence of the magnetic field \cite{Furukawa98}. For \(q=3\), we find that, at \(1/6\) and \(5/6\) lattice filling, SC and SDW are nearly degenerate when both are at perfect nesting, while SDW is strongly favored at half-filling. A small symmetry-allowed detuning from perfect nesting in the SDW therefore favors the pairing instability at \(1/6\) and \(5/6\) filling, which necessarily breaks a subset of the MTG symmetries \cite{ShafferWangSantos21}. We find that the resulting SC state is a fully gapped chiral topological phase with Chern number $\mathcal{C} = \pm6$ that preserves a \(\mathbb{Z}_{3}\) subgroup of the MTG. Surprisingly, this phase realized for \(q=3\) also possesses an emergent self-similarity symmetry due to the RG flow approaching a special self-similar fixed trajectory that exists as another consequence of the MTG symmetries. We identify this self-similar fixed trajectory for all \(q\), implying that long-range self-similar HSC states can be competing instabilities at flux values beyond those studied numerically in this work.

\section{Results}

\subsection{Hofstadter-Hubbard VHS Patch Model and Interactions}\label{Sec:Model}

We consider the nearest neighbor square lattice repulsive HH Hamiltonian
\begin{align}
\label{H0}
H &=-\sum_{\langle \mathbf{r} \mathbf{r}' \rangle \sigma} t\, e^{2 \pi i A_{\mathbf{r} \mathbf{r}'}} c_{\mathbf{r}\sigma}^\dagger c_{\mathbf{r}' \sigma} + h.c.
-\mu \sum_{\mathbf{r} \sigma} c_{\mathbf{r}\sigma}^\dagger c_{\mathbf{r} \sigma}+\nonumber\\
&+U\sum_{\mathbf{r}}
n_{\mathbf{r}\uparrow}\,n_{\mathbf{r} \downarrow}=H_{0} + H_{\textrm{int}}
\,,
\end{align}
with \(U>0\) where \(\mu\) is the chemical potential, $n_{\bs{r}\sigma}$ is the number operator with spin \(\sigma=\uparrow,\downarrow\) at site \(\mathbf{r}=(x, y)\in\mathbb{Z}^2\), and \(A_{\mathbf{r} \mathbf{r}'}=\int_{\mathbf{r}}^{ \mathbf{r}'}\mathbf{A}\cdot d\mathbf{r}/\Phi_0=\frac{p}{q}x(1-\delta_{yy'})\) corresponding to a flux per unit cell \(\Phi=2\pi p/q\) that is a rational multiple of the flux quantum \(\Phi_0\). We work in the Landau gauge with vector potential \(\mathbf{A}=xB\hat{y}\) and set the lattice constant \(a=1\). 
Note that while time-reversal symmetry (TRS) is broken by the orbital effect, we neglect the Zeeman splitting in our analysis and retain the full \(\mathrm{SU}(2)\) spin rotation symmetry, implying that our weak coupling analysis in applicable in the regime $E_{Z}~ \ll~ \Delta \ll W$, where $E_{Z}$ is the Zeeman splitting and $\Delta$ is the characteristic energy scale of the electron instabilities. The interesting regime of spin  polarized Hofstadter bands case merits a separate discussion which is outside the scope of this work.

In addition to TRS, the vector potential breaks the translation symmetry \(T_x\) along the \(x\) direction. However, the magnetic translation \(\hat{T}_x=T_x e^{2\pi i  aBy/\Phi_0}\) remains a symmetry of the Hamiltonian. \(\hat{T}_x\) and the unbroken translation \(T_y=\hat{T}_y\) along the \(y\) direction generate the non-Abelian magnetic translation group (MTG) satisfying \(\hat{T}_{x} \hat{T}_{y} = \omega_q^p \hat{T}_{y} \hat{T}_{x}\) with \(\omega_q=e^{ 2\pi i/q}\) being the \(q^{th}\) root of unity. 
Point group symmetries of the original Hamiltonian in the absence of the magnetic field similarly give rise to their magnetic versions with appropriate gauge transformations of the vector potential. For example, this includes inversion symmetry that guarantees a logarithmic pairing instability, as well as the original \(C_4\) symmetry that becomes \(\hat{C}_4=C_4 e^{-2\pi i xyB/\Phi_0}\), where the additional gauge transformation rotates \(\mathbf{A}=xB\hat{\mathbf{y}}\rightarrow yB\hat{\mathbf{x}}\). The \(\hat{C}_4\) symmetry will play a role below when we consider the instabilities of the \(\pi\)-flux Hofstadter Hamiltonian.

The commutation relations imply that \(\hat{T}_{x}^q\) and \(\hat{T}_{y}\) commute with each other and the Hamiltonian, effectively enlarging the unit cell along the \(x\) direction. 
We correspondingly define operators \(c_{\mathbf{R},s,\sigma}=c_{s\hat{\mathbf{x}}+\mathbf{R},\sigma}\) with \(s=0,\dots,q-1\) being the sub-lattice index defined modulo \(q\) and \(\mathbf{R}=(q j ,y)\) with \(j,y\in\mathbb{Z}\) labeling the extended unit lattice cites. Bloch's theorem then applies to these operators and we can write the Hofstadter Hamiltonian \(H_0\) in momentum space using
\(c_{\mathbf{k}s\sigma}=\frac{1}{\sqrt{N}}\sum_{\mathbf{R}}e^{-i\mathbf{k}\cdot (s\hat{\mathbf{x}}+\mathbf{R})}c_{s\hat{\mathbf{x}}+\mathbf{R},\sigma}\,,\) with \(N\) being the total number of unit cells and where the quasi-momentum \(\mathbf{k}\) is defined on the folded magnetic Brillouin zone (MBZ) $\mathbf{k} = (k_x, k_y) \in [-\pi/q, \pi/q) \times [-\pi,\pi)$. In this basis
\begin{align}
H_0=-&\sum_{\mathbf{k}s}(2t\cos(k_y+sQ)+\mu)c_{\mathbf{k}s\sigma}^\dagger c_{\mathbf{k}s\sigma}-\nonumber\\
 -&\sum_{\mathbf{k}\langle s s'\rangle}t e^{-ik_x(s-s')}c_{\mathbf{k}s\sigma}^\dagger c_{\mathbf{k}s'\sigma}
 \,,
\end{align}
and the magnetic translation symmetries act as \(\hat{T}_x c_{\mathbf{k}s\sigma}\hat{T}_x^\dagger=e^{-ik_x}c_{\mathbf{k}+\mathbf{Q},s+1,\sigma}\) and \(\hat{T}_y c_{\mathbf{k}s\sigma}\hat{T}_y^\dagger=e^{-ik_y}c_{\mathbf{k}s\sigma}\), with \(\mathbf{Q}=\frac{2\pi p}{q}\hat{\mathbf{y}}\).

The Hofstadter Hamiltonian \(H_0\) can then be diagonalized as \(H_0=\sum_{\mathbf{k}\alpha\sigma}\varepsilon_\alpha(\mathbf{k})d^{\dagger}_{\mathbf{k}\alpha\sigma}d_{\mathbf{k}\alpha\sigma}\) using a unitary transformation
\[d_{\mathbf{k}\alpha\sigma}=\sum_{s}\mathcal{U}^s_\alpha(\mathbf{k})c_{\mathbf{k}s} 
\,.
\label{Us}\]
Note that there is a large freedom in choosing the \(\mathrm{U}(1)\) phases in \(\mathcal{U}^s_\alpha(\mathbf{k})\). For concreteness, we take \(\mathcal{U}^{s+1}_\alpha(\mathbf{k+Q})=\mathcal{U}^s_\alpha(\mathbf{k})\),
which endures a canonical transformation under MTG for the band operators:
\(\hat{T}_x d_{\mathbf{k}\alpha\sigma}\hat{T}_x^\dagger=e^{-ik_x}d_{\mathbf{k+Q},\alpha,\sigma}\) and \(\hat{T}_y d_{\mathbf{k}\alpha\sigma}\hat{T}_y^\dagger=e^{-ik_y}d_{\mathbf{k}\alpha\sigma}\).
Furthermore, we fix the remaining gauge freedom by taking \(\mathcal{U}^1_\alpha(\mathbf{k}) \in \mathbb{R}\).
This choice makes it clear that the \(\hat{T}_x\) symmetry implies a $q$-fold degeneracy of each band, $\varepsilon_{\alpha}(\mathbf{k}) = \varepsilon_{\alpha}(\mathbf{k+Q})$. This means we can further restrict the quasi-momentum to a reduced magnetic Brillouin zone (rMBZ) \(\mathbf{p} = (p_x, p_y) \in [-\pi/q, \pi/q)^2\) and define \(d_{\mathbf{p}\ell\alpha\sigma}=d_{\mathbf{p}+\ell\mathbf{Q},\alpha\sigma}\) where \(\ell=0,\dots,q-1\) is the magnetic patch index defined modulo \(q\) as defined in \cite{ShafferWangSantos21} (see Fig. \ref{fig:VHS} (c-d)). We also refer to \(\ell\) as the rMBZ magnetic flavor index to distinguish it from the VHS indices introduced below.

Unlike earlier mean-field analyses of the fermionic HH model \cite{HongSalk99,HongSSLeeSalk00,Maska02, MoSudbo02, ZhaiOktel10, Iskin15a, MishraShankar16,SohalFradkin20} (see also \cite{ZhangFoster22} who studied SC in the related Aubry-Andr\'e model), here we investigate the instabilities driven by repulsive on-site interactions due to diverging DOS at the VHSs. In the square lattice Hofstadter model, the VHSs occur at electron fillings that are odd multiples of \(1/(2q)\) (counting spin), i.e. in half-filled Hofstadter bands. 
In each band there are a total of \(2q\) VHSs occurring at momenta \(\mathbf{K}_{\ell,\mathrm{v}}=\left((1+\mathrm{v})\frac{\pi}{q},\mathrm{v}\frac{\pi}{q}\right)+\ell\mathbf{Q}\) which we label with the VHS index \(\mathrm{v}=0, 1\) \cite{Naumis16}. Note that the VHSs thus lie at the images of the original VHSs of the square lattice at zero flux under a rescaling of the momentum by \(1/q\), which is a consequence of the self-similarity property of the Hofstadter spectrum \cite{Thouless83, wang_classification_2020} that also implies that the Fermi surfaces are composed of \(q\) touching squares for any Hofstadter band (see Fig. \ref{fig:VHS} (a), (c-d)).

Within this weak-coupling framework we can project the interactions onto the Fermi surfaces formed by a single band \(\alpha\), neglecting all other bands and expand the dispersions around patches centered at the VHS momenta \(\mathbf{K}_{\ell,\mathrm{v}}\), obtaining a VHS patch model that we will analyse in in Sec. \ref{Sec:RG} using fermionic RG \cite{schulz1987superconductivity,dzyaloshinskiui1987maximal,maiti_superconductivity_2013,Nandkishore2012,Lin-Nandkishore-2019}. 
We thus define the patch model operators \(d_{\mathbf{p}\ell \mathrm{v}\alpha\sigma} = d_{\mathbf{p}+\mathbf{K}_{\ell,\mathrm{v}}, \alpha,\sigma}\) with \(\mathbf{p}\) a small momentum expanded around a patch centered at \(\mathbf{K}_{\ell,\mathrm{v}}\). 
For bookkeeping purposes, we include a redundancy in our description and allow \(\mathrm{v}=-1\) with the identification \(\mathbf{K}_{\ell,-1}\equiv\mathbf{K}_{\ell-1,1}\) which makes the VHS and magnetic flavor indices conserved quantities in Feynman diagrams we use in the RG analysis.

We then project \(H_{int}\) in Eq. (\ref{H0}) onto the patches obtaining an effective interaction Hamiltonian
\begin{widetext}
\[H_{int}\rightarrow H_{int}^{(\alpha)}=\frac{1}{2}\sum_{\substack{\ell m n\\ \mathrm{u} \mathrm{v} \mathrm{w}, \sigma\sigma'}} g^{(\alpha;\,\ell,\mathrm{u})}_{m,\mathrm{v};n,\mathrm{w}} d^\dagger_{\ell+n,\mathrm{u}+\mathrm{w},\alpha,\sigma}d^\dagger_{-n,-\mathrm{w},\alpha,\sigma'}d_{-m,-\mathrm{v},\alpha,\sigma'}d_{\ell+m,\mathrm{u}+\mathrm{v},\alpha,\sigma}\label{Hint}
\,,
\]
where \(\ell,m,n=0,\dots q-1\) are magnetic flavor indices, \(\mathrm{u},\mathrm{v},\mathrm{w}=0,\pm 1\) are the VHS indices, and
\[g^{(\alpha;\,\ell,\mathrm{u})}_{m,\mathrm{v};n,\mathrm{w}}=U\sum_s \mathcal{U}_\alpha^{s}\left(\mathbf{K}_{\ell+n,\mathrm{u}+\mathrm{w}}\right)\mathcal{U}_\alpha^{s}\left(\mathbf{K}_{-n,-\mathrm{w}}\right)\mathcal{U}_\alpha^{s*}\left(\mathbf{K}_{-m,-\mathrm{v}}\right)\mathcal{U}_\alpha^{s*}\left(\mathbf{K}_{\ell+m,\mathrm{v}}\right)\]
\end{widetext}
are the coupling constants corresponding to interactions between electrons with total momenta \(\mathrm{u}(\pi,\pi)/q+\ell\mathbf{Q}\), dressed by form factors originating from the unitary transformation Eq. (\ref{Us}). Henceforth we will consider a fixed band \(\alpha\) and drop the index where it is clear from context.

As there are \(2q\) VHSs, the number of coupling constants grows quickly with \(q\), which manifests the MTG action in momentum space.
Taking hermiticity, MTG symmetries, and redundancy of the VHS indices into account, there are a total of \(\mathcal{O}(q^2)\) independent coupling constants that can be classified into five processes according to their VHS indices:
\begin{align}\label{gs}
    g^{(\ell)1}_{mn}&=g^{(\ell,0)}_{m,0;n,0} 
    &g^{(\ell)1'}_{mn}=g^{(\ell,0)}_{m,1;n,1}\nonumber\\
    g^{(\ell)2}_{mn}&=g^{(\ell,1)}_{m,0;n,0}\nonumber
    &g^{(\ell)3}_{mn}=g^{(\ell,1)}_{m,0;n,-1}\nonumber\\
    g^{(\ell)4}_{mn}&=g^{(\ell,0)}_{m,0;n,1}
    \,,
\end{align}
as shown in Fig. \ref{fig:VHS} (b). \(g_1\) and \(g_{1'}\) correspond to intra-patch processes for \(\mathrm{v}=0,\pm1\) VHSs respectively, \(g_2\) (\(g_3\)) is an inter-patch process without (with) exchange, and \(g_4\) is a pair-hopping process. Note that in the absence of TRS, not all coupling constants are necessarily real. In addition to relations imposed by hermiticity, the coupling constants also satisfy \(g^{(\ell)j}_{mn}=g^{(\ell+2),j}_{m-1,n-1}\) as a consequence of the  MTG action on the fermion operators \(\hat{T}_x d_{\mathbf{p}\ell\mathrm{v}\sigma}\hat{T}_x^\dagger=e^{-ip_x}d_{\mathbf{p},\ell+1,\mathrm{v} \sigma}\). In particular, for odd \(q\) all coupling constants can be expressed in terms of \(g^{(0)j}_{mn}\). For even \(q\), all coupling constants can be expressed in terms of either \(g^{(0)j}_{mn}\) or \(g^{(1)j}_{mn}\), with an additional relation \(g^{(\ell)j}_{mn}=g^{(\ell)j}_{m-q/2,n-q/2}\); see the supplementary material (SM) for further relations satisfied by the coupling constants.
By virtue of the MTG symmetries the coupling constants Eq.\eqref{gs} thus organize into processes that resemble those in zero magnetic field. As we will see in Sec. \ref{Sec:RG}, this has the important implication that the RG equations exhibit a degree of self-similarity that we will elucidate below.

\subsection{RG Analysis}\label{Sec:RG}

In this section we extend the RG analysis developed previously for the half-filled square lattice \cite{schulz1987superconductivity,dzyaloshinskiui1987maximal,Furukawa98} and the quarter-filled hexagonal lattice \cite{Nandkishore2012} with 2 and 3 VHSs, respectively, to the patch model with \(2q\) VHSs presented above. The competing instability channels fall into two classes: particle-particle channels with momentum transfers \(\ell\mathbf{Q}\); and particle-hole channels with momentum transfers \((\pi,\pi)/q+\ell\mathbf{Q}\). Due to the MTG symmetries, all the susceptibilities are independent of the magnetic flavor indices \(\ell\), and the two relevant susceptibilities are \(\Pi_{pp}(\ell\mathbf{Q})\approx\nu_0\ln^2\Lambda/T\) and \(\Pi_{ph}((\pi,\pi)/q+\ell\mathbf{Q})\approx d_{ph}\nu_0\ln^2\Lambda/T\) where \(\Lambda\) is the high energy cutoff, \(T\) is the temperature and \(\nu_0\ln\Lambda/E\) is the DOS at energy \(E\) above the VHS \cite{Furukawa98, Nandkishore2012,Lin-Nandkishore-2020}. Here we introduce the standard phenomenological detuning parameter \(d_{ph}=\Pi_{ph}/\Pi_{pp}\in[0,1]\) to account for possibly imperfect nesting in the particle-hole channels due to additional symmetry allowed terms that break particle-hole symmetry at half-filling or for chemical potentials slightly away from the VHSs \cite{Furukawa98}.

Performing the one loop RG procedure (see Supplementary Material) and keeping only the most diverging \(\ln^2\) corrections, we obtain the flow equations for the coupling constants. The full expressions are given in the SM and can be represented symbolically as:
\begin{widetext}
\begin{align}\label{Gflow}
\dot{g}^{(\ell)1}_{mn}&=-g^{(\ell)1}_{mk}g^{(\ell)1}_{kn}-g^{(\ell)4}_{mk}g^{(\ell)4*}_{nk}\\
\dot{g}^{(\ell)1'}_{mn}&=-g^{(\ell)1'}_{mk}g^{(\ell)1'}_{kn}-g^{(\ell)4*}_{km}g^{(\ell)4}_{kn}\nonumber\\
\dot{g}^{(\ell)2}_{mn}&=d_{ph}\left(g^{(\ell+n-k)2}_{mk}g^{(\ell+m-k)2}_{kn}+g^{(\ell+n-k)4*}_{k,m-1}g^{(\ell+m-k)4}_{k,n-1}\right)\nonumber\\
\dot{g}^{(\ell)3}_{mn}&=2d_{ph}g^{(\ell+m+n+k)3}_{-n-k,-m-k}\left(g^{(k)2}_{m,-n-k}-g^{(k)3}_{mn}\right)+d_{ph}g^{(\ell+m+n+k)4}_{-n-k,-m-k}\left(g^{(k)4*}_{n,-m-k}-2g^{(k)4*}_{n,m-1}\right)+d_{ph}g^{(\ell+m+n+k)4}_{-n-k,-n-\ell-1}g^{(k)4*}_{n,m-1}\nonumber\\
\dot{g}^{(\ell)4}_{mn}&=-g^{(\ell)1}_{mk}g^{(\ell)4}_{kn}-g^{(\ell)4}_{mk}g^{(\ell)1'}_{kn}+d_{ph}\left(g^{(\ell+n-k)2}_{k-\ell-m-n,-\ell-n}g^{(\ell+m-k)4}_{kn}+g^{(\ell+n-k)4}_{mk}g^{(\ell+m-k-1)2}_{k+1,n+1}\right)-\nonumber\\
&+d_{ph}g^{(\ell+m+n+k)4}_{-n-k,-m-k}\left(g^{(k-1)2}_{-m-k+1,n+1}-2g^{(k-1)3}_{1-k-m,-k-n}\right)+d_{ph}g^{(\ell+m+n+k)4}_{-n-k,-n-\ell-1}g^{(k-1)3}_{1-k-m,-k-n}+\nonumber\\
&+d_{ph}g^{(\ell+m+n+k)3}_{-n-k,-m-k}g^{(k)4}_{-m-k,n}+d_{ph}\left(g^{(\ell+m+n+k),2}_{-n-k,-n-\ell}-2g^{(\ell+m+n+k)3}_{-n-k,-m-k}\right)g^{(k)4}_{mn}\nonumber
\,.
\end{align}
\end{widetext}
The dot denotes the derivative with respect to the running RG time \(t=\Pi_{pp}(E)=\nu_0\ln^2\Lambda/E\), with high energy modes integrated above the energy scale \(E\).
For \(q=1\), i.e. zero flux, Eq.\ref{Gflow}
reduces to the standard result for the half-filled square lattice in \cite{schulz1987superconductivity} (in this case \(g_{1'}=g_1\) by \(C_4\) symmetry that is otherwise broken at non-zero flux).  Recall that in that case repulsive Hubbard interactions lead to degenerate d-wave SC and SDW orders, with the degeneracy being lifted either by imperfect nesting or subleading terms in RG \cite{schulz1987superconductivity}.

For $q \neq 1$ the RG equations \ref{Gflow} in principle allow for a large number of fixed trajectories that characterize the instabilities of the Hofstadter metal.
Despite the apparent complexity of these equations, by grouping the coupling constants into the \(g^{(\ell)1}_{mn},\, g^{(\ell)1'}_{mn},\, g^{(\ell)2}_{mn},\, g^{(\ell)3}_{mn},\, g^{(\ell)4}_{mn}\) processes according to the VHS patch index structure, we see that the form of these equations is similar to the RG equations in the absence of the magnetic flux, i.e. for \(q=1\). In particular,
it can be verified that they admit a fixed point trajectory characterized by \(g^{(\ell)j}_{mn}=g_j/\sqrt{q}\),
i.e. coupling constants independent of the magnetic flavor indices and depending only on the VHS patch indices. Plugging this ansatz into the RG equations \ref{Gflow}, one can directly verify that \(g_j\) 
satisfy the same set of equations as for \(q=1\). As a consequence, the resulting instability and its properties such as critical exponents are identical to those in the \(q=1\) system, and  we thus refer to such solutions as self-similar fixed trajectories.
We note that this property extends to all classes of Hofstadter systems beyond the one studied here, given that the MTG symmetries are preserved and provided the weak coupling regime is valid. 

Though we show that the self-similar solutions exist, at the beginning of the RG flow local interactions produce bare couplings $g^{(\ell)j}_{mn}(t =0)$ that in general have a dependence on the magnetic flavour.
It is therefore not a given that the self-similar trajectory is reached by the RG flow, and we find for example that it is not reached with repulsive Hubbard interactions for \(q=2\). We do find, on the other hand, that with the same repulsive Hubbard interactions in top and bottom Hofstadter bands for \(q=3\), the coupling constants do tend asymptotically to this self-similar fixed trajectory.
The existence of such non-trivial self-similar behavior in the RG equations and their relation to unconventional SC is one of the main results of this work.

\subsection{Vertices and Susceptibilities}\label{Sec:RG:B}

\begin{figure*}[t]
\centering
\includegraphics[width=0.99\textwidth]{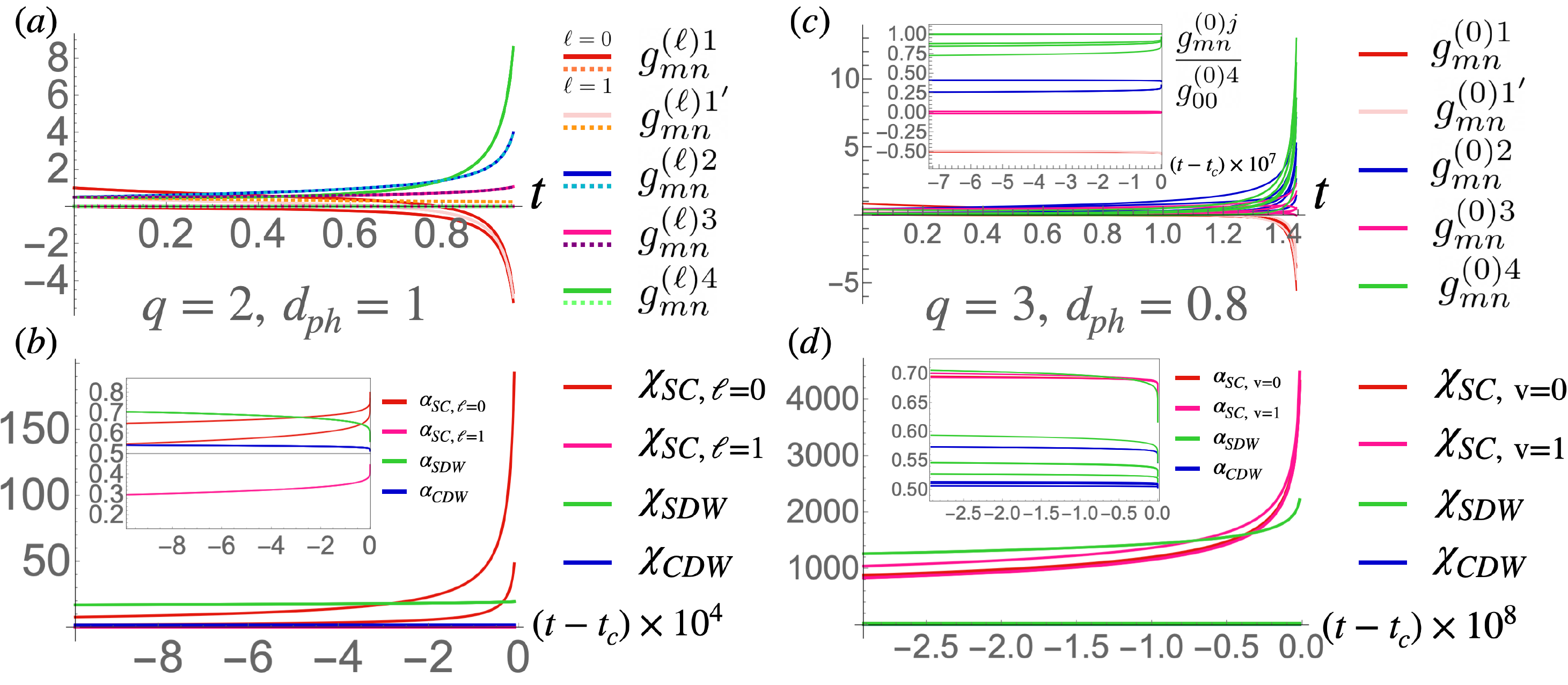}
\caption{RG flow of coupling constants and susceptibilities. The flow of the coupling constants \(g_{mn}^{(\ell)j}\) with \(\ell=0,1\) (solid and dashed lines respectively), \(j=1,1',2,3,4\) (red, light red, blue, magenta and green respectively), and \(m,n=0,\dots,q-1\) are shown for (a) \(q=2\) at \(1/4\) filling at perfect nesting \(d_{ph}=1\); and for (c) \(q=3\) at \(1/6\) filling with \(d_{ph}=0.8\) (\(U=1\) in arbitrary units in all plots). The instability occurs at \(t_c=0.98\) and \(t_c=1.46\) for \(q=2\) and \(3\) respectively. The flows for \(q=2\) and \(3\) are otherwise qualitatively similar, and both are similar to the flow in the absence of the magnetic field: note that while all coupling constants are initially positive or vanishing, \(g_{mn}^{(0)1}\) and \(g_{mn}^{(0)1'}\) eventually change sign, leading to effective attraction in the pairing channel. The inset in (c) shows the \(q=3\) flow normalized by \(g^{(0)4}_{00}\) which shows that the self-similar fixed trajectory \(g_{mn}^{(\ell)j}=g_j/\sqrt{q}\) is reached at the end of the flow, as indicated by curves of the same color approaching the same value (we also find \(g_1=g_{1'}\)).
(b) \(q=2\) RG flow of the susceptibilities \(\chi_I\) with \(I\) corresponding to SC with Cooper pairs with zero momentum (\(\ell=0\), red) or momentum \(\mathbf{Q}=\frac{2\pi p}{q}\hat{\mathbf{y}}\) (\(\ell=1\), magenta), SDW (green) or CDW (blue). Initially \(\chi_{SDW}\) is the fastest growing susceptibility, but eventually The \(\ell=0\) SC susceptibility takes over. The inset shows the corresponding critical exponents \(\alpha_I=\left(1-\log_{t_c-t}\chi_I\right)/2\) for the same range of RG times \(t\). The largest exponent at the end of the flow is \(\alpha_{SC,\ell=0}(t_c)\approx0.73\). (d) Shows that analogous plots for \(q=3\), but in this case the \(\ell=0\) and \(1\) SC channels are degenerate so only the former is plotted; in this case red and magenta colors indicate the suscpetibilities at \(\mathrm{v}=0\) and \(1\) VHS points respectively, which contribute to the same SC channel. The largest exponent at the end of the flow is \(\alpha_{SC}(t_c)\approx0.65\). Color online.
}
\label{fig:RGflow}
\end{figure*}

Under the RG flow some of the coupling constants diverge at some finite RG time \(t_c\), indicating an instability of the Fermi surface (see Fig. \ref{fig:RGflow} (a) and (c)). To study these instabilities, we introduce the following test vertices and study their flow:
\begin{eqnarray}\label{eq:Vertices}
H_{SC}&=&\Delta^{(\ell)}_{m;\mathrm{v}}i\sigma^y_{\sigma\sigma'}d^\dagger_{\ell+m,\mathrm{v},\sigma}d^\dagger_{-m,-\mathrm{v},\sigma'}+h.c.\\
H_{CDW}&=&\rho^{[\ell]}_{m;\mathrm{v}}d^\dagger_{\ell+m,-\mathrm{v},\sigma}d_{m,1+\mathrm{v},\sigma}\nonumber\\
H_{SDW}&=&\mathbf{M}^{[\ell]}_{m;\mathrm{v}}\cdot\boldsymbol{\sigma}_{\sigma\sigma'}d^\dagger_{\ell+m,-\mathrm{v},\sigma}d_{m,1+\mathrm{v},\sigma'}\nonumber
\end{eqnarray}
with summation over the indices implied. \(\Delta^{(\ell)}_{m;\mathrm{v}}\), \(\rho^{[\ell]}_{m;\mathrm{v}}\), and \(\mathbf{M}^{[\ell]}_{m;\mathrm{v}}\) are the SC, CDW, and SDW order parameters respectively with momentum transfers \(\ell\mathbf{Q}\) for SC and \((\pi,\pi)/q+\ell\mathbf{Q}\) for the density waves. Note that by hermiticity, \(\rho^{[\ell]}_{m;0}=\rho^{[1-\ell]*}_{m+\ell;1}\), and similarly \(M^{[\ell,j]}_{m;0}=M^{[1-\ell,j]*}_{m+\ell;1}\), which therefore belong to the same channel in the RG flow.
As shown in the SM, the CDW and SDW flow equations decouple into \(q^2\) channels each: \(\tilde{\rho}^{[\ell]}_{k;\mathrm{v}}=\sum_{m}\omega_q^{mk}\rho^{[\ell]}_{m;\mathrm{v}}\) for CDW and similarly for SDW. This is consistent with the fact that being charge-0 instabilities, CDW and SDW transform as 1D irreducible representations (irreps) of the MTG. Similar CDW orders have been found numerically in a real-space mean field analysis of a spinless fermionic HH model on a hexagonal lattice, though their MTG irreps have not been established \cite{MishraShankar16}. Note also that we do not find the previously proposed \((\pi,\pi)\) SDW as a potential instability \cite{HongSalk99,HongSSLeeSalk00}.

As shown in \cite{ShafferWangSantos21}, unlike the charge-0 orders, the charge-2 SC orders transform according to \(q\) or \(q/2\) irreps of the MTG for odd and even \(q\) respectively. This means that in general there are \(q\) or \(q/2\) degenerate flows in the SC channels corresponding to each choice of \(\ell\) in Eq. (\ref{eq:Vertices}), with even and odd \(\ell\) being non-degenerate for even \(q\). In addition, when \(q\) is even the SC order parameter decouples into \(\Delta^{(\ell,\pm)}_{m;\mathrm{v}}=\Delta^{(\ell)}_{m;\mathrm{v}}\pm\Delta^{(\ell)}_{m+q/2;\mathrm{v}}\), with \(\Delta^{(\ell,+)}_{m;\mathrm{v}}\) (even under \(\hat{T}_x^{q/2}\)) and \(\Delta^{(\ell,-)}_{m;\mathrm{v}}\) (odd under \(\hat{T}_x^{q/2}\)) flowing independently, consistent with the fact that there are four \(q/2\)-dimensional irreps in this case \cite{ShafferWangSantos21}.

The vertex RG flow equation are shown schematically in Fig. \ref{fig:RGflowVertexEq}. Observe that the coupling constants \(g^{(\ell)1}_{mn}\) and \(g^{(\ell)1'}_{mn}\) only contribute to the flow of the SC vertices, \(g^{(\ell)3}_{mn}\) only contributes to the CDW flow, while \(g^{(\ell)2}_{mn}\) contributes to the flow of both CDW and SDW vertices.  \(g^{(\ell)4}_{mn}\), on the other hand, contributes to all the channels, with a similar structure in the flow of the coupling constants themselves in Eq. (\ref{Gflow}). SC is thus generally favored by negative  \(g^{(\ell)1}_{mn}\) and \(g^{(\ell)1'}_{mn}\). Importantly, although these are positive initially in the repulsive Hubbard model, they can potentially change sign due to the \(|g_4|^2\) term in their flow in Eq. (\ref{Gflow}). We indeed find this to be the case for \(q=2\) and \(q=3\), as shown in Figs. \ref{fig:RGflow} (a) and (c).

\begin{figure}[t]
\centering
\includegraphics[width=0.45\textwidth]{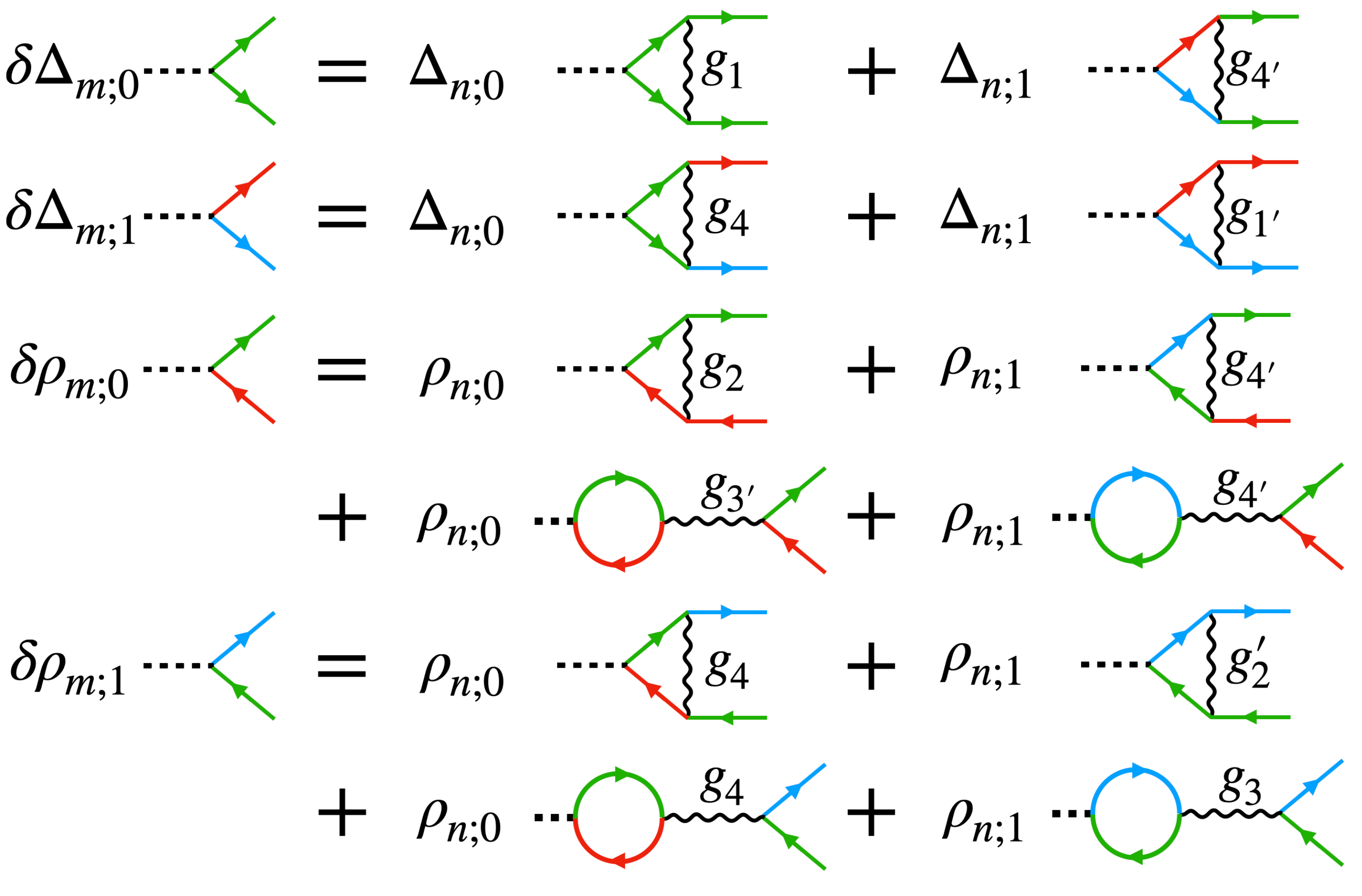}
\caption{The 1 loop Feynman diagrams contributing to the SC and CDW vertex corrections. The VHS index structure is shown (green for \(\mathrm{v}=0\), red/blue for \(\mathrm{v}=\pm1\) respectively). SDW Diagrams for SDW vertex corrections are the same as the CDW diagrams with \(M\) instead of \(\rho\), except for the loop diagrams in the second lines that vanish for SDW vertices. The propagators additionally carry
the magnetic flavor indices, not shown in the figure (these can be found in the SM).}
\label{fig:RGflowVertexEq}
\end{figure}

In order to establish which instability actually takes place, we additionally consider the flow of the susceptibilities \(\chi_I\) where \(I=\Delta^{(\ell)}_{m;\mathrm{v}},\, \tilde{\rho}^{[\ell]}_{k;\mathrm{v}},\, \tilde{M}^{[\ell]}_{k;\mathrm{v}}\) corresponding to the instability. The susceptibilities flow as \(\dot{\chi}_I=d_I \left|I(t)/I(0)\right|^2\) with \(d_\Delta=1\) and else \(d_I=d_{ph}\) \cite{maiti_superconductivity_2013,Lin-Nandkishore-2020}. The leading instability corresponds to \(\chi_I\) that diverges most strongly at \(t_c\), around which they generally diverge as \(\chi_I(t)\propto(t_c-t)^{1-2\alpha_I}\)
with some critical exponent \(\alpha_I\) that needs to be larger than \(1/2\) for the instability to occur. A representative flow for \(q=2\) (\(q=3\)) is shown in Fig. \ref{fig:RGflow} (b) (Fig. \ref{fig:RGflow} (d)). 
In that case we find that SC is the leading instability with critical exponent \(\alpha_{SC}\approx0.73\) (\(\alpha_{SC}\approx0.65\)). The exponent is computed as the final value of \(\alpha_I(t)=\left(1-\log_{t_c-t}\chi_I\right)/2\), with a representative plot of \(\alpha_I(t)\) shown in the inserts of Fig. \ref{fig:RGflow} (b) and (d).

\subsection{Resulting Instabilities}\label{Sec:RG:C}

We studied the RG equations for \(p/q=1/2, 1/3\) and \(2/3\), with the results summarized in Table \ref{table:Results}. For zero flux, \(q=1\) and we recover the results for the square lattice with repulsive interactions  at half-filling \cite{schulz1987superconductivity, dzyaloshinskiui1987maximal, Furukawa98}. In that case it is found that a d-wave SC and SDW are degenerate within one loop at perfect nesting in the SDW channel, \(d_{ph}=1\), with d-wave SC winning for any \(d_{ph}<1\). This has been interpreted as SDW fluctuations leading to an effective attraction in the d-wave SC channel.
In the flow of the coupling constants this is reflected in the initial growth of \(g_4\) that pushes \(g_1\) to become negative and eventually diverge. We observe a qualitatively similar RG flow for \(q=2\) and \(3\) as seen in Fig. \ref{fig:RGflow} (a) and (c). We now analyse the resulting instabilities for those cases.

\begin{figure*}
\centering
\includegraphics[width=0.99\textwidth]{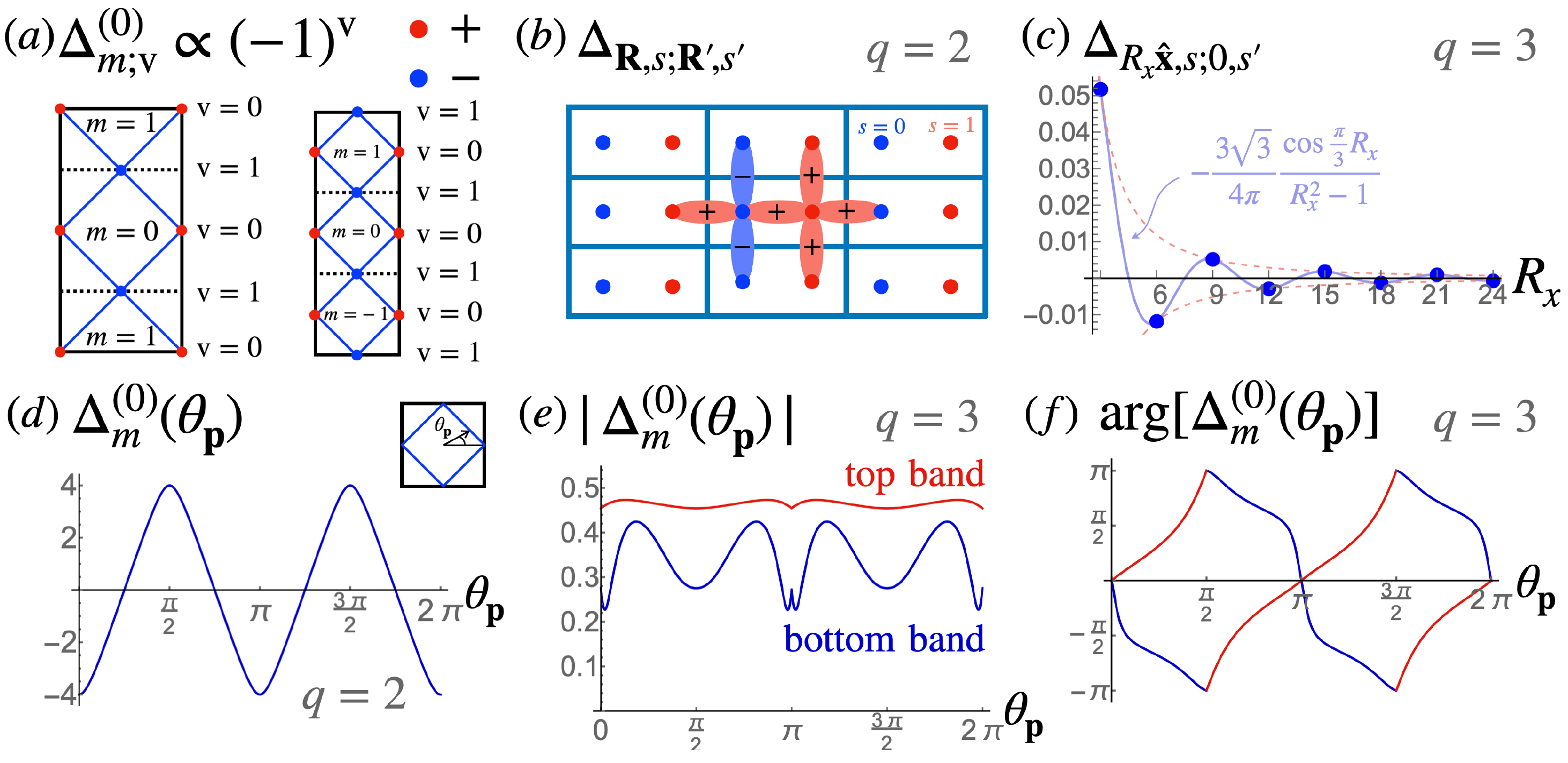}
\caption{Properties of the gap functions. (a) Gap functions at the VHS obtained from the RG analysis for \(q=2\) at perfect nesting (left) and for \(q=3\) at \(d_{ph}=0.8\) in the top and bottom Hofstadter bands (right). In both cases the gap function changes sign between the two VHSs \(\mathrm{v}=0,1\). Here we focus on pairing with zero total momentum \(\ell=0\), with pairings for \(\ell\neq0\) determined by MTG symmetries. (b) Real space structure of the gap function for \(q=2\) even under \(\hat{T}_1\) and \(\hat{T}_2\) and odd under \(\hat{C}_4\), shown within a single magnetic unit cell (the pattern repeats in all cells). (c) Profile of the gap function \(\Delta_{R_x\hat{\mathbf{x}},s;0,s'}\) for \(q=3\) as a function of the horizontal magnetic unit cell separation \(R_x\) between Cooper pairs (with lattice constant \(a=1\)). Note that the gap function oscillates between each unit cell and decays as \(1/R_x^2\) at long distances. See the SM for more details. (d) The projection onto the Fermi surface of the gap function for \(q=2\) shown in (b) as a function of the angle \(\theta_{\mathbf{p}}\) along the Fermi surface within the rMBZ (note that \(\Delta_m^{(\ell)}\) are equal within each patch \(m\)). Note that the gap crosses zero, indicating nodes in the fermionic spectrum. (e-f) The projection onto the Fermi surface of the model gap function for \(q=3\) for the top (red) and bottom (blue) bands that agrees with the gap function found in the RG analysis (color online). Note that the magnitude of the gap function never vanishes as shown in (e), implying that the fermionic spectrum is fully gapped (the sharp features at \(\theta_\mathbf{p}=0,\pi\) are due to the corners of the Fermi surface). The phase of the projected gap functions, however, winds by \(\pm 4\pi\) around the Fermi surface in the top and bottom bands respectively, as shown in (f), implying each \(\Delta_m^{(\ell)}\) contributes \(\pm 2\) to the Chern number. Plots (c-e) are given in arbitrary units as the magnitude of the gap function is not determined within the weak coupling theory.}
\label{fig:GapFunctions}
\end{figure*}

Unlike the \(q=1\) case, for \(q=2\) we find that an SC instability occurs already at perfect nesting in both Hofstadter bands (with critical exponent \(\alpha_{SC}\approx0.77>0.5\)). As shown in \cite{ShafferWangSantos21}, in this case the SC orders belong to one of four one dimensional irreducible representations (irreps) of the MTG determined by the gap function being even or odd under \(\hat{T}_y\) and \(\hat{T}_x\). The SC phase that wins in our RG calculation is even under both \(\hat{T}_y\) and \(\hat{T}_x\), which corresponds to \(\Delta^{(1)}_{m;\mathrm{v}}=0\) and \(\Delta^{(0)}_{0;\mathrm{v}}=\Delta^{(0)}_{1;\mathrm{v}}\) respectively. Furthermore, we find that \(\Delta^{(0)}_{m;0}=-\Delta^{(0)}_{m;1}\) (see Fig. \ref{fig:GapFunctions} (a)), which implies that the gap function is odd under the magnetic \(\hat{C}_4\) rotation. We note that this is an exceptional case, as for \(q>2\) the gap function necessarily breaks one of the MTG symmetries, and must either break the \(\hat{C}_4\) symmetry or break the remaining MTG symmetry \cite{ShafferWangSantos21}. Only when the gap function is both even or both odd under \(\hat{T}_x\) and \(\hat{T}_y\), as in the present case, can it also have a well-defined \(\hat{C}_4\) symmetry.

The RG analysis only determines the gap function at the VHSs, so our approach does not determine the gap function \(\Delta^{(0)}_{m}(\mathbf{p})\) along the entire Fermi surface (with \(\Delta^{(0)}_{m;\mathrm{v}}=\Delta^{(0)}_{m}(\mathbf{K}_{0,\mathrm{v}})\)). In principle, this issue can be addressed by using a method that extends the RG calculation to the entire Fermi surface, for example a function RG calculation or a two-step RG approach combined with a random phase approximation type calculation (see e.g. \cite{RaghuKivelsonScalapino2010}); however, this is an involved computation that is beyond the scope of this work.
For \(q=2\) we can circumvent this issue by using the fact that a \(\hat{T}_x\) and \(\hat{T}_y\) symmetric gap function odd under \(\hat{C}_4\), which we refer to as a \(d\)-wave gap function, has a unique nearest-neighbor form in the \(c_{\mathbf{k}s}\) basis, namely:
\[\Delta^{(d)}_{ss'}(\mathbf{k})=\Delta_0\left(\cos k_x \sigma^x_{ss'}-\cos k_y \sigma^z_{ss'}\right)\]
The anti-symmetry of this order parameter under \(\hat{C}_4\) symmetry can be checked directly by using
\[\hat{C}_4 c_{\mathbf{p}+\ell\mathbf{Q},s\sigma}\hat{C}_4^\dagger=\frac{1}{q}\sum_{s'\ell'}\omega_q^{-p(ss'+\ell s'+\ell's)}c_{\bar{\mathbf{p}}+\ell'\mathbf{Q},s'\sigma}\label{C4}\]
where \(\bar{\mathbf{p}}=(-p_y,p_x)\) (one can also check that the RHS in Eq. (\ref{C4}) is an eigenstate of \(\hat{T}_x\)).
Fig. \ref{fig:GapFunctions} (b) shows the corresponding gap function \(\Delta_{\mathbf{R},s;\mathbf{R}',s'}\) in real-space in the \(c_{\mathbf{R}s}\) basis. The gap function \(\Delta^{(0)}_{m}(\mathbf{p})\) is then obtained by projecting \(\Delta^{(d)}_{ss'}(\mathbf{k})\) onto the band basis \(d_{\mathbf{k}\alpha}\) (see SM for details). Importantly, the resulting gap is nodal (see Fig. \ref{fig:GapFunctions} (d)).

\begin{figure*}[t]
\centering
\includegraphics[width=0.99\textwidth]{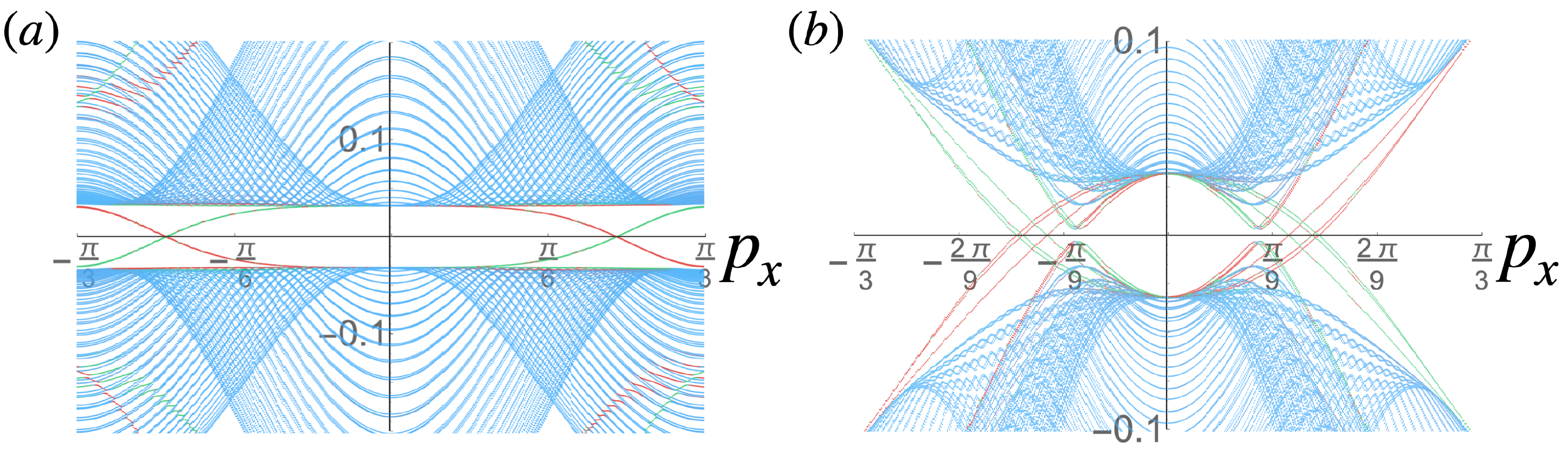}
\caption{Edge modes in the BdG spectrum of the Hofstadter SC for \(q=3\). Cylindrical boundary conditions open in the \(y\) direction were taken for the self-similar \(\hat{T}_x\hat{T}_y\) symmetric gap function Eq. (\ref{Delta3}) at (a) \(5/6\) and (b) \(1/6\) filling (chemical potential \(\mu=\pm 2.44\) respectively, with \(t=1\) and \(\Delta_0=0.02\) in (a) and \(0.2\) in (b), taking 100 extended unit cells along the \(y\) direction; see \hyperref[Methods]{Methods} for details of the BdG Hamiltonian and SM for more details of the calculation). The spectra are colored according to a weighted inverse participation ratio with green and red indicating states localized to the top and bottom edges of the cylinder respectively, while blue indicates bulk states. In (a) there are pairs of crossing edge modes at zero energy around \(p_x=\pm 2\pi/9\), and we find that each is three-fold degenerate, corresponding to Chern number \(6\). In (b) there are six right-moving and six left-moving zero energy edge modes are located around \(p_x=\pm \pi/6\), giving a total Chern number \(-6\). Observe that the edge modes of the same color move in opposite direction in (a) and  (b). Localized edge modes at higher energies that do not cross zero energy are the normal state edge modes that connect to higher energy Hofstadter bands not shown in the figure.}
\label{fig:EdgeModes}
\end{figure*}

A gap function of this form has been considered as a toy model of a nodal \(d\)-wave superconductor in a magnetic field in \cite{MoritaHatsugai01}, but without a microscopic justification or a consideration of its symmetries presented here (indeed, the gap function in that model does not transform as a proper irreducible representation of the MTG for \(q>2\)). The \(\pi\)-flux superconductor on a square lattice has also previously been studied using quantum Monte Carlo at half-filling, i.e. at the Dirac nodes of the normal spectrum, where a so-called \(ds^*\) SC phase has been found \cite{Guo2018Unconventional}. The corresponding gap function, which we simply refer to as \(s\)-wave, has the form \(\Delta^{(s)}_{ss'}(\mathbf{k})=\Delta_0\left(\cos k_x \sigma^x_{ss'}+\cos k_y \sigma^z_{ss'}\right)\) and we find that it is precisely the \(\hat{T}_x\), \(\hat{T}_y\) symmetric gap that is even under \(\hat{C}_4\), and therefore distinct from the phase we find in RG at VHS fillings.

\begin{table*}[htp]
\begin{center}
\begin{tabular}{ c c c ccc c ccc}
\hline\hline\noalign{\smallskip}
\multirow{2}{*}{}  &  \(q=2\), \(d_{ph}=1\) & \multirow{2}{*}{}  & \multicolumn{3}{c}{\(q=3\), \(d_{ph}=1\)} &\multirow{2}{*}{}  & \multicolumn{3}{c}{\(q=3\), \(d_{ph}=0.8\)} \\
\noalign{\smallskip}\cline{2-2} \cline{4-6} \cline{8-10}\noalign{\smallskip}
 & \(1/4,3/4\) & & \(1/6\) & \(1/2\) & \(5/6\) & & \(1/6\) & \(1/2\) & \(5/6\)\\
\hline\noalign{\smallskip}
\(I\) & \(\Delta^{(d)}\) & & \quad \(\tilde{M}^{[0]}_0,\, \tilde{M}^{[1]}_0\) \quad & \quad \(\tilde{M}^{[0]}_2,\, \tilde{M}^{[1]}_1\) \quad& \quad \(\tilde{M}^{[0]}_0,\, \tilde{M}^{[1]}_0\) \quad& &\(\Delta\)&  \(\tilde{M}^{[0]}_2,\, \tilde{M}^{[1]}_1\)& \(\Delta\)\\
\(\alpha_I\) & 0.77 & & 0.68 & 0.71 & 0.68 & & 0.65 & 0.65 & 0.65\\
\multirow{2}{*}{\(Symmetries\)}\quad & \multirow{2}{*}{\quad\(\hat{T}_x\), \(\hat{T}_y\), \(\hat{C}_4(-1)\)} &\quad\quad & \quad\(\hat{T}_x(\omega_3^{-1/2})\) \quad & \quad\(\hat{T}_x(\omega_3^{-1/2})\)\quad & \quad\(\hat{T}_x(\omega_3^{-1/2})\) \quad & \quad\quad & \multirow{2}{*}{\(\hat{T}_x\hat{T}_y(\omega_3^n)\), \(\hat{S}\) \quad}& \(\hat{T}_x(\omega_3^{-1/2})\)\quad & \multirow{2}{*}{\quad\(\hat{T}_x\hat{T}_y(\omega_3^n)\), \(\hat{S}\)}\\
&  &   & \(\hat{T}_y(\omega_3^{-1/2})\) &  \(\hat{T}_y(\omega_3^{3/2})\) &  \(\hat{T}_y(\omega_3^{-1/2})\) &  &  &  \(\hat{T}_y(\omega_3^{3/2})\)  & \\

\noalign{\smallskip}
\hline\hline
\end{tabular}
\caption{Summary of instabilities \(I=\Delta, \tilde{M}^{[\ell]}_k,\) and \(\tilde{\rho}^{[\ell]}_k,\) (SC, SDW, and CDW respectively) found in the RG analysis for \(q=2\) (column two) and \(q=3\) at (\(d_{ph}=1\), next three columns) and away from (\(d_{ph}=0.8\), last three columns) perfect nesting in the particle-hole channels. For \(q=3\) the subcolumns indicate the filling corresponding to the VHSs at which the instabilities are found (for \(q=2\) the same instability occurs at both \(1/4\) and \(3/4\) VHS fillings). Second row indicates the critical exponent \(\alpha_I\) of the corresponding instability and the last row shows its symmetry; values in parentheses indicate the phase picked up by the order parameter under the symmetry, e.g. \(\Delta^{(d)}\xrightarrow{\hat{C}_4}-\Delta^{(d)}\). Recall that \(\omega_q=e^{2\pi i /q}\).
}
\label{table:Results}
\end{center} 
\end{table*}

For \(q=3\), the \(\tilde{M}^{[0]}_2\) and \(\tilde{M}^{[1]}_1\) SDW susceptibilities (degenerate by hermiticity) diverge first in the middle band, while SC and \(\tilde{M}^{[0]}_0\) (or the degenerate \(\tilde{M}^{[1]}_0\), again by hermiticity) SDW diverge first in the top and bottom bands. This suggests strongly competing instabilities in the top and bottom bands that likely remain degenerate at perfect nesting as in the \(q=1\) case, and a small detuning from perfect nesting generally favors SC instabilities. We find that for \(d_{ph}=0.8\), SC is a clear winner in the top and bottom bands at \(1/6\) and \(5/6\) fillings, but SDW remains the apparent leading instability at half-filling. Remarkably, we find that when SC is the winning instability, the RG flow approaches the self-similar fixed trajectory \(g^{(\ell)j}_{mn}=g_j/\sqrt{q}\) within numerical accuracy, as shown in the inset in Fig. \ref{fig:RGflow} (c). We therefore expect the results for the \(q=1\) case to generalize in this case. Observe that this is unlike the \(q=2\) case for which the self-similar fixed trajectory is not reached.

Indeed, the SC phase we find in the top and bottom bands satisfies \(\Delta^{(0)}_{m;\mathrm{v}}=\Delta^{(0)}_{m+1;\mathrm{v}}\) and \(\Delta^{(0)}_{m;0}=-\Delta^{(0)}_{m;1}\), similar to the \(q=2\) and \(q=1\) cases. Unlike those cases, however, there is no natural interpretation of these relations in terms of MTG and \(\hat{C}_4\) symmetries. As shown in \cite{ShafferWangSantos21}, in this case the gap function transforms according to a 3D irrep of the MTG and necessarily breaks at least one of \(\hat{T}_x\) or \(\hat{T}_y\), and any \(\hat{C}_4\) symmetric gap breaks all of the MTG symmetries. In order to determine the symmetries of the resulting degenerate ground states it is necessary to include fourth order terms in the Ginzburg-Landau free energy, which goes beyond the 1 loop RG analysis. Computing the fourth order term using an approximation scheme outlined in the SM, we find that there are three degenerate ground states symmetric under \(\omega_q^m\hat{T}_x\hat{T}_y\) with \(m=0,1,2\), and \(\hat{C}_4\) is therefore broken. This determines the rest of the \(\Delta^{(\ell)}_{m;\mathrm{v}}\) order parameters for \(\ell\neq0\), so below we will focus on the form of \(\Delta^{(0)}_{m;\mathrm{v}}\) only.

The \(\Delta^{(0)}_{m;\mathrm{v}}=\Delta^{(0)}_{m+1;\mathrm{v}}\) condition extended to \(\Delta^{(0)}_{m}(\mathbf{p})=\Delta^{(0)}_{m+1}(\mathbf{p})\) on the full rMBZ implies an additional symmetry that emerges under the RG flow, which we refer to as a self-similarity symmetry \(\hat{S}\). This symmetry acts on the gap function as
\[\Delta(\mathbf{p})\xrightarrow{\hat{S}}\hat{T}_x(\mathbf{p})\Delta(\mathbf{p})\hat{T}_x(-\mathbf{p})\label{S}\]
(in contrast to the canonical action of \(\hat{T}_x\) itself, which acts as \(\Delta(\mathbf{p})\xrightarrow{\hat{T}_x}\hat{T}_x(\mathbf{p})\Delta(\mathbf{p})\hat{T}_x^T(-\mathbf{p})\) \cite{ShafferWangSantos21}). Stated another way, \(\hat{S}\) acts as \(\hat{T}_x\) on the particle sector but as \(\hat{T}_x^{-1}\) on the hole sector in the Nambu space of the Bogoliubov-de Gennes (BdG) formalism. We refer to this symmetry as self-similarity because in momentum space it implies that the gap function is independent of the magnetic flavor index and thus repeats three times (or \(q\) times generalized to other \(q\)).

Though the self-similarity symmetry \(\hat{S}\) acts in a simple way in momentum space, its action on the gap function
in the sub-lattice basis \(c_{\mathbf{k}s}\) is not trivial and it takes
\(\Delta^{(\ell)}_{ss'}(\mathbf{k})\xrightarrow{\hat{S}} \Delta^{(\ell)}_{s-1,s'+1}(\mathbf{k+Q})\). In the real space basis \(c_{\mathbf{R}s}\), the action of this symmetry has a highly non-local character: \(\Delta_{\mathbf{R}s;\mathbf{R}'s'}\xrightarrow{\hat{S}}e^{-i\mathbf{Q}\cdot(\mathbf{R}-\mathbf{R}')}\sum_{X\in q\mathbb{Z}}\mathrm{sinc}\left[\frac{\pi}{q}(X+2)\right]\Delta_{\mathbf{R},s+1;\mathbf{R}'+X\hat{\mathbf{x}},s'-1}\),
where \(\mathrm{sinc}(x)=\sin x/x\) (see SM for details of the change of basis transformation). In particular, if \(\Delta_{\mathbf{R}s;\mathbf{R}'s'}\) is symmetric under \(\hat{S}\), it decays as \(1/(R_x-R'_x)^2\), implying a long-range order
and an obstruction to constructing fully localized Wannier states of the BdG Hamiltonian (see Fig. \ref{fig:GapFunctions} (c)).

As for \(q=2\), our method does not determine the form of the gap function along the whole Fermi surface, and either a chiral or a nodal form of the gap within the rMBZ matches the \(\Delta^{(0)}_{m;0}=-\Delta^{(0)}_{m;1}\) relation. In this case symmetry does not completely fix the form of the gap function, but we find that the simplest form of the extended gap function respecting the \(\hat{S}\) symmetry and matching the RG result at VHSs can be obtained in the sub-lattice basis:
\[\Delta^{(0)}_{ss'}(\mathbf{k})=\Delta_0\left[1-\cos k_x-\cos (k_y-(s-s')Q)\right]\label{Delta3}\]
Though as mentioned above this gap function cannot be written down in real space using nearest neighbor terms, it can be constructed using an extended \(s\)-wave gap function \(\Delta^{(S)}_{\mathbf{r}\mathbf{r'}}=\Delta_0(\delta_{\mathbf{rr'}}-\sum_\mathbf{a}\delta_{\mathbf{r,r'+\mathbf{a}}}/2)\) where \(\mathbf{a}\) is summed over all nearest neighbors of the square lattice. The real space order parameter can then be obtained by repeatedly applying the \(\hat{S}\) symmetry, \(\Delta_{\mathbf{R}s;\mathbf{R'}s'}=\sum_j\hat{S}^j\left[\Delta^{(S)}_{\mathbf{r}\mathbf{r'}}\right]\). We then obtain the extension \(\Delta^{(0)}_{m}(\mathbf{p})\) by projecting onto the band basis \(d_{\mathbf{p}\ell\alpha}\) (see SM), and find that the resulting order parameter is fully gapped and chiral, \(\Delta^{(0)}_{m}(\mathbf{p})\sim e^{\pm 2i\theta}\) with \(\pm\) for the upper and lower bands respectively, contributing a Chern number of \(\pm 2\) (see Fig. \ref{fig:GapFunctions} (e-f)). An important consequence of the \(\hat{S}\) symmetry is the three fold degeneracy of the BdG spectrum of the fermionic excitations, which therefore implies that the total Chern number of this phase is \(\pm 6\). We verify this numerically for the \(\hat{T}_x\hat{T}_y\) symmetric gap function by computing the BdG spectrum with cylindrical boundary conditions periodic in the \(x\) direction and open in the \(y\) direction (taking advantage of the gap function being short-ranged in the latter). The resulting spectrum is shown in Fig. \ref{fig:EdgeModes}.

\section{Discussion}\label{Sec:Disc}

To summarize, we have investigated the nature of electronic instabilities on the square lattice Hofstadter-Hubbard model using a weak coupling renormalization group analysis to characterize competing electronic orders when the Fermi level is brought near a manifold of $2q$ VHSs and the flux per unit cell is $\Phi = 2\pi p/q$.
The RG analysis allows for the treatment of competing instabilities on equal footing, revealing how the progressive elimination of high energy modes renormalizes the bare repulsive interactions and opens low energy instability channels.
One of the main results of our analysis is the demonstration of the existence of self-similar fixed trajectories of the RG flow related to the RG equations at zero field. 
Remarkably, we find that the self-similar fixed trajectory is reached by the RG flow for \(q=3\) (but not for \(q=2\)) when the SC instability occurs.
The existence of a self-similar structure in the RG flow of Hofstadter systems is a novel result that illustrates the power of the magnetic translation group in constraining the low energy instabilities.

We analysed the RG equations for two representative cases, with the results summarized in Table \ref{table:Results}. First, for $p/q = 1/2$ corresponding to the TR-symmetric $\pi$-flux phase we have identified nodal d-wave superconducting instabilities near $1/4$ and $3/4$ fillings. The nodal order parameters are odd under the magnetic rotation $\hat{C}_4$ and have unusual real space structure (see Fig. \ref{fig:GapFunctions} (b)) giving rise to a gapless spectrum of Bogoliubov quasiparticles that manifest themselves in a linear-in-temperature specific heat. Importantly, unlike the zero flux case, the SC instability is leading even when the nesting is perfect in the density wave channels.
Second, for $p/q = 1/3, 2/3$ corresponding to $\pm 2\pi/3$ flux lattices our analysis uncovers the existence of a novel chiral topological superconductors near $1/6$ and $5/6$ fillings. 
These TRS broken paired states break $\hat{C}_4$ symmetry while preserving a $\mathbb{Z}_3$ subgroup of the MTG, thus realizing a $\mathbb{Z}_3$ Hofstadter superconductor classified in Ref. \cite{ShafferWangSantos21}.
Having a gapped bulk spectrum, these novel phases are characterized by a bulk Chern number topological invariant $\mathcal{C} = \pm 6$, which accounts for a chiral phase with $6$ net chiral Majorana edge modes. A universal experimental signature of such phases is a quantized thermal Hall coefficient $\kappa_{xy}/T = 6 \times (\pi^2 k^{2}_{B}/3h)$.
Even more remarkably, the chiral phases occur when the system flows to the self-similar trajectory of the RG equations and as a result possess a self-similarity symmetry \(\hat{S}\) defined in Eq. (\ref{S}) that forces the real-space order parameters to be long-ranged, providing another experimental signature of these phases. Moreover, since the self-similar trajectory is present for all \(q\), the self-similar HSC instability is viable for all values of the magnetic flux.
The prediction of unconventional nodal and self-similar topological superconductivity in partially filled Hofstadter bands form intrinsic electronic interactions are the two main results of this work.

In addition, we found several closely competing spin density wave instabilities that break MTG symmetries and that may be of experimental interest in their own right. Below the transition temperature, these states can coexist with the HSC states and can give rise to rich and complex phase diagram similar to those of high \(T_c\) superconductors \cite{FradkinKivelsonTranquada15}. Moreover, the multi-component nature of the HSC order parameters implies that vestigial density wave orders may appear in the vicinity of the SC instability and can provide an experimental signature of these phases \cite{FernandesSchmalian19}.

Recently, Hofstadter systems have experienced a renaissance caused by the advent of 2D moir\'e superlattices realizing large magnetic fluxes in laboratory accessible magnetic fields. For nearly four decades, Hofstadter bands have been predominantly studied as platforms for the quantum Hall effect, following the seminal work of Thouless and collaborators \cite{TKNN} that showed that it is a consequence of the topology of filled Hofstadter bands. However, the connection between Hofstadter systems and the quantum Hall effect is but one aspect of the physics embodied by fractal electronic bands.
This work invites a broader view on the potentialities of Hofstadter quantum materials.
Rather surprisingly, our RG analysis predicts that superconductivity can be driven by repulsive interactions in Hofstadter systems, surprising not just because of the role played by electronic interactions, but also because it  implies the formation of Cooper pairs in large magnetic fields that cause a strong orbital effect commonly viewed as detrimental for superconductivity.
Our analysis therefore establishes a new microscopic mechanism for the realization of reentrant superconductivity in Hofstadter materials, which could be within near-term experimental reach in moir\'e superlattices.

In particular, our theoretical findings on the square lattice Hofstadter-Hubbard model may directly inform the realization of reentrant Hofstadter superconductivity in a number of experimental platforms, including optical lattices \cite{mueller2004artificial, gerbier2010gauge, Aidelsburger11, hauke2012non, celi2014synthetic} and twisted cuprate moir\'e systems \cite{Zhu21,zhao2021emergent}.
Moreover, the RG framework developed here for the square lattice can be directly generalized to other Hofstadter systems. A particularly interesting direction is to extend this formalism to effective lattice models describing the band structure of magic angle twisted bilayer graphene where $2\pi/3$ and $\pi$ flux lattices can be realized at accessible magnetic fields $B \sim 8$ T and $B \sim 12$ T, respectively. Similar fields would be required away from the magic angle in which case the bands are not as flat and our weak-coupling analysis may apply more directly. In that regard, the experimental observation \cite{das2021observation} of reentrant behavior in twisted bilayer graphene and other moir\'e systems with small Zeeman splitting ($\lesssim 2$ meV) may offer a promising route to search for emergence of Hofstadter superconductivity, enabled by the competition of electronic orders in the complex manifold of Van Hove singularities present in moir\'e Hofstadter superlattices.

The RG theory can also be extended to the case of spin polarized bands for materials in which the Zeeman splitting is strong. In that regime triplet Hofstadter superconductivity may become possible. Recent observations of triplet SC in twisted trilayer graphene \cite{CaoTTG21,ChristosSachdev22} as well as Bernal stacked bilayer graphene \cite{ZhouYoung22} indicate that this may be another promising route to realizing HSCs. Of course in all these systems, including TBG, strong correlation effects may play an important role, which are known to affect the Hofstadter spectrum \cite{AndrewsSoluyanov20,WangVafek21,ShefferStern21} and have been seen to lead to fractional and ferromagnetic states in experiment in the Hofstadter regime \cite{Wang15,Spanton18,Saito2021}. Recently, Hofstadter superconductivity has also been studied in the strong coupling limit using a mean field theory \cite{TuNeupert18}. Including these strong coupling effects in the RG framework
is likely necessary to properly study Hofstadter superconductivity in magic angle TBG due to the presence of flat bands. This is a challenging task we leave for a future study, but we expect a non-trivial interplay of HSC with these strongly correlated states that can give rise to even more unconventional phases.

\section{Methods}\label{Methods}

In this work we extended the standard parquet RG analysis of VHS patch models previously used to study SC from repulsive interactions on square and hexagonal lattices \cite{schulz1987superconductivity,dzyaloshinskiui1987maximal,Furukawa98,Nandkishore2012,maiti_superconductivity_2013,Lin-Nandkishore-2020,classen_competing_2020} to the HH model with half-filled Hofstadter bands. The details of this calculation are presented in the accompanying Supplementary Material. In this analysis we introduce test vertices corresponding to all possible instabilities of the Fermi surface and study their RG flow. The resulting flows are shown in Fig. \ref{fig:RGflow}. The chief advantage of this method is that it allows us to go beyond mean-field and study all possible instabilities on equal footing, letting the system decide which instability wins. Throughout the RG analysis we make extensive use of the MTG symmetries to identify different channels of the RG flow.

Since the RG calculation only determines the order parameter at the VHS points, it is necessary to extend it in some way to determine the nature of the resulting phase (chiral or nodal). In principle, one needs to extend the RG calculation to the whole BZ, which is computationally prohibitive already for moderate \(q\). Even solving the self-consistent gap equation for a constant Hubbard interaction numerically is quite challenging. We therefore adopt a simpler approach and construct an ansatz gap function in real space in the \(c_{\mathbf{r}\sigma}\) basis first (e.g. standard \(s\)- or \(d\)-wave gap functions with up to nearest neighbor terms, etc.) consistent with the symmetries of the ground state,  and then projecting onto the Hofstadter band of interest via \(d_{\mathbf{k}\alpha\sigma}=\sum_{s}\mathcal{U}^s_\alpha(\mathbf{k})c_{\mathbf{k}s\alpha}\) with the band index \(\alpha\) fixed. The details of this projection are presented in the SM and the resulting gap function extensions are presented in Fig. \ref{fig:GapFunctions}.

With the gap function extension we can then study the Bogoliubov-De Gennes (BdG) spectrum of the fermionic excitations of the system, including its topological properties. The BdG spectrum is also needed to derive Ginzburg-Landau free energy. As shown in \cite{ShafferWangSantos21}, due to the HSC order parameter belonging to a multi-dimensional irreducible representation of the MTG for \(q>2\), it is necessary to expand the free energy up to fourth order in powers of the order parameter (the one-loop approximation in the RG being equivalent to a second order approximation of the free energy). We use this in order to establish the MTG symmetry of the \(q=3\) HSC phases, as outlines in the SM. The BdG Hamiltonian with the gap function extension expressed in real space also allows us to study the system on a cylinder (i.e with periodic boundary conditions along the \(x\) direction but open in the \(y\) direction) in order to identify the topological edge modes. The resulting edge modes are shown in Fig. \ref{fig:EdgeModes}, with the details of the calculation presented in the SM.

\section*{Data Availability}

Data sharing not applicable to this article as no datasets were generated or analysed during the current study.

\section*{Code Availability}

All numerical codes in this paper are publicly accessible at \cite{code}.

\bibliographystyle{naturemag}
\bibliography{bibliography}

\begin{thebibliography}{100}
\expandafter\ifx\csname url\endcsname\relax
  \def\url#1{\texttt{#1}}\fi
\expandafter\ifx\csname urlprefix\endcsname\relax\def\urlprefix{URL }\fi
\providecommand{\bibinfo}[2]{#2}
\providecommand{\eprint}[2][]{\url{#2}}

\bibitem{GinzburgLandau50}
\bibinfo{author}{Ginzburg, V.} \& \bibinfo{author}{Landau, L.}
\newblock \bibinfo{title}{Theory of superconductivity}.
\newblock \emph{\bibinfo{journal}{Zh. Eksp. Teor. Fiz.;(USSR)}}
  \textbf{\bibinfo{volume}{20}} (\bibinfo{year}{1950}).

\bibitem{BCS57}
\bibinfo{author}{Bardeen, J.}, \bibinfo{author}{Cooper, L.~N.} \&
  \bibinfo{author}{Schrieffer, J.~R.}
\newblock \bibinfo{title}{Theory of superconductivity}.
\newblock \emph{\bibinfo{journal}{Phys. Rev.}} \textbf{\bibinfo{volume}{108}},
  \bibinfo{pages}{1175--1204} (\bibinfo{year}{1957}).
\newblock \urlprefix\url{https://link.aps.org/doi/10.1103/PhysRev.108.1175}.

\bibitem{Abrikosov57}
\bibinfo{author}{Abrikosov, A.~A.}
\newblock \bibinfo{title}{On the magnetic properties of superconductors of the
  second group}.
\newblock \emph{\bibinfo{journal}{Sov. Phys. JETP}}
  \textbf{\bibinfo{volume}{5}}, \bibinfo{pages}{1174--1182}
  (\bibinfo{year}{1957}).

\bibitem{RasoltTesanovich92}
\bibinfo{author}{Rasolt, M.} \& \bibinfo{author}{Te\v{s}anovi\'{c}, Z.}
\newblock \bibinfo{title}{Theoretical aspects of superconductivity in very high
  magnetic fields}.
\newblock \emph{\bibinfo{journal}{Reviews of Modern Physics}}
  \textbf{\bibinfo{volume}{64}}, \bibinfo{pages}{709--754}
  (\bibinfo{year}{1992}).
\newblock \urlprefix\url{https://link.aps.org/doi/10.1103/RevModPhys.64.709}.
\newblock \bibinfo{note}{Publisher: American Physical Society}.

\bibitem{chaudhary2021Quantum}
\bibinfo{author}{Chaudhary, G.}, \bibinfo{author}{MacDonald, A.~H.} \&
  \bibinfo{author}{Norman, M.~R.}
\newblock \bibinfo{title}{Quantum {Hall} superconductivity from moir\'e
  {Landau} levels}.
\newblock \emph{\bibinfo{journal}{Phys. Rev. Research}}
  \textbf{\bibinfo{volume}{3}}, \bibinfo{pages}{033260} (\bibinfo{year}{2021}).
\newblock
  \urlprefix\url{https://link.aps.org/doi/10.1103/PhysRevResearch.3.033260}.

\bibitem{cao2018unconventional}
\bibinfo{author}{Cao, Y.} \emph{et~al.}
\newblock \bibinfo{title}{Unconventional superconductivity in magic-angle
  graphene superlattices}.
\newblock \emph{\bibinfo{journal}{Nature}} \textbf{\bibinfo{volume}{556}},
  \bibinfo{pages}{43--50} (\bibinfo{year}{2018}).
\newblock \urlprefix\url{http://www.nature.com/articles/nature26160}.
\newblock \bibinfo{note}{Number: 7699 Publisher: Nature Publishing Group}.

\bibitem{Azbel64}
\bibinfo{author}{Azbel, M.~Y.}
\newblock \bibinfo{title}{Energy spectrum of a conduction electron in a
  magnetic field}.
\newblock \emph{\bibinfo{journal}{Sov. Phys. JETP}}
  \textbf{\bibinfo{volume}{19}}, \bibinfo{pages}{634--645}
  (\bibinfo{year}{1964}).

\bibitem{Hofstadter76}
\bibinfo{author}{Hofstadter, D.~R.}
\newblock \bibinfo{title}{Energy levels and wave functions of {Bloch} electrons
  in rational and irrational magnetic fields}.
\newblock \emph{\bibinfo{journal}{Physical Review B}}
  \textbf{\bibinfo{volume}{14}}, \bibinfo{pages}{2239--2249}
  (\bibinfo{year}{1976}).
\newblock \urlprefix\url{https://link.aps.org/doi/10.1103/PhysRevB.14.2239}.
\newblock \bibinfo{note}{Publisher: American Physical Society}.

\bibitem{Dean13}
\bibinfo{author}{Dean, C.~R.} \emph{et~al.}
\newblock \bibinfo{title}{Hofstadter's butterfly and the fractal quantum {Hall}
  effect in moir\'{e} superlattices}.
\newblock \emph{\bibinfo{journal}{Nature}} \textbf{\bibinfo{volume}{497}},
  \bibinfo{pages}{598--602} (\bibinfo{year}{2013}).
\newblock \urlprefix\url{http://www.nature.com/articles/nature12186}.
\newblock \bibinfo{note}{Number: 7451 Publisher: Nature Publishing Group}.

\bibitem{Ponomarenko13}
\bibinfo{author}{Ponomarenko, L.~A.} \emph{et~al.}
\newblock \bibinfo{title}{Cloning of {Dirac} fermions in graphene
  superlattices}.
\newblock \emph{\bibinfo{journal}{Nature}} \textbf{\bibinfo{volume}{497}},
  \bibinfo{pages}{594--597} (\bibinfo{year}{2013}).
\newblock \urlprefix\url{https://doi.org/10.1038/nature12187}.

\bibitem{Hunt13}
\bibinfo{author}{Hunt, B.} \emph{et~al.}
\newblock \bibinfo{title}{Massive {Dirac} {Fermions} and {Hofstadter}
  {Butterfly} in a van der {Waals} {Heterostructure}}.
\newblock \emph{\bibinfo{journal}{Science}} \textbf{\bibinfo{volume}{340}},
  \bibinfo{pages}{1427--1430} (\bibinfo{year}{2013}).
\newblock \urlprefix\url{https://science.sciencemag.org/content/340/6139/1427}.
\newblock \bibinfo{note}{Publisher: American Association for the Advancement of
  Science Section: Report}.

\bibitem{Forsythe18}
\bibinfo{author}{Forsythe, C.} \emph{et~al.}
\newblock \bibinfo{title}{Band structure engineering of {2D} materials using
  patterned dielectric superlattices}.
\newblock \emph{\bibinfo{journal}{Nature Nanotechnology}}
  \textbf{\bibinfo{volume}{13}}, \bibinfo{pages}{566--571}
  (\bibinfo{year}{2018}).
\newblock \urlprefix\url{http://www.nature.com/articles/s41565-018-0138-7}.
\newblock \bibinfo{note}{Number: 7 Publisher: Nature Publishing Group}.

\bibitem{Wang15}
\bibinfo{author}{Wang, L.} \emph{et~al.}
\newblock \bibinfo{title}{Evidence for a fractional fractal quantum {Hall}
  effect in graphene superlattices}.
\newblock \emph{\bibinfo{journal}{Science}} \textbf{\bibinfo{volume}{350}},
  \bibinfo{pages}{1231--1234} (\bibinfo{year}{2015}).
\newblock \urlprefix\url{https://science.sciencemag.org/content/350/6265/1231}.
\newblock \bibinfo{note}{Publisher: American Association for the Advancement of
  Science Section: Report}.

\bibitem{Spanton18}
\bibinfo{author}{Spanton, E.~M.} \emph{et~al.}
\newblock \bibinfo{title}{Observation of fractional {Chern} insulators in a van
  der {Waals} heterostructure}.
\newblock \emph{\bibinfo{journal}{Science}} \textbf{\bibinfo{volume}{360}},
  \bibinfo{pages}{62--66} (\bibinfo{year}{2018}).
\newblock \urlprefix\url{https://science.sciencemag.org/content/360/6384/62}.
\newblock \bibinfo{note}{Publisher: American Association for the Advancement of
  Science Section: Report}.

\bibitem{Saito21}
\bibinfo{author}{Saito, Y.} \emph{et~al.}
\newblock \bibinfo{title}{Hofstadter subband ferromagnetism and symmetry-broken
  {Chern} insulators in twisted bilayer graphene}.
\newblock \emph{\bibinfo{journal}{Nature Physics}}
  \textbf{\bibinfo{volume}{17}}, \bibinfo{pages}{478--481}
  (\bibinfo{year}{2021}).
\newblock \urlprefix\url{http://www.nature.com/articles/s41567-020-01129-4}.
\newblock \bibinfo{note}{Number: 4 Publisher: Nature Publishing Group}.

\bibitem{wang_classification_2020}
\bibinfo{author}{Wang, J.} \& \bibinfo{author}{Santos, L.~H.}
\newblock \bibinfo{title}{Classification of topological phase transitions and
  van {Hove} singularity steering mechanism in graphene superlattices}.
\newblock \emph{\bibinfo{journal}{Phys. Rev. Lett.}}
  \textbf{\bibinfo{volume}{125}}, \bibinfo{pages}{236805}
  (\bibinfo{year}{2020}).
\newblock
  \urlprefix\url{https://link.aps.org/doi/10.1103/PhysRevLett.125.236805}.

\bibitem{Herzog-Arbeitman_Hofstadter2020}
\bibinfo{author}{Herzog-Arbeitman, J.}, \bibinfo{author}{Song, Z.-D.},
  \bibinfo{author}{Regnault, N.} \& \bibinfo{author}{Bernevig, B.~A.}
\newblock \bibinfo{title}{Hofstadter topology: Noncrystalline topological
  materials at high flux}.
\newblock \emph{\bibinfo{journal}{Phys. Rev. Lett.}}
  \textbf{\bibinfo{volume}{125}}, \bibinfo{pages}{236804}
  (\bibinfo{year}{2020}).
\newblock
  \urlprefix\url{https://link.aps.org/doi/10.1103/PhysRevLett.125.236804}.

\bibitem{Spanton2018}
\bibinfo{author}{Spanton, E.~M.} \emph{et~al.}
\newblock \bibinfo{title}{Observation of fractional {Chern} insulators in a van
  der {Waals} heterostructure}.
\newblock \emph{\bibinfo{journal}{Science}} \textbf{\bibinfo{volume}{360}},
  \bibinfo{pages}{62--66} (\bibinfo{year}{2018}).
\newblock \urlprefix\url{https://science.sciencemag.org/content/360/6384/62}.
\newblock \eprint{https://science.sciencemag.org/content/360/6384/62.full.pdf}.

\bibitem{sharpe2019emergent}
\bibinfo{author}{Sharpe, A.~L.} \emph{et~al.}
\newblock \bibinfo{title}{Emergent ferromagnetism near three-quarters filling
  in twisted bilayer graphene}.
\newblock \emph{\bibinfo{journal}{Science}} \textbf{\bibinfo{volume}{365}},
  \bibinfo{pages}{605--608} (\bibinfo{year}{2019}).
\newblock \urlprefix\url{http://www.science.org/doi/10.1126/science.aaw3780}.
\newblock \bibinfo{note}{Publisher: American Association for the Advancement of
  Science}.

\bibitem{serlin2020intrinsic}
\bibinfo{author}{Serlin, M.} \emph{et~al.}
\newblock \bibinfo{title}{Intrinsic quantized anomalous {Hall} effect in a
  moir\'{e} heterostructure}.
\newblock \emph{\bibinfo{journal}{Science}} \textbf{\bibinfo{volume}{367}},
  \bibinfo{pages}{900--903} (\bibinfo{year}{2020}).
\newblock \urlprefix\url{http://www.science.org/doi/10.1126/science.aay5533}.
\newblock \bibinfo{note}{Publisher: American Association for the Advancement of
  Science}.

\bibitem{Saito2021}
\bibinfo{author}{Saito, Y.} \emph{et~al.}
\newblock \bibinfo{title}{Hofstadter subband ferromagnetism and symmetry-broken
  {Chern} insulators in twisted bilayer graphene}.
\newblock \emph{\bibinfo{journal}{Nature Physics}}
  \textbf{\bibinfo{volume}{17}}, \bibinfo{pages}{478--481}
  (\bibinfo{year}{2021}).
\newblock \urlprefix\url{https://doi.org/10.1038/s41567-020-01129-4}.

\bibitem{xie2021fractional}
\bibinfo{author}{Xie, Y.} \emph{et~al.}
\newblock \bibinfo{title}{Fractional {Chern} insulators in magic-angle twisted
  bilayer graphene}.
\newblock \emph{\bibinfo{journal}{Nature}} \textbf{\bibinfo{volume}{600}},
  \bibinfo{pages}{439--443} (\bibinfo{year}{2021}).
\newblock \urlprefix\url{https://doi.org/10.1038/s41586-021-04002-3}.

\bibitem{nuckolls2020strongly}
\bibinfo{author}{Nuckolls, K.~P.} \emph{et~al.}
\newblock \bibinfo{title}{Strongly correlated {Chern} insulators in magic-angle
  twisted bilayer graphene}.
\newblock \emph{\bibinfo{journal}{Nature}} \textbf{\bibinfo{volume}{588}},
  \bibinfo{pages}{610--615} (\bibinfo{year}{2020}).
\newblock \urlprefix\url{http://www.nature.com/articles/s41586-020-3028-8}.
\newblock \bibinfo{note}{Number: 7839 Publisher: Nature Publishing Group}.

\bibitem{wu_chern_2021}
\bibinfo{author}{Wu, S.}, \bibinfo{author}{Zhang, Z.},
  \bibinfo{author}{Watanabe, K.}, \bibinfo{author}{Taniguchi, T.} \&
  \bibinfo{author}{Andrei, E.~Y.}
\newblock \bibinfo{title}{Chern insulators, van {Hove} singularities and
  topological flat bands in magic-angle twisted bilayer graphene}.
\newblock \emph{\bibinfo{journal}{Nature Materials}} \bibinfo{pages}{1--7}
  (\bibinfo{year}{2021}).
\newblock \urlprefix\url{https://www.nature.com/articles/s41563-020-00911-2}.
\newblock \bibinfo{note}{Publisher: Nature Publishing Group}.

\bibitem{das2021symmetry}
\bibinfo{author}{Das, I.} \emph{et~al.}
\newblock \bibinfo{title}{Symmetry-broken {Chern} insulators and {Rashba}-like
  {Landau}-level crossings in magic-angle bilayer graphene}.
\newblock \emph{\bibinfo{journal}{Nature Physics}}
  \textbf{\bibinfo{volume}{17}}, \bibinfo{pages}{710--714}
  (\bibinfo{year}{2021}).
\newblock \urlprefix\url{http://www.nature.com/articles/s41567-021-01186-3}.
\newblock \bibinfo{note}{Number: 6 Publisher: Nature Publishing Group}.

\bibitem{choi2021correlation}
\bibinfo{author}{Choi, Y.} \emph{et~al.}
\newblock \bibinfo{title}{Correlation-driven topological phases in magic-angle
  twisted bilayer graphene}.
\newblock \emph{\bibinfo{journal}{Nature}} \textbf{\bibinfo{volume}{589}},
  \bibinfo{pages}{536--541} (\bibinfo{year}{2021}).
\newblock \urlprefix\url{https://doi.org/10.1038/s41586-020-03159-7}.

\bibitem{park2021flavour}
\bibinfo{author}{Park, J.~M.}, \bibinfo{author}{Cao, Y.},
  \bibinfo{author}{Watanabe, K.}, \bibinfo{author}{Taniguchi, T.} \&
  \bibinfo{author}{Jarillo-Herrero, P.}
\newblock \bibinfo{title}{Flavour {Hund}'s coupling, {Chern} gaps and charge
  diffusivity in moir\'{e} graphene}.
\newblock \emph{\bibinfo{journal}{Nature}} \textbf{\bibinfo{volume}{592}},
  \bibinfo{pages}{43--48} (\bibinfo{year}{2021}).
\newblock \urlprefix\url{http://www.nature.com/articles/s41586-021-03366-w}.
\newblock \bibinfo{note}{Number: 7852 Publisher: Nature Publishing Group}.

\bibitem{stepanov2021competing}
\bibinfo{author}{Stepanov, P.} \emph{et~al.}
\newblock \bibinfo{title}{Competing zero-field {Chern} insulators in
  superconducting twisted bilayer graphene}.
\newblock \emph{\bibinfo{journal}{Phys. Rev. Lett.}}
  \textbf{\bibinfo{volume}{127}}, \bibinfo{pages}{197701}
  (\bibinfo{year}{2021}).
\newblock
  \urlprefix\url{https://link.aps.org/doi/10.1103/PhysRevLett.127.197701}.

\bibitem{pierce2021unconventional}
\bibinfo{author}{Pierce, A.~T.} \emph{et~al.}
\newblock \bibinfo{title}{Unconventional sequence of correlated {Chern}
  insulators in magic-angle twisted bilayer graphene}.
\newblock \emph{\bibinfo{journal}{Nature Physics}}
  \textbf{\bibinfo{volume}{17}}, \bibinfo{pages}{1210--1215}
  (\bibinfo{year}{2021}).
\newblock \urlprefix\url{http://www.nature.com/articles/s41567-021-01347-4}.
\newblock \bibinfo{note}{Number: 11 Publisher: Nature Publishing Group}.

\bibitem{yu2021correlated}
\bibinfo{author}{Yu, J.} \emph{et~al.}
\newblock \bibinfo{title}{Correlated {Hofstadter} spectrum and flavour phase
  diagram in magic-angle twisted bilayer graphene}.
\newblock \emph{\bibinfo{journal}{Nature Physics}}
  \textbf{\bibinfo{volume}{18}}, \bibinfo{pages}{825--831}
  (\bibinfo{year}{2022}).
\newblock \urlprefix\url{https://doi.org/10.1038/s41567-022-01589-w}.

\bibitem{ShafferWangSantos21}
\bibinfo{author}{Shaffer, D.}, \bibinfo{author}{Wang, J.} \&
  \bibinfo{author}{Santos, L.~H.}
\newblock \bibinfo{title}{Theory of {Hofstadter} superconductors}.
\newblock \emph{\bibinfo{journal}{Phys. Rev. B}}
  \textbf{\bibinfo{volume}{104}}, \bibinfo{pages}{184501}
  (\bibinfo{year}{2021}).
\newblock \urlprefix\url{https://link.aps.org/doi/10.1103/PhysRevB.104.184501}.

\bibitem{Zak64_1}
\bibinfo{author}{Zak, J.}
\newblock \bibinfo{title}{Magnetic {Translation} {Group}}.
\newblock \emph{\bibinfo{journal}{Physical Review}}
  \textbf{\bibinfo{volume}{134}}, \bibinfo{pages}{A1602--A1606}
  (\bibinfo{year}{1964}).
\newblock \urlprefix\url{https://link.aps.org/doi/10.1103/PhysRev.134.A1602}.
\newblock \bibinfo{note}{Publisher: American Physical Society}.

\bibitem{Zak64_2}
\bibinfo{author}{Zak, J.}
\newblock \bibinfo{title}{Magnetic {Translation} {Group}. {II}. {Irreducible}
  {Representations}}.
\newblock \emph{\bibinfo{journal}{Physical Review}}
  \textbf{\bibinfo{volume}{134}}, \bibinfo{pages}{A1607--A1611}
  (\bibinfo{year}{1964}).
\newblock \urlprefix\url{https://link.aps.org/doi/10.1103/PhysRev.134.A1607}.
\newblock \bibinfo{note}{Publisher: American Physical Society}.

\bibitem{Brown64}
\bibinfo{author}{Brown, E.}
\newblock \bibinfo{title}{Bloch {Electrons} in a {Uniform} {Magnetic} {Field}}.
\newblock \emph{\bibinfo{journal}{Physical Review}}
  \textbf{\bibinfo{volume}{133}}, \bibinfo{pages}{A1038--A1044}
  (\bibinfo{year}{1964}).
\newblock \urlprefix\url{https://link.aps.org/doi/10.1103/PhysRev.133.A1038}.
\newblock \bibinfo{note}{Publisher: American Physical Society}.

\bibitem{Agterberg20}
\bibinfo{author}{Agterberg, D.~F.} \emph{et~al.}
\newblock \bibinfo{title}{The {Physics} of {Pair}-{Density} {Waves}: {Cuprate}
  {Superconductors} and {Beyond}}.
\newblock \emph{\bibinfo{journal}{Annual Review of Condensed Matter Physics}}
  \textbf{\bibinfo{volume}{11}}, \bibinfo{pages}{231--270}
  (\bibinfo{year}{2020}).
\newblock
  \urlprefix\url{https://www.annualreviews.org/doi/10.1146/annurev-conmatphys-031119-050711}.
\newblock \bibinfo{note}{Publisher: Annual Reviews}.

\bibitem{Maska02}
\bibinfo{author}{Ma\ifmmode~\acute{s}\else \'{s}\fi{}ka, M.~M.}
\newblock \bibinfo{title}{Reentrant superconductivity in a strong applied field
  within the tight-binding model}.
\newblock \emph{\bibinfo{journal}{Physical Review B}}
  \textbf{\bibinfo{volume}{66}}, \bibinfo{pages}{054533}
  (\bibinfo{year}{2002}).
\newblock \urlprefix\url{https://link.aps.org/doi/10.1103/PhysRevB.66.054533}.
\newblock \bibinfo{note}{Publisher: American Physical Society}.

\bibitem{MoSudbo02}
\bibinfo{author}{Mo, S.} \& \bibinfo{author}{Sudb\o, A.}
\newblock \bibinfo{title}{Fermion-pairing on a square lattice in extreme
  magnetic fields}.
\newblock \emph{\bibinfo{journal}{Physica C: Superconductivity}}
  \textbf{\bibinfo{volume}{383}}, \bibinfo{pages}{279--286}
  (\bibinfo{year}{2002}).
\newblock
  \urlprefix\url{https://www.sciencedirect.com/science/article/pii/S092145340201345X}.

\bibitem{ZhaiOktel10}
\bibinfo{author}{Zhai, H.},
  \bibinfo{author}{Umucal\ifmmode\imath\else\i\fi{}lar, R.~O.} \&
  \bibinfo{author}{Oktel, M.~O.}
\newblock \bibinfo{title}{Pairing and vortex lattices for interacting fermions
  in optical lattices with a large magnetic field}.
\newblock \emph{\bibinfo{journal}{Phys. Rev. Lett.}}
  \textbf{\bibinfo{volume}{104}}, \bibinfo{pages}{145301}
  (\bibinfo{year}{2010}).
\newblock
  \urlprefix\url{https://link.aps.org/doi/10.1103/PhysRevLett.104.145301}.

\bibitem{Iskin15a}
\bibinfo{author}{Iskin, M.}
\newblock \bibinfo{title}{Stripe-ordered superfluid and supersolid phases in
  the attractive {Hofstadter}-{Hubbard} model}.
\newblock \emph{\bibinfo{journal}{Physical Review A}}
  \textbf{\bibinfo{volume}{91}}, \bibinfo{pages}{011601}
  (\bibinfo{year}{2015}).
\newblock \urlprefix\url{https://link.aps.org/doi/10.1103/PhysRevA.91.011601}.
\newblock \bibinfo{note}{Publisher: American Physical Society}.

\bibitem{JeonJain19}
\bibinfo{author}{Jeon, G.~S.}, \bibinfo{author}{Jain, J.~K.} \&
  \bibinfo{author}{Liu, C.-X.}
\newblock \bibinfo{title}{Topological superconductivity in {Landau} levels}.
\newblock \emph{\bibinfo{journal}{Physical Review B}}
  \textbf{\bibinfo{volume}{99}}, \bibinfo{pages}{094509}
  (\bibinfo{year}{2019}).
\newblock \urlprefix\url{https://link.aps.org/doi/10.1103/PhysRevB.99.094509}.
\newblock \bibinfo{note}{Publisher: American Physical Society}.

\bibitem{SohalFradkin20}
\bibinfo{author}{Sohal, R.} \& \bibinfo{author}{Fradkin, E.}
\newblock \bibinfo{title}{Intertwined order in fractional {Chern} insulators
  from finite-momentum pairing of composite fermions}.
\newblock \emph{\bibinfo{journal}{Physical Review B}}
  \textbf{\bibinfo{volume}{101}}, \bibinfo{pages}{245154}
  (\bibinfo{year}{2020}).
\newblock \urlprefix\url{https://link.aps.org/doi/10.1103/PhysRevB.101.245154}.
\newblock \bibinfo{note}{Publisher: American Physical Society}.

\bibitem{SchirmerJain22}
\bibinfo{author}{Schirmer, J.}, \bibinfo{author}{Liu, C.-X.} \&
  \bibinfo{author}{Jain, J.~K.}
\newblock \bibinfo{title}{Phase diagram of superconductivity in the integer
  quantum {Hall} regime}.
\newblock \emph{\bibinfo{journal}{arXiv:2204.11737 [cond-mat]}}
  (\bibinfo{year}{2022}).
\newblock \urlprefix\url{http://arxiv.org/abs/2204.11737}.
\newblock \bibinfo{note}{ArXiv: 2204.11737}.

\bibitem{vanHove1953theoccurrence}
\bibinfo{author}{Van~Hove, L.}
\newblock \bibinfo{title}{The occurrence of singularities in the elastic
  frequency distribution of a crystal}.
\newblock \emph{\bibinfo{journal}{Phys. Rev.}} \textbf{\bibinfo{volume}{89}},
  \bibinfo{pages}{1189--1193} (\bibinfo{year}{1953}).
\newblock \urlprefix\url{https://link.aps.org/doi/10.1103/PhysRev.89.1189}.

\bibitem{schulz1987superconductivity}
\bibinfo{author}{Schulz, H.~J.}
\newblock \bibinfo{title}{Superconductivity and {Antiferromagnetism} in the
  {Two}-{Dimensional} {Hubbard} {Model}: {Scaling} {Theory}}.
\newblock \emph{\bibinfo{journal}{Europhysics Letters (EPL)}}
  \textbf{\bibinfo{volume}{4}}, \bibinfo{pages}{609--615}
  (\bibinfo{year}{1987}).
\newblock \urlprefix\url{https://doi.org/10.1209/0295-5075/4/5/016}.
\newblock \bibinfo{note}{Publisher: IOP Publishing}.

\bibitem{dzyaloshinskiui1987maximal}
\bibinfo{author}{Dzyaloshinski, I.}
\newblock \bibinfo{title}{Maximal increase of the superconducting transition
  temperature due to the presence of van {Hoff} singularities}.
\newblock \emph{\bibinfo{journal}{JETP Lett}} \textbf{\bibinfo{volume}{46}}
  (\bibinfo{year}{1987}).

\bibitem{markiewicz1997survey}
\bibinfo{author}{Markiewicz, R.~S.}
\newblock \bibinfo{title}{A survey of the {Van} {Hove} scenario for high-tc
  superconductivity with special emphasis on pseudogaps and striped phases}.
\newblock \emph{\bibinfo{journal}{Journal of Physics and Chemistry of Solids}}
  \textbf{\bibinfo{volume}{58}}, \bibinfo{pages}{1179--1310}
  (\bibinfo{year}{1997}).
\newblock
  \urlprefix\url{https://www.sciencedirect.com/science/article/pii/S0022369797000255}.

\bibitem{Nandkishore2012}
\bibinfo{author}{Nandkishore, R.}, \bibinfo{author}{Levitov, L.~S.} \&
  \bibinfo{author}{Chubukov, A.~V.}
\newblock \bibinfo{title}{Chiral superconductivity from repulsive interactions
  in doped graphene}.
\newblock \emph{\bibinfo{journal}{Nature Physics}}
  \textbf{\bibinfo{volume}{8}}, \bibinfo{pages}{158--163}
  (\bibinfo{year}{2012}).
\newblock \urlprefix\url{https://doi.org/10.1038/nphys2208}.

\bibitem{WangFunctionalRG2012}
\bibinfo{author}{Wang, W.-S.} \emph{et~al.}
\newblock \bibinfo{title}{Functional renormalization group and variational
  {Monte} {Carlo} studies of the electronic instabilities in graphene near
  $\frac{1}{4}$ doping}.
\newblock \emph{\bibinfo{journal}{Phys. Rev. B}} \textbf{\bibinfo{volume}{85}},
  \bibinfo{pages}{035414} (\bibinfo{year}{2012}).
\newblock \urlprefix\url{https://link.aps.org/doi/10.1103/PhysRevB.85.035414}.

\bibitem{KieselCompeting2012}
\bibinfo{author}{Kiesel, M.~L.}, \bibinfo{author}{Platt, C.},
  \bibinfo{author}{Hanke, W.}, \bibinfo{author}{Abanin, D.~A.} \&
  \bibinfo{author}{Thomale, R.}
\newblock \bibinfo{title}{Competing many-body instabilities and unconventional
  superconductivity in graphene}.
\newblock \emph{\bibinfo{journal}{Phys. Rev. B}} \textbf{\bibinfo{volume}{86}},
  \bibinfo{pages}{020507} (\bibinfo{year}{2012}).
\newblock \urlprefix\url{https://link.aps.org/doi/10.1103/PhysRevB.86.020507}.

\bibitem{GonzalezGraphene2008}
\bibinfo{author}{Gonz\'alez, J.}
\newblock \bibinfo{title}{Kohn-{Luttinger} superconductivity in graphene}.
\newblock \emph{\bibinfo{journal}{Phys. Rev. B}} \textbf{\bibinfo{volume}{78}},
  \bibinfo{pages}{205431} (\bibinfo{year}{2008}).
\newblock \urlprefix\url{https://link.aps.org/doi/10.1103/PhysRevB.78.205431}.

\bibitem{isobe2018unconventional}
\bibinfo{author}{Isobe, H.}, \bibinfo{author}{Yuan, N.~F.} \&
  \bibinfo{author}{Fu, L.}
\newblock \bibinfo{title}{Unconventional {Superconductivity} and {Density}
  {Waves} in {Twisted} {Bilayer} {Graphene}}.
\newblock \emph{\bibinfo{journal}{Physical Review X}}
  \textbf{\bibinfo{volume}{8}}, \bibinfo{pages}{041041} (\bibinfo{year}{2018}).
\newblock \urlprefix\url{https://link.aps.org/doi/10.1103/PhysRevX.8.041041}.
\newblock \bibinfo{note}{Publisher: American Physical Society}.

\bibitem{Sherkunov-Betouras2018}
\bibinfo{author}{Sherkunov, Y.} \& \bibinfo{author}{Betouras, J.~J.}
\newblock \bibinfo{title}{Electronic phases in twisted bilayer graphene at
  magic angles as a result of {Van} {Hove} singularities and interactions}.
\newblock \emph{\bibinfo{journal}{Phys. Rev. B}} \textbf{\bibinfo{volume}{98}},
  \bibinfo{pages}{205151} (\bibinfo{year}{2018}).
\newblock \urlprefix\url{https://link.aps.org/doi/10.1103/PhysRevB.98.205151}.

\bibitem{Liu-Chiral2018}
\bibinfo{author}{Liu, C.-C.}, \bibinfo{author}{Zhang, L.-D.},
  \bibinfo{author}{Chen, W.-Q.} \& \bibinfo{author}{Yang, F.}
\newblock \bibinfo{title}{Chiral spin density wave and $d+id$ superconductivity
  in the magic-angle-twisted bilayer graphene}.
\newblock \emph{\bibinfo{journal}{Phys. Rev. Lett.}}
  \textbf{\bibinfo{volume}{121}}, \bibinfo{pages}{217001}
  (\bibinfo{year}{2018}).
\newblock
  \urlprefix\url{https://link.aps.org/doi/10.1103/PhysRevLett.121.217001}.

\bibitem{Kennes-Strongcorrelations-2018}
\bibinfo{author}{Kennes, D.~M.}, \bibinfo{author}{Lischner, J.} \&
  \bibinfo{author}{Karrasch, C.}
\newblock \bibinfo{title}{Strong correlations and $d+\mathit{id}$
  superconductivity in twisted bilayer graphene}.
\newblock \emph{\bibinfo{journal}{Phys. Rev. B}} \textbf{\bibinfo{volume}{98}},
  \bibinfo{pages}{241407} (\bibinfo{year}{2018}).
\newblock \urlprefix\url{https://link.aps.org/doi/10.1103/PhysRevB.98.241407}.

\bibitem{You-Vishwanath-2019}
\bibinfo{author}{You, Y.-Z.} \& \bibinfo{author}{Vishwanath, A.}
\newblock \bibinfo{title}{Superconductivity from valley fluctuations and
  approximate {S}{O}(4) symmetry in a weak coupling theory of twisted bilayer
  graphene}.
\newblock \emph{\bibinfo{journal}{npj Quantum Materials}}
  \textbf{\bibinfo{volume}{4}}, \bibinfo{pages}{16} (\bibinfo{year}{2019}).
\newblock \urlprefix\url{https://doi.org/10.1038/s41535-019-0153-4}.

\bibitem{Lin-Nandkishore-2020}
\bibinfo{author}{Lin, Y.-P.} \& \bibinfo{author}{Nandkishore, R.~M.}
\newblock \bibinfo{title}{Parquet renormalization group analysis of
  weak-coupling instabilities with multiple high-order {Van} {Hove} points
  inside the {Brillouin} zone}.
\newblock \emph{\bibinfo{journal}{Phys. Rev. B}}
  \textbf{\bibinfo{volume}{102}}, \bibinfo{pages}{245122}
  (\bibinfo{year}{2020}).
\newblock \urlprefix\url{https://link.aps.org/doi/10.1103/PhysRevB.102.245122}.

\bibitem{hsu_topological_2020}
\bibinfo{author}{Hsu, Y.-T.}, \bibinfo{author}{Wu, F.} \&
  \bibinfo{author}{Das~Sarma, S.}
\newblock \bibinfo{title}{Topological superconductivity, ferromagnetism, and
  valley-polarized phases in moir\'e systems: Renormalization group analysis
  for twisted double bilayer graphene}.
\newblock \emph{\bibinfo{journal}{Phys. Rev. B}}
  \textbf{\bibinfo{volume}{102}}, \bibinfo{pages}{085103}
  (\bibinfo{year}{2020}).
\newblock \urlprefix\url{https://link.aps.org/doi/10.1103/PhysRevB.102.085103}.

\bibitem{classen_competing_2020}
\bibinfo{author}{Classen, L.}, \bibinfo{author}{Chubukov, A.~V.},
  \bibinfo{author}{Honerkamp, C.} \& \bibinfo{author}{Scherer, M.~M.}
\newblock \bibinfo{title}{Competing orders at higher-order {Van} {Hove}
  points}.
\newblock \emph{\bibinfo{journal}{Physical Review B}}
  \textbf{\bibinfo{volume}{102}}, \bibinfo{pages}{125141}
  (\bibinfo{year}{2020}).
\newblock \urlprefix\url{https://link.aps.org/doi/10.1103/PhysRevB.102.125141}.
\newblock \bibinfo{note}{Publisher: American Physical Society}.

\bibitem{Chichinadze2020Nematicsuperconductivity}
\bibinfo{author}{Chichinadze, D.~V.}, \bibinfo{author}{Classen, L.} \&
  \bibinfo{author}{Chubukov, A.~V.}
\newblock \bibinfo{title}{Nematic superconductivity in twisted bilayer
  graphene}.
\newblock \emph{\bibinfo{journal}{Phys. Rev. B}}
  \textbf{\bibinfo{volume}{101}}, \bibinfo{pages}{224513}
  (\bibinfo{year}{2020}).
\newblock \urlprefix\url{https://link.aps.org/doi/10.1103/PhysRevB.101.224513}.

\bibitem{shankar_renormalization-group_1994}
\bibinfo{author}{Shankar, R.}
\newblock \bibinfo{title}{Renormalization-group approach to interacting
  fermions}.
\newblock \emph{\bibinfo{journal}{Reviews of Modern Physics}}
  \textbf{\bibinfo{volume}{66}}, \bibinfo{pages}{129--192}
  (\bibinfo{year}{1994}).

\bibitem{polchinski1992effective}
\bibinfo{author}{Polchinski, J.}
\newblock \bibinfo{title}{Effective field theory and the {Fermi} surface}.
\newblock \emph{\bibinfo{journal}{arXiv preprint hep-th/9210046}}
  (\bibinfo{year}{1992}).

\bibitem{maiti_superconductivity_2013}
\bibinfo{author}{Maiti, S.} \& \bibinfo{author}{Chubukov, A.~V.}
\newblock \bibinfo{title}{Superconductivity from repulsive interaction}.
\newblock \emph{\bibinfo{journal}{AIP Conference Proceedings}}
  \textbf{\bibinfo{volume}{1550}}, \bibinfo{pages}{3--73}
  (\bibinfo{year}{2013}).
\newblock \urlprefix\url{https://aip.scitation.org/doi/abs/10.1063/1.4818400}.
\newblock \bibinfo{note}{Publisher: American Institute of Physics}.

\bibitem{MishraShankar16}
\bibinfo{author}{Mishra, A.}, \bibinfo{author}{Hassan, S.~R.} \&
  \bibinfo{author}{Shankar, R.}
\newblock \bibinfo{title}{Effects of interaction in the {Hofstadter} regime of
  the honeycomb lattice}.
\newblock \emph{\bibinfo{journal}{Physical Review B}}
  \textbf{\bibinfo{volume}{93}}, \bibinfo{pages}{125134}
  (\bibinfo{year}{2016}).
\newblock \urlprefix\url{https://link.aps.org/doi/10.1103/PhysRevB.93.125134}.
\newblock \bibinfo{note}{Publisher: American Physical Society}.

\bibitem{HongSalk99}
\bibinfo{author}{Hong, S.-P.} \& \bibinfo{author}{Suck~Salk, S.-H.}
\newblock \bibinfo{title}{Harper's equation for two-dimensional systems of
  antiferromagnetically correlated electrons}.
\newblock \emph{\bibinfo{journal}{Physical Review B}}
  \textbf{\bibinfo{volume}{60}}, \bibinfo{pages}{9550--9554}
  (\bibinfo{year}{1999}).
\newblock \urlprefix\url{https://link.aps.org/doi/10.1103/PhysRevB.60.9550}.
\newblock \bibinfo{note}{Publisher: American Physical Society}.

\bibitem{HongSSLeeSalk00}
\bibinfo{author}{Hong, S.-P.}, \bibinfo{author}{Lee, S.-S.} \&
  \bibinfo{author}{Suck~Salk, S.-H.}
\newblock \bibinfo{title}{Effects of magnetic field on the two-dimensional
  systems of antiferromagnetically correlated electrons based on the {Hubbard}
  model {Hamiltonian} with easy axis: {Aharonov}-{Bohm} and {Zeeman} effects}.
\newblock \emph{\bibinfo{journal}{Physical Review B}}
  \textbf{\bibinfo{volume}{62}}, \bibinfo{pages}{14880--14885}
  (\bibinfo{year}{2000}).
\newblock \urlprefix\url{https://link.aps.org/doi/10.1103/PhysRevB.62.14880}.
\newblock \bibinfo{note}{Publisher: American Physical Society}.

\bibitem{Kol1993}
\bibinfo{author}{Kol, A.} \& \bibinfo{author}{Read, N.}
\newblock \bibinfo{title}{Fractional quantum {Hall} effect in a periodic
  potential}.
\newblock \emph{\bibinfo{journal}{Phys. Rev. B}} \textbf{\bibinfo{volume}{48}},
  \bibinfo{pages}{8890--8898} (\bibinfo{year}{1993}).
\newblock \urlprefix\url{https://link.aps.org/doi/10.1103/PhysRevB.48.8890}.

\bibitem{Wen91}
\bibinfo{author}{Wen, X.~G.}
\newblock \bibinfo{title}{Non-{Abelian} statistics in the fractional quantum
  {Hall} states}.
\newblock \emph{\bibinfo{journal}{Physical Review Letters}}
  \textbf{\bibinfo{volume}{66}}, \bibinfo{pages}{802--805}
  (\bibinfo{year}{1991}).
\newblock \urlprefix\url{https://link.aps.org/doi/10.1103/PhysRevLett.66.802}.
\newblock \bibinfo{note}{Publisher: American Physical Society}.

\bibitem{MollerCooper09}
\bibinfo{author}{M\"{o}ller, G.} \& \bibinfo{author}{Cooper, N.~R.}
\newblock \bibinfo{title}{Composite {Fermion} {Theory} for {Bosonic} {Quantum}
  {Hall} {States} on {Lattices}}.
\newblock \emph{\bibinfo{journal}{Physical Review Letters}}
  \textbf{\bibinfo{volume}{103}}, \bibinfo{pages}{105303}
  (\bibinfo{year}{2009}).
\newblock
  \urlprefix\url{https://link.aps.org/doi/10.1103/PhysRevLett.103.105303}.
\newblock \bibinfo{note}{Publisher: American Physical Society}.

\bibitem{Moller2015}
\bibinfo{author}{M\"oller, G.} \& \bibinfo{author}{Cooper, N.~R.}
\newblock \bibinfo{title}{Fractional {Chern} insulators in
  {Harper}-{Hofstadter} bands with higher {Chern} number}.
\newblock \emph{\bibinfo{journal}{Phys. Rev. Lett.}}
  \textbf{\bibinfo{volume}{115}}, \bibinfo{pages}{126401}
  (\bibinfo{year}{2015}).
\newblock
  \urlprefix\url{https://link.aps.org/doi/10.1103/PhysRevLett.115.126401}.

\bibitem{ScaffidiSimon14}
\bibinfo{author}{Scaffidi, T.} \& \bibinfo{author}{Simon, S.~H.}
\newblock \bibinfo{title}{Exact solutions of fractional {Chern} insulators:
  {Interacting} particles in the {Hofstadter} model at finite size}.
\newblock \emph{\bibinfo{journal}{Physical Review B}}
  \textbf{\bibinfo{volume}{90}}, \bibinfo{pages}{115132}
  (\bibinfo{year}{2014}).
\newblock \urlprefix\url{https://link.aps.org/doi/10.1103/PhysRevB.90.115132}.
\newblock \bibinfo{note}{Publisher: American Physical Society}.

\bibitem{MotrukZaletelMong16}
\bibinfo{author}{Motruk, J.}, \bibinfo{author}{Zaletel, M.~P.},
  \bibinfo{author}{Mong, R. S.~K.} \& \bibinfo{author}{Pollmann, F.}
\newblock \bibinfo{title}{Density matrix renormalization group on a cylinder in
  mixed real and momentum space}.
\newblock \emph{\bibinfo{journal}{Physical Review B}}
  \textbf{\bibinfo{volume}{93}}, \bibinfo{pages}{155139}
  (\bibinfo{year}{2016}).
\newblock \urlprefix\url{https://link.aps.org/doi/10.1103/PhysRevB.93.155139}.
\newblock \bibinfo{note}{Publisher: American Physical Society}.

\bibitem{Lee2018Emergent}
\bibinfo{author}{Lee, J.~Y.}, \bibinfo{author}{Wang, C.},
  \bibinfo{author}{Zaletel, M.~P.}, \bibinfo{author}{Vishwanath, A.} \&
  \bibinfo{author}{He, Y.-C.}
\newblock \bibinfo{title}{Emergent multi-flavor {${\mathrm{QED}}_{3}$} at the
  plateau transition between fractional {Chern} insulators: Applications to
  graphene heterostructures}.
\newblock \emph{\bibinfo{journal}{Phys. Rev. X}} \textbf{\bibinfo{volume}{8}},
  \bibinfo{pages}{031015} (\bibinfo{year}{2018}).
\newblock \urlprefix\url{https://link.aps.org/doi/10.1103/PhysRevX.8.031015}.

\bibitem{Sohal-2018}
\bibinfo{author}{Sohal, R.}, \bibinfo{author}{Santos, L.~H.} \&
  \bibinfo{author}{Fradkin, E.}
\newblock \bibinfo{title}{Chern-{Simons} composite fermion theory of fractional
  {Chern} insulators}.
\newblock \emph{\bibinfo{journal}{Phys. Rev. B}} \textbf{\bibinfo{volume}{97}},
  \bibinfo{pages}{125131} (\bibinfo{year}{2018}).
\newblock \urlprefix\url{https://link.aps.org/doi/10.1103/PhysRevB.97.125131}.

\bibitem{AndrewsMoller18}
\bibinfo{author}{Andrews, B.} \& \bibinfo{author}{M\"{o}ller, G.}
\newblock \bibinfo{title}{Stability of fractional {Chern} insulators in the
  effective continuum limit of {Harper}-{Hofstadter} bands with {Chern} number
  {${\textbar}{C}{\textbar}{\textgreater}1$}}.
\newblock \emph{\bibinfo{journal}{Physical Review B}}
  \textbf{\bibinfo{volume}{97}}, \bibinfo{pages}{035159}
  (\bibinfo{year}{2018}).
\newblock \urlprefix\url{https://link.aps.org/doi/10.1103/PhysRevB.97.035159}.
\newblock \bibinfo{note}{Publisher: American Physical Society}.

\bibitem{AndrewsSoluyanov20}
\bibinfo{author}{Andrews, B.} \& \bibinfo{author}{Soluyanov, A.}
\newblock \bibinfo{title}{Fractional quantum {Hall} states for moir\'{e}
  superstructures in the {Hofstadter} regime}.
\newblock \emph{\bibinfo{journal}{Physical Review B}}
  \textbf{\bibinfo{volume}{101}}, \bibinfo{pages}{235312}
  (\bibinfo{year}{2020}).
\newblock \urlprefix\url{https://link.aps.org/doi/10.1103/PhysRevB.101.235312}.
\newblock \bibinfo{note}{Publisher: American Physical Society}.

\bibitem{Thouless83}
\bibinfo{author}{Thouless, D.~J.}
\newblock \bibinfo{title}{Bandwidths for a quasiperiodic tight-binding model}.
\newblock \emph{\bibinfo{journal}{Phys. Rev. B}} \textbf{\bibinfo{volume}{28}},
  \bibinfo{pages}{4272--4276} (\bibinfo{year}{1983}).
\newblock \urlprefix\url{https://link.aps.org/doi/10.1103/PhysRevB.28.4272}.

\bibitem{HofstatterDemlerLukin02}
\bibinfo{author}{Hofstetter, W.}, \bibinfo{author}{Cirac, J.~I.},
  \bibinfo{author}{Zoller, P.}, \bibinfo{author}{Demler, E.} \&
  \bibinfo{author}{Lukin, M.~D.}
\newblock \bibinfo{title}{High-temperature superfluidity of fermionic atoms in
  optical lattices}.
\newblock \emph{\bibinfo{journal}{Phys. Rev. Lett.}}
  \textbf{\bibinfo{volume}{89}}, \bibinfo{pages}{220407}
  (\bibinfo{year}{2002}).
\newblock
  \urlprefix\url{https://link.aps.org/doi/10.1103/PhysRevLett.89.220407}.

\bibitem{mueller2004artificial}
\bibinfo{author}{Mueller, E.~J.}
\newblock \bibinfo{title}{Artificial electromagnetism for neutral atoms:
  {Escher} staircase and {Laughlin} liquids}.
\newblock \emph{\bibinfo{journal}{Physical Review A}}
  \textbf{\bibinfo{volume}{70}}, \bibinfo{pages}{041603}
  (\bibinfo{year}{2004}).
\newblock \urlprefix\url{https://link.aps.org/doi/10.1103/PhysRevA.70.041603}.
\newblock \bibinfo{note}{Publisher: American Physical Society}.

\bibitem{Lewenstein07}
\bibinfo{author}{Lewenstein, M.} \emph{et~al.}
\newblock \bibinfo{title}{Ultracold atomic gases in optical lattices: mimicking
  condensed matter physics and beyond}.
\newblock \emph{\bibinfo{journal}{Advances in Physics}}
  \textbf{\bibinfo{volume}{56}}, \bibinfo{pages}{243--379}
  (\bibinfo{year}{2007}).
\newblock \urlprefix\url{https://doi.org/10.1080/00018730701223200}.
\newblock \bibinfo{note}{Publisher: Taylor \& Francis \_eprint:
  https://doi.org/10.1080/00018730701223200}.

\bibitem{Goldman10}
\bibinfo{author}{Goldman, N.} \emph{et~al.}
\newblock \bibinfo{title}{Realistic {Time}-{Reversal} {Invariant} {Topological}
  {Insulators} with {Neutral} {Atoms}}.
\newblock \emph{\bibinfo{journal}{Physical Review Letters}}
  \textbf{\bibinfo{volume}{105}}, \bibinfo{pages}{255302}
  (\bibinfo{year}{2010}).
\newblock
  \urlprefix\url{https://link.aps.org/doi/10.1103/PhysRevLett.105.255302}.
\newblock \bibinfo{note}{Publisher: American Physical Society}.

\bibitem{gerbier2010gauge}
\bibinfo{author}{Gerbier, F.} \& \bibinfo{author}{Dalibard, J.}
\newblock \bibinfo{title}{Gauge fields for ultracold atoms in optical
  superlattices}.
\newblock \emph{\bibinfo{journal}{New Journal of Physics}}
  \textbf{\bibinfo{volume}{12}}, \bibinfo{pages}{033007}
  (\bibinfo{year}{2010}).
\newblock \urlprefix\url{https://doi.org/10.1088/1367-2630/12/3/033007}.
\newblock \bibinfo{note}{Publisher: IOP Publishing}.

\bibitem{Aidelsburger11}
\bibinfo{author}{Aidelsburger, M.} \emph{et~al.}
\newblock \bibinfo{title}{Experimental realization of strong effective magnetic
  fields in an optical lattice}.
\newblock \emph{\bibinfo{journal}{Phys. Rev. Lett.}}
  \textbf{\bibinfo{volume}{107}}, \bibinfo{pages}{255301}
  (\bibinfo{year}{2011}).
\newblock
  \urlprefix\url{https://link.aps.org/doi/10.1103/PhysRevLett.107.255301}.

\bibitem{hauke2012non}
\bibinfo{author}{Hauke, P.} \emph{et~al.}
\newblock \bibinfo{title}{Non-abelian gauge fields and topological insulators
  in shaken optical lattices}.
\newblock \emph{\bibinfo{journal}{Phys. Rev. Lett.}}
  \textbf{\bibinfo{volume}{109}}, \bibinfo{pages}{145301}
  (\bibinfo{year}{2012}).
\newblock
  \urlprefix\url{https://link.aps.org/doi/10.1103/PhysRevLett.109.145301}.

\bibitem{Miyake13}
\bibinfo{author}{Miyake, H.}, \bibinfo{author}{Siviloglou, G.~A.},
  \bibinfo{author}{Kennedy, C.~J.}, \bibinfo{author}{Burton, W.~C.} \&
  \bibinfo{author}{Ketterle, W.}
\newblock \bibinfo{title}{Realizing the {Harper} hamiltonian with
  laser-assisted tunneling in optical lattices}.
\newblock \emph{\bibinfo{journal}{Phys. Rev. Lett.}}
  \textbf{\bibinfo{volume}{111}}, \bibinfo{pages}{185302}
  (\bibinfo{year}{2013}).
\newblock
  \urlprefix\url{https://link.aps.org/doi/10.1103/PhysRevLett.111.185302}.

\bibitem{AidelsburgerBloch13}
\bibinfo{author}{Aidelsburger, M.} \emph{et~al.}
\newblock \bibinfo{title}{Realization of the {Hofstadter} hamiltonian with
  ultracold atoms in optical lattices}.
\newblock \emph{\bibinfo{journal}{Phys. Rev. Lett.}}
  \textbf{\bibinfo{volume}{111}}, \bibinfo{pages}{185301}
  (\bibinfo{year}{2013}).
\newblock
  \urlprefix\url{https://link.aps.org/doi/10.1103/PhysRevLett.111.185301}.

\bibitem{celi2014synthetic}
\bibinfo{author}{Celi, A.} \emph{et~al.}
\newblock \bibinfo{title}{Synthetic gauge fields in synthetic dimensions}.
\newblock \emph{\bibinfo{journal}{Phys. Rev. Lett.}}
  \textbf{\bibinfo{volume}{112}}, \bibinfo{pages}{043001}
  (\bibinfo{year}{2014}).
\newblock
  \urlprefix\url{https://link.aps.org/doi/10.1103/PhysRevLett.112.043001}.

\bibitem{Niemeyer99}
\bibinfo{author}{Niemeyer, M.}, \bibinfo{author}{Freericks, J.~K.} \&
  \bibinfo{author}{Monien, H.}
\newblock \bibinfo{title}{Strong-coupling perturbation theory for the
  two-dimensional {Bose}-{Hubbard} model in a magnetic field}.
\newblock \emph{\bibinfo{journal}{Physical Review B}}
  \textbf{\bibinfo{volume}{60}}, \bibinfo{pages}{2357--2362}
  (\bibinfo{year}{1999}).
\newblock \urlprefix\url{https://link.aps.org/doi/10.1103/PhysRevB.60.2357}.
\newblock \bibinfo{note}{Publisher: American Physical Society}.

\bibitem{Balents05}
\bibinfo{author}{Balents, L.}, \bibinfo{author}{Bartosch, L.},
  \bibinfo{author}{Burkov, A.}, \bibinfo{author}{Sachdev, S.} \&
  \bibinfo{author}{Sengupta, K.}
\newblock \bibinfo{title}{Putting competing orders in their place near the
  {Mott} transition}.
\newblock \emph{\bibinfo{journal}{Physical Review B}}
  \textbf{\bibinfo{volume}{71}}, \bibinfo{pages}{144508}
  (\bibinfo{year}{2005}).
\newblock \urlprefix\url{https://link.aps.org/doi/10.1103/PhysRevB.71.144508}.
\newblock \bibinfo{note}{Publisher: American Physical Society}.

\bibitem{SorensenDemlerLukin05}
\bibinfo{author}{S\o{}rensen, A.~S.}, \bibinfo{author}{Demler, E.} \&
  \bibinfo{author}{Lukin, M.~D.}
\newblock \bibinfo{title}{Fractional quantum {Hall} states of atoms in optical
  lattices}.
\newblock \emph{\bibinfo{journal}{Phys. Rev. Lett.}}
  \textbf{\bibinfo{volume}{94}}, \bibinfo{pages}{086803}
  (\bibinfo{year}{2005}).
\newblock
  \urlprefix\url{https://link.aps.org/doi/10.1103/PhysRevLett.94.086803}.

\bibitem{HafeziDemlerLukin07}
\bibinfo{author}{Hafezi, M.}, \bibinfo{author}{S{\o}rensen, A.~S.},
  \bibinfo{author}{Demler, E.} \& \bibinfo{author}{Lukin, M.~D.}
\newblock \bibinfo{title}{Fractional quantum {Hall} effect in optical
  lattices}.
\newblock \emph{\bibinfo{journal}{Physical Review A}}
  \textbf{\bibinfo{volume}{76}}, \bibinfo{pages}{023613}
  (\bibinfo{year}{2007}).
\newblock \urlprefix\url{https://link.aps.org/doi/10.1103/PhysRevA.76.023613}.
\newblock \bibinfo{note}{Publisher: American Physical Society}.

\bibitem{Oktel07}
\bibinfo{author}{Oktel, M.~O.}, \bibinfo{author}{Ni\ifmmode \mbox{\c{t}}\else
  \c{t}\fi{}\ifmmode~\u{a}\else \u{a}\fi{}, M.} \& \bibinfo{author}{Tanatar,
  B.}
\newblock \bibinfo{title}{Mean-field theory for {Bose}-{Hubbard} model under a
  magnetic field}.
\newblock \emph{\bibinfo{journal}{Phys. Rev. B}} \textbf{\bibinfo{volume}{75}},
  \bibinfo{pages}{045133} (\bibinfo{year}{2007}).
\newblock \urlprefix\url{https://link.aps.org/doi/10.1103/PhysRevB.75.045133}.

\bibitem{PowellDasSarma11}
\bibinfo{author}{Powell, S.}, \bibinfo{author}{Barnett, R.},
  \bibinfo{author}{Sensarma, R.} \& \bibinfo{author}{Das~Sarma, S.}
\newblock \bibinfo{title}{Bogoliubov theory of interacting bosons on a lattice
  in a synthetic magnetic field}.
\newblock \emph{\bibinfo{journal}{Physical Review A}}
  \textbf{\bibinfo{volume}{83}}, \bibinfo{pages}{013612}
  (\bibinfo{year}{2011}).
\newblock \urlprefix\url{https://link.aps.org/doi/10.1103/PhysRevA.83.013612}.
\newblock \bibinfo{note}{Publisher: American Physical Society}.

\bibitem{OrthHofstetter13}
\bibinfo{author}{Orth, P.~P.} \emph{et~al.}
\newblock \bibinfo{title}{Correlated topological phases and exotic magnetism
  with ultracold fermions}.
\newblock \emph{\bibinfo{journal}{Journal of Physics B: Atomic, Molecular and
  Optical Physics}} \textbf{\bibinfo{volume}{46}}, \bibinfo{pages}{134004}
  (\bibinfo{year}{2013}).
\newblock \urlprefix\url{https://doi.org/10.1088/0953-4075/46/13/134004}.
\newblock \bibinfo{note}{Publisher: IOP Publishing}.

\bibitem{Wang14}
\bibinfo{author}{Wang, L.}, \bibinfo{author}{Hung, H.-H.} \&
  \bibinfo{author}{Troyer, M.}
\newblock \bibinfo{title}{Topological phase transition in the
  {Hofstadter}-{Hubbard} model}.
\newblock \emph{\bibinfo{journal}{Physical Review B}}
  \textbf{\bibinfo{volume}{90}}, \bibinfo{pages}{205111}
  (\bibinfo{year}{2014}).
\newblock \urlprefix\url{https://link.aps.org/doi/10.1103/PhysRevB.90.205111}.
\newblock \bibinfo{note}{Publisher: American Physical Society}.

\bibitem{Peotta15}
\bibinfo{author}{Peotta, S.} \& \bibinfo{author}{T\"orm\"a, P.}
\newblock \bibinfo{title}{Superfluidity in topologically nontrivial flat
  bands}.
\newblock \emph{\bibinfo{journal}{Nature Communications}}
  \textbf{\bibinfo{volume}{6}}, \bibinfo{pages}{8944} (\bibinfo{year}{2015}).
\newblock \urlprefix\url{https://www.nature.com/articles/ncomms9944}.
\newblock \bibinfo{note}{Number: 1 Publisher: Nature Publishing Group}.

\bibitem{UmucalilarIskin17}
\bibinfo{author}{Umucal\ifmmode\imath\else\i\fi{}lar, R.~O.} \&
  \bibinfo{author}{Iskin, M.}
\newblock \bibinfo{title}{{BCS} {Theory} of {Time}-{Reversal}-{Symmetric}
  {Hofstadter}-{Hubbard} {Model}}.
\newblock \emph{\bibinfo{journal}{Physical Review Letters}}
  \textbf{\bibinfo{volume}{119}}, \bibinfo{pages}{085301}
  (\bibinfo{year}{2017}).
\newblock
  \urlprefix\url{https://link.aps.org/doi/10.1103/PhysRevLett.119.085301}.
\newblock \bibinfo{note}{Publisher: American Physical Society}.

\bibitem{Zeng19}
\bibinfo{author}{Zeng, C.}, \bibinfo{author}{Stanescu, T.},
  \bibinfo{author}{Zhang, C.}, \bibinfo{author}{Scarola, V.} \&
  \bibinfo{author}{Tewari, S.}
\newblock \bibinfo{title}{Majorana {Corner} {Modes} with {Solitons} in an
  {Attractive} {Hubbard}-{Hofstadter} {Model} of {Cold} {Atom} {Optical}
  {Lattices}}.
\newblock \emph{\bibinfo{journal}{Physical Review Letters}}
  \textbf{\bibinfo{volume}{123}}, \bibinfo{pages}{060402}
  (\bibinfo{year}{2019}).
\newblock
  \urlprefix\url{https://link.aps.org/doi/10.1103/PhysRevLett.123.060402}.
\newblock \bibinfo{note}{Publisher: American Physical Society}.

\bibitem{Lin21}
\bibinfo{author}{Lin, L.} \& \bibinfo{author}{Wu, X.}
\newblock \bibinfo{title}{Numerical solution of large scale
  {Hartree}-{Fock}-{Bogoliubov} equations}.
\newblock \emph{\bibinfo{journal}{ESAIM: Mathematical Modelling and Numerical
  Analysis}} \textbf{\bibinfo{volume}{55}}, \bibinfo{pages}{763--787}
  (\bibinfo{year}{2021}).
\newblock
  \urlprefix\url{http://www.esaim.m2an.org/articles/m2an/abs/2021/04/m2an200003/m2an200003.html}.
\newblock \bibinfo{note}{Number: 3 Publisher: EDP Sciences}.

\bibitem{yu2019high}
\bibinfo{author}{Yu, Y.} \emph{et~al.}
\newblock \bibinfo{title}{High-temperature superconductivity in monolayer
  {Bi$_2$Sr$_2$CaCu$_2$O}$_{8+\delta}$}.
\newblock \emph{\bibinfo{journal}{Nature}} \textbf{\bibinfo{volume}{575}},
  \bibinfo{pages}{156--163} (\bibinfo{year}{2019}).
\newblock \urlprefix\url{http://www.nature.com/articles/s41586-019-1718-x}.
\newblock \bibinfo{note}{Number: 7781 Publisher: Nature Publishing Group}.

\bibitem{BilleKlemm01}
\bibinfo{author}{Bille, A.}, \bibinfo{author}{Klemm, R.~A.} \&
  \bibinfo{author}{Scharnberg, K.}
\newblock \bibinfo{title}{Models of c-axis twist {Josephson} tunneling}.
\newblock \emph{\bibinfo{journal}{Physical Review B}}
  \textbf{\bibinfo{volume}{64}}, \bibinfo{pages}{174507}
  (\bibinfo{year}{2001}).
\newblock \urlprefix\url{https://link.aps.org/doi/10.1103/PhysRevB.64.174507}.
\newblock \bibinfo{note}{Publisher: American Physical Society}.

\bibitem{can2021high}
\bibinfo{author}{Can, O.} \emph{et~al.}
\newblock \bibinfo{title}{High-temperature topological superconductivity in
  twisted double-layer copper oxides}.
\newblock \emph{\bibinfo{journal}{Nature Physics}}
  \textbf{\bibinfo{volume}{17}}, \bibinfo{pages}{519--524}
  (\bibinfo{year}{2021}).
\newblock \urlprefix\url{http://www.nature.com/articles/s41567-020-01142-7}.
\newblock \bibinfo{note}{Number: 4 Publisher: Nature Publishing Group}.

\bibitem{volkov2020magic}
\bibinfo{author}{Volkov, P.~A.}, \bibinfo{author}{Wilson, J.~H.} \&
  \bibinfo{author}{Pixley, J.~H.}
\newblock \bibinfo{title}{Magic angles and current-induced topology in twisted
  nodal superconductors}.
\newblock \emph{\bibinfo{journal}{arXiv:2012.07860 [cond-mat]}}
  (\bibinfo{year}{2020}).
\newblock \urlprefix\url{http://arxiv.org/abs/2012.07860}.
\newblock \bibinfo{note}{ArXiv: 2012.07860}.

\bibitem{VolkovKimPixley21}
\bibinfo{author}{Volkov, P.~A.} \emph{et~al.}
\newblock \bibinfo{title}{Josephson effects in twisted nodal superconductors}.
\newblock \emph{\bibinfo{journal}{arXiv:2108.13456 [cond-mat]}}
  (\bibinfo{year}{2021}).
\newblock \urlprefix\url{http://arxiv.org/abs/2108.13456}.
\newblock \bibinfo{note}{ArXiv: 2108.13456}.

\bibitem{SongVishwanath21}
\bibinfo{author}{Song, X.-Y.}, \bibinfo{author}{Zhang, Y.-H.} \&
  \bibinfo{author}{Vishwanath, A.}
\newblock \bibinfo{title}{Doping a moir\'{e} {Mott} {Insulator}: {A} t-{J}
  model study of twisted cuprates}.
\newblock \emph{\bibinfo{journal}{arXiv:2109.08142 [cond-mat]}}
  (\bibinfo{year}{2021}).
\newblock \urlprefix\url{http://arxiv.org/abs/2109.08142}.
\newblock \bibinfo{note}{ArXiv: 2109.08142}.

\bibitem{Zhu21}
\bibinfo{author}{Zhu, Y.} \emph{et~al.}
\newblock \bibinfo{title}{Presence of $s$-wave pairing in josephson junctions
  made of twisted ultrathin
  {${\mathrm{Bi}}_{2}{\mathrm{Sr}}_{2}{\mathrm{CaCu}}_{2}{\mathrm{O}}_{8+x}$}
  flakes}.
\newblock \emph{\bibinfo{journal}{Phys. Rev. X}} \textbf{\bibinfo{volume}{11}},
  \bibinfo{pages}{031011} (\bibinfo{year}{2021}).
\newblock \urlprefix\url{https://link.aps.org/doi/10.1103/PhysRevX.11.031011}.

\bibitem{zhao2021emergent}
\bibinfo{author}{Zhao, S. Y.~F.} \emph{et~al.}
\newblock \bibinfo{title}{Emergent {Interfacial} {Superconductivity} between
  {Twisted} {Cuprate} {Superconductors}}.
\newblock \emph{\bibinfo{journal}{arXiv:2108.13455 [cond-mat]}}
  (\bibinfo{year}{2021}).
\newblock \urlprefix\url{http://arxiv.org/abs/2108.13455}.
\newblock \bibinfo{note}{ArXiv: 2108.13455}.

\bibitem{Furukawa98}
\bibinfo{author}{Furukawa, N.}, \bibinfo{author}{Rice, T.~M.} \&
  \bibinfo{author}{Salmhofer, M.}
\newblock \bibinfo{title}{Truncation of a two-dimensional {Fermi} surface due
  to quasiparticle gap formation at the saddle points}.
\newblock \emph{\bibinfo{journal}{Phys. Rev. Lett.}}
  \textbf{\bibinfo{volume}{81}}, \bibinfo{pages}{3195--3198}
  (\bibinfo{year}{1998}).
\newblock \urlprefix\url{https://link.aps.org/doi/10.1103/PhysRevLett.81.3195}.

\bibitem{ZhangFoster22}
\bibinfo{author}{Zhang, X.} \& \bibinfo{author}{Foster, M.~S.}
\newblock \bibinfo{title}{Enhanced {Amplitude} for {Superconductivity} due to
  {Spectrum}-wide {Wave} {Function} {Criticality} in {Quasiperiodic} and
  {Power}-law {Random} {Hopping} {Models}}.
\newblock \bibinfo{type}{Tech. Rep.} \bibinfo{number}{arXiv:2204.02996},
  \bibinfo{institution}{arXiv} (\bibinfo{year}{2022}).
\newblock \urlprefix\url{http://arxiv.org/abs/2204.02996}.
\newblock \bibinfo{note}{ArXiv:2204.02996 [cond-mat] type: article}.

\bibitem{Naumis16}
\bibinfo{author}{Naumis, G.~G.}
\newblock \bibinfo{title}{Topological map of the {Hofstadter} butterfly: {Fine}
  structure of {Chern} numbers and {Van} {Hove} singularities}.
\newblock \emph{\bibinfo{journal}{Physics Letters A}}
  \textbf{\bibinfo{volume}{380}}, \bibinfo{pages}{1772--1780}
  (\bibinfo{year}{2016}).
\newblock
  \urlprefix\url{https://www.sciencedirect.com/science/article/pii/S0375960116300019}.

\bibitem{Lin-Nandkishore-2019}
\bibinfo{author}{Lin, Y.-P.} \& \bibinfo{author}{Nandkishore, R.~M.}
\newblock \bibinfo{title}{Chiral twist on the high-${T}_{c}$ phase diagram in
  moir\'e heterostructures}.
\newblock \emph{\bibinfo{journal}{Phys. Rev. B}}
  \textbf{\bibinfo{volume}{100}}, \bibinfo{pages}{085136}
  (\bibinfo{year}{2019}).
\newblock \urlprefix\url{https://link.aps.org/doi/10.1103/PhysRevB.100.085136}.

\bibitem{RaghuKivelsonScalapino2010}
\bibinfo{author}{Raghu, S.}, \bibinfo{author}{Kivelson, S.~A.} \&
  \bibinfo{author}{Scalapino, D.~J.}
\newblock \bibinfo{title}{Superconductivity in the repulsive {Hubbard} model:
  An asymptotically exact weak-coupling solution}.
\newblock \emph{\bibinfo{journal}{Phys. Rev. B}} \textbf{\bibinfo{volume}{81}},
  \bibinfo{pages}{224505} (\bibinfo{year}{2010}).
\newblock \urlprefix\url{https://link.aps.org/doi/10.1103/PhysRevB.81.224505}.

\bibitem{MoritaHatsugai01}
\bibinfo{author}{Morita, Y.} \& \bibinfo{author}{Hatsugai, Y.}
\newblock \bibinfo{title}{Duality in the {Azbel}-{Hofstadter} problem and
  two-dimensional $\mathit{d}$-wave superconductivity with a magnetic field}.
\newblock \emph{\bibinfo{journal}{Phys. Rev. Lett.}}
  \textbf{\bibinfo{volume}{86}}, \bibinfo{pages}{151--154}
  (\bibinfo{year}{2001}).
\newblock \urlprefix\url{https://link.aps.org/doi/10.1103/PhysRevLett.86.151}.

\bibitem{Guo2018Unconventional}
\bibinfo{author}{Guo, H.} \emph{et~al.}
\newblock \bibinfo{title}{Unconventional pairing symmetry of interacting
  {Dirac} fermions on a $\ensuremath{\pi}$-flux lattice}.
\newblock \emph{\bibinfo{journal}{Phys. Rev. B}} \textbf{\bibinfo{volume}{97}},
  \bibinfo{pages}{155146} (\bibinfo{year}{2018}).
\newblock \urlprefix\url{https://link.aps.org/doi/10.1103/PhysRevB.97.155146}.

\bibitem{FradkinKivelsonTranquada15}
\bibinfo{author}{Fradkin, E.}, \bibinfo{author}{Kivelson, S.~A.} \&
  \bibinfo{author}{Tranquada, J.~M.}
\newblock \bibinfo{title}{Colloquium: {Theory} of intertwined orders in high
  temperature superconductors}.
\newblock \emph{\bibinfo{journal}{Reviews of Modern Physics}}
  \textbf{\bibinfo{volume}{87}}, \bibinfo{pages}{457--482}
  (\bibinfo{year}{2015}).
\newblock \urlprefix\url{https://link.aps.org/doi/10.1103/RevModPhys.87.457}.
\newblock \bibinfo{note}{Publisher: American Physical Society}.

\bibitem{FernandesSchmalian19}
\bibinfo{author}{Fernandes, R.~M.}, \bibinfo{author}{Orth, P.~P.} \&
  \bibinfo{author}{Schmalian, J.}
\newblock \bibinfo{title}{Intertwined {Vestigial} {Order} in {Quantum}
  {Materials}: {Nematicity} and {Beyond}}.
\newblock \emph{\bibinfo{journal}{Annual Review of Condensed Matter Physics}}
  \textbf{\bibinfo{volume}{10}}, \bibinfo{pages}{133--154}
  (\bibinfo{year}{2019}).
\newblock
  \urlprefix\url{https://www.annualreviews.org/doi/10.1146/annurev-conmatphys-031218-013200}.
\newblock \bibinfo{note}{Publisher: Annual Reviews}.

\bibitem{TKNN}
\bibinfo{author}{Thouless, D.~J.}, \bibinfo{author}{Kohmoto, M.},
  \bibinfo{author}{Nightingale, M.~P.} \& \bibinfo{author}{den Nijs, M.}
\newblock \bibinfo{title}{Quantized {Hall} {Conductance} in a
  {Two}-{Dimensional} {Periodic} {Potential}}.
\newblock \emph{\bibinfo{journal}{Physical Review Letters}}
  \textbf{\bibinfo{volume}{49}}, \bibinfo{pages}{405--408}
  (\bibinfo{year}{1982}).
\newblock \urlprefix\url{https://link.aps.org/doi/10.1103/PhysRevLett.49.405}.
\newblock \bibinfo{note}{Publisher: American Physical Society}.

\bibitem{das2021observation}
\bibinfo{author}{Das, I.} \emph{et~al.}
\newblock \bibinfo{title}{Observation of re-entrant correlated insulators and
  interaction driven {Fermi} surface reconstructions at one magnetic flux
  quantum per moir\'{e} unit cell in magic-angle twisted bilayer graphene}
  (\bibinfo{year}{2021}).
\newblock \urlprefix\url{https://arxiv.org/abs/2111.11341}.

\bibitem{CaoTTG21}
\bibinfo{author}{Cao, Y.}, \bibinfo{author}{Park, J.~M.},
  \bibinfo{author}{Watanabe, K.}, \bibinfo{author}{Taniguchi, T.} \&
  \bibinfo{author}{Jarillo-Herrero, P.}
\newblock \bibinfo{title}{Large {Pauli} {Limit} {Violation} and {Reentrant}
  {Superconductivity} in {Magic}-{Angle} {Twisted} {Trilayer} {Graphene}}.
\newblock \emph{\bibinfo{journal}{arXiv:2103.12083 [cond-mat]}}
  (\bibinfo{year}{2021}).
\newblock \urlprefix\url{http://arxiv.org/abs/2103.12083}.
\newblock \bibinfo{note}{ArXiv: 2103.12083}.

\bibitem{ChristosSachdev22}
\bibinfo{author}{Christos, M.}, \bibinfo{author}{Sachdev, S.} \&
  \bibinfo{author}{Scheurer, M.~S.}
\newblock \bibinfo{title}{Correlated {Insulators}, {Semimetals}, and
  {Superconductivity} in {Twisted} {Trilayer} {Graphene}}.
\newblock \emph{\bibinfo{journal}{Physical Review X}}
  \textbf{\bibinfo{volume}{12}}, \bibinfo{pages}{021018}
  (\bibinfo{year}{2022}).
\newblock \urlprefix\url{https://link.aps.org/doi/10.1103/PhysRevX.12.021018}.
\newblock \bibinfo{note}{Publisher: American Physical Society}.

\bibitem{ZhouYoung22}
\bibinfo{author}{Zhou, H.} \emph{et~al.}
\newblock \bibinfo{title}{Isospin magnetism and spin-polarized
  superconductivity in {Bernal} bilayer graphene}.
\newblock \emph{\bibinfo{journal}{Science}} \textbf{\bibinfo{volume}{375}},
  \bibinfo{pages}{774--778} (\bibinfo{year}{2022}).
\newblock
  \urlprefix\url{http://www.science.org/doi/full/10.1126/science.abm8386}.
\newblock \bibinfo{note}{Publisher: American Association for the Advancement of
  Science}.

\bibitem{WangVafek21}
\bibinfo{author}{Wang, X.} \& \bibinfo{author}{Vafek, O.}
\newblock \bibinfo{title}{Narrow bands in magnetic field and strong-coupling
  {Hofstadter} spectra}.
\newblock \emph{\bibinfo{journal}{arXiv:2112.08620 [cond-mat]}}
  (\bibinfo{year}{2021}).
\newblock \urlprefix\url{http://arxiv.org/abs/2112.08620}.
\newblock \bibinfo{note}{ArXiv: 2112.08620}.

\bibitem{ShefferStern21}
\bibinfo{author}{Sheffer, Y.} \& \bibinfo{author}{Stern, A.}
\newblock \bibinfo{title}{Chiral {Magic}-{Angle} {Twisted} {Bilayer} {Graphene}
  in a {Magnetic} {Field}: {Landau} {Level} {Correspondence}, {Exact}
  {Wavefunctions} and {Fractional} {Chern} {Insulators}}.
\newblock \emph{\bibinfo{journal}{arXiv:2106.10650 [cond-mat]}}
  (\bibinfo{year}{2021}).
\newblock \urlprefix\url{http://arxiv.org/abs/2106.10650}.
\newblock \bibinfo{note}{ArXiv: 2106.10650}.

\bibitem{TuNeupert18}
\bibinfo{author}{Tu, W.-L.}, \bibinfo{author}{Schindler, F.},
  \bibinfo{author}{Neupert, T.} \& \bibinfo{author}{Poilblanc, D.}
\newblock \bibinfo{title}{Competing orders in the {Hofstadter} $t-{J}$ model}.
\newblock \emph{\bibinfo{journal}{Physical Review B}}
  \textbf{\bibinfo{volume}{97}}, \bibinfo{pages}{035154}
  (\bibinfo{year}{2018}).
\newblock \urlprefix\url{https://link.aps.org/doi/10.1103/PhysRevB.97.035154}.
\newblock \bibinfo{note}{Publisher: American Physical Society}.

\bibitem{code}
\bibinfo{author}{Shaffer, D.}, \bibinfo{author}{Wang, J.} \&
  \bibinfo{author}{Santos, L.~H.}
\newblock \bibinfo{title}{Unconventional self-similar {Hofstadter}
  superconductivity from repulsive interactions}.
\newblock
  \urlprefix\url{https://github.com/dshaffer90/NCOMMS-22-20832-submission}.

\end{thebibliography}

\section*{Acknowledgments}
We thank Claudio Chamon, Dmitri Chichinadze, Ben Feldman, Gil Refael and J\"org Schmalian for helpful discussions.  
This research was partially supported by the U.S. Department of Energy, Office of Science, Basic Energy Sciences, under Award DE-SC0023327
and by startup funds at Emory University.

\section*{Author Contributions}
L.H.S. conceived and designed the project. D.S. carried out the RG calculations. D.S and J.W. performed the numerical analysis of the RG instabilities and analyzed the properties of the paired states. L.H.S supervised the project. All authors contributed to the writing of the paper.

\section*{Competing Interests}
The authors declare no competing interests.

\pagebreak
\widetext
\pagebreak

\title{Unconventional Self-Similar Hofstadter Superconductivity from Repulsive Interactions}

\author{Daniel Shaffer}
\affiliation
{
Department  of  Physics,  Emory  University,  400 Dowman Drive, Atlanta,  GA  30322,  USA
}

\author{Jian Wang}
\affiliation
{
Department  of  Physics,  Emory  University,  400 Dowman Drive, Atlanta,  GA  30322,  USA
}

\author{Luiz H. Santos}
 \email{luiz.santos@emory.edu}
\affiliation
{
Department  of  Physics,  Emory  University,  400 Dowman Drive, Atlanta,  GA  30322,  USA
}

\date{\today}

\widetext
\begin{center}
\textbf{\large Supplementary Material for:\\
\textit{Unconventional Self-Similar Hofstadter Superconductivity from Repulsive Interactions}}
\end{center}
\setcounter{equation}{0}
\setcounter{figure}{0}
\setcounter{table}{0}
\setcounter{page}{1}
\makeatletter
\renewcommand{\theequation}{S\arabic{equation}}
\renewcommand{\figurename}{Supplementary Fig.}
\renewcommand{\bibnumfmt}[1]{[S#1]}
\renewcommand{\thesection}{\Roman{section}}

\section{Details of the RG Calculation}\label{A}

\begin{figure}[h!]
\centering
\includegraphics[width=0.75 \textwidth]{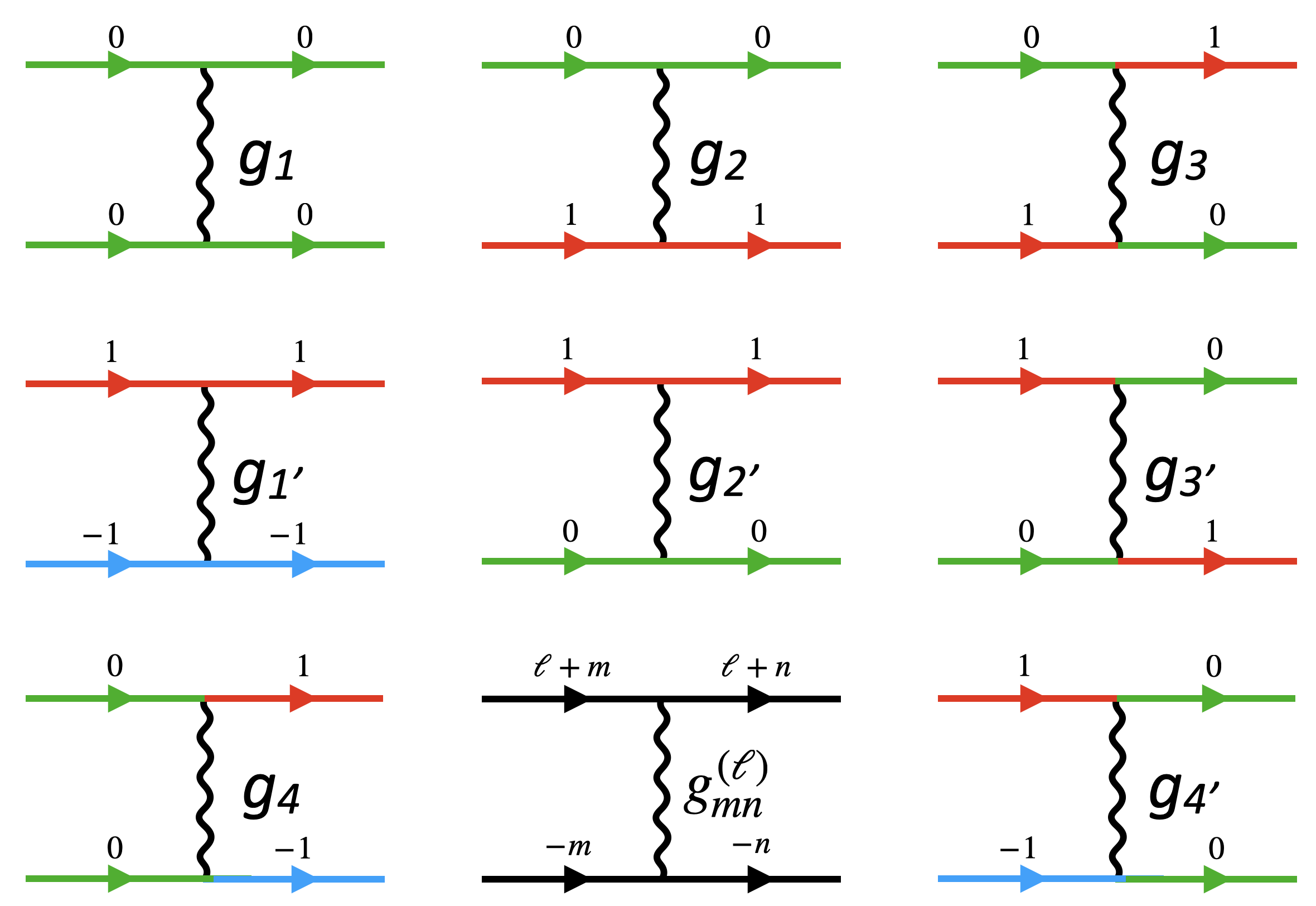}
\caption{Feynman diagrams representing the interaction processes in Eq. (4) and (5). The colored diagrams show the intrapatch processes \(g_j\) (green lines correspond to \(\mathrm{v}=0\), red to \(\mathrm{v}=1\) and blue to \(\mathrm{v}=-1\)). The black diagram shows the interpatch processes with \(\ell\) labeling the total momentum of the incoming and outgoing pairs (which is conserved), while \(m\) and \(n\) label relative momenta of incoming and outgoing pairs respectively (so the momentum transfer is labeled by \(m-n\)). Color online.} 
\label{fig:g}
\end{figure}

Here we present some details of the diagrammatic RG calculations.  Fig. \ref{fig:g} shows all the relevant Feynman diagrams representing the projected interactions in Eqs. (4-5). In addition to the diagrams in Fig. 1(b) in the main text, there are additional processes \(g_{2'}\), \(g_{3'}\) and \(g_{4'}\) related by commutation relations to the \(g_{2}\) and \(g_{3}\) processes and by hermiticity to the \(g_{4}\) process, respectively.
As mentioned in the main text, hermiticity and commutation relations, along with MTG symmetries, impose several relations on the projected interactions:
\begin{align}\label{Grelations}
    g^{(\ell)1}_{mn}&=g^{(\ell),1*}_{nm}=g^{(\ell)1}_{-\ell-m,-\ell-n}\\
    g^{(\ell)1'}_{mn}&=g^{(\ell),1'*}_{nm}=g^{(\ell)1'}_{-1-\ell-m,-1-\ell-n}\nonumber\\
    g^{(\ell)2}_{mn}&=g^{(\ell),2*}_{nm}=g^{(\ell)2'}_{-\ell-m,-\ell-n}\nonumber\\
    g^{(\ell)3}_{mn}&=g^{(\ell),3*}_{-\ell-n,-\ell-m}=g^{(\ell)3'}_{-\ell-m,-\ell-n}\nonumber\\
    g^{(\ell)4}_{mn}&=g^{(\ell),4}_{-\ell-m,-\ell-n-1}=g^{(\ell)4'*}_{nm}\nonumber
\end{align}
These are in addition to the MTG-imposed relation \(g^{(\ell),j}_{mn}=g^{(\ell+2),j}_{m-1,n-1}\); note the TRS is broken for \(q>2\), which means that some coupling constants may be complex. There is a total of \(2q(q+1)+1\) complex coupling constants for odd \(q\) and \(2q(q+2)\) for even \(q\); counting only independent \emph{real} parameters, this amounts to \(4q^2+2\) parameters for odd \(q\) and \(4q^2+2q\) for even \(q\).

\begin{figure*}[t]
\centering
\includegraphics[width=1 \textwidth]{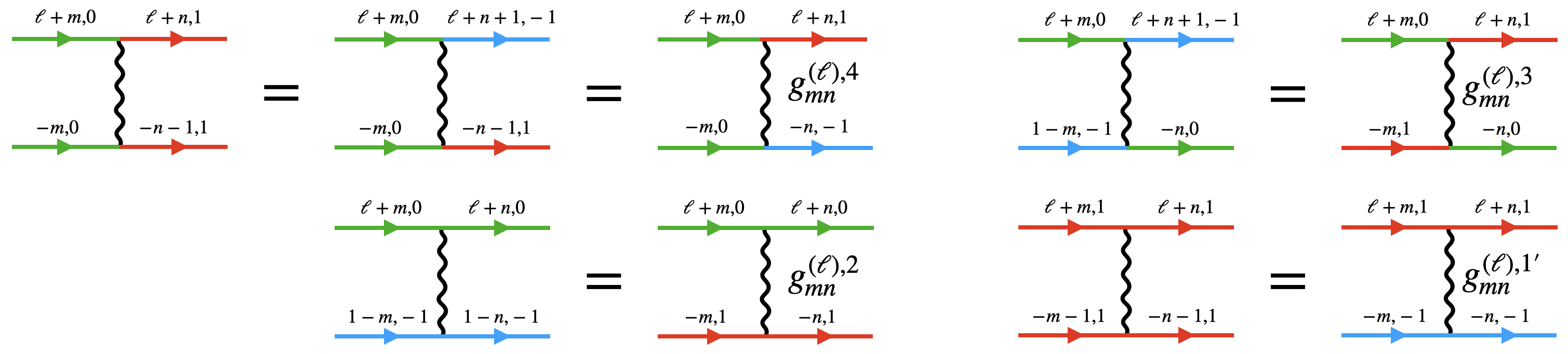}
\caption{Redundant diagrams introduced to separately conserve patch and intrapatch indices. In all RG equations, the diagrams on the left are replaced with the diagrams on the right whenever they occur to avoid introducing additional redundant coupling constants. Green, red, and blue colors correspond to VHS indices \(\mathrm{v}=0,1,-1\) respectively (color online).} 
\label{fig:redundantGs}
\end{figure*}

In addition, recall that we introduced a redundant VHS index \(\mathrm{v}=-1\), with the relation \(\mathbf{K}_{\ell,-1}=\mathbf{K}_{\ell-1,1}\). This is done in order to avoid diagrams such as the \(g_4\) process in the top left in Fig. \ref{fig:redundantGs} in which neither the sum of rMBZ magnetic flavor indices \(\ell,m,n=0,\dots,q-1\) nor the sum of the VHS indices \(\mathrm{v, u, w}\) is equal for incoming and outgoing pairs. This allows us to treat both rMBZ and VHS indices as conserved quantities, but introduces some redundant diagrams also shown in Fig. \ref{fig:redundantGs}. Such redundant diagrams are to be replaced with diagrams in Fig. 1(b) whenever they appear in the diagrammatic expansion.

In order to compute the RG equations, we need the particle-particle bubble, which also serves as the RG time \(t=\Pi_{pp}^{(0)}\) with
\begin{align}
\Pi_{pp}^{(\mathrm{v})}&=-i T\sum_{\omega}\int G_{n,\mathrm{v}}(i\omega,\mathbf{p})G_{\ell-n,\mathrm{v}}(-i\omega,-\mathbf{p})\frac{d^2 p}{(2\pi)^2}=\nu_0 \log^2 \frac{\Lambda}{E}
\end{align}
The particle-hole bubble is similarly defined as
\begin{align}
\Pi_{ph}^{(\mathrm{v})}&=i T\sum_{\omega}\int G_{n,\mathrm{v}}(i\omega,\mathbf{p})G_{\ell+n,\mathrm{v}+1}(i\omega,\mathbf{p})\frac{d^2 p}{(2\pi)^2}=\nu_0 \log^2 \frac{\Lambda}{E}    
\end{align}
(note that neither bubble depends on the choice of \(n\) and \(\ell\) by MTG symmetry; further note that the inter-VHS particle-particle and intra-VHS particle-hole bubbles vanish). Here
\[G_{n,\mathrm{v}}(i\omega,\mathbf{p})=\frac{1}{i\omega-\varepsilon_{n,\mathrm{v}}(\mathbf{p})}\]
is the Green's function for the \(n^{th}\) magnetic flavor and \(\mathrm{v}^{th}\) VHS, and \(E\) is the energy scale down to which the high energy modes have been integrated out to. The dispersion expanded around the VHS points is \(\varepsilon_{\ell,\mathrm{v}}(\mathbf{p})\approx \pm (-1)^\mathrm{v}\frac{p_x^2-p_y^2}{2m}-\mu\) (the \(\pm\) depends on which Hofstadter band the chemical potential is in, but does not alter the calculations).
The extra logarithm comes from the diverging DOS at the VHSs. We then define \(d_{pp}^{(\mathrm{v})}=\frac{d\Pi_{pp}^{(\mathrm{v})}}{d\Pi_{pp}^{(0)}}\approx\frac{\Pi_{pp}^{(\mathrm{v})}}{\Pi_{pp}^{(0)}}\) and \(d_{ph}^{(\mathrm{v})}=\frac{d\Pi_{ph}^{(\mathrm{v})}}{d\Pi_{pp}^{(0)}}\approx\frac{\Pi_{ph}^{(\mathrm{v})}}{\Pi_{pp}^{(0)}}\). Note that due to the \(\hat{C}_4\) symmetry \(d_{pp}^{(0)}=d_{pp}^{(1)}=1\) and \(d_{ph}^{(0)}=d_{ph}^{(1)}\). We thus drop the superscripts.

\begin{figure}[b]
\centering
\includegraphics[width=0.75\textwidth]{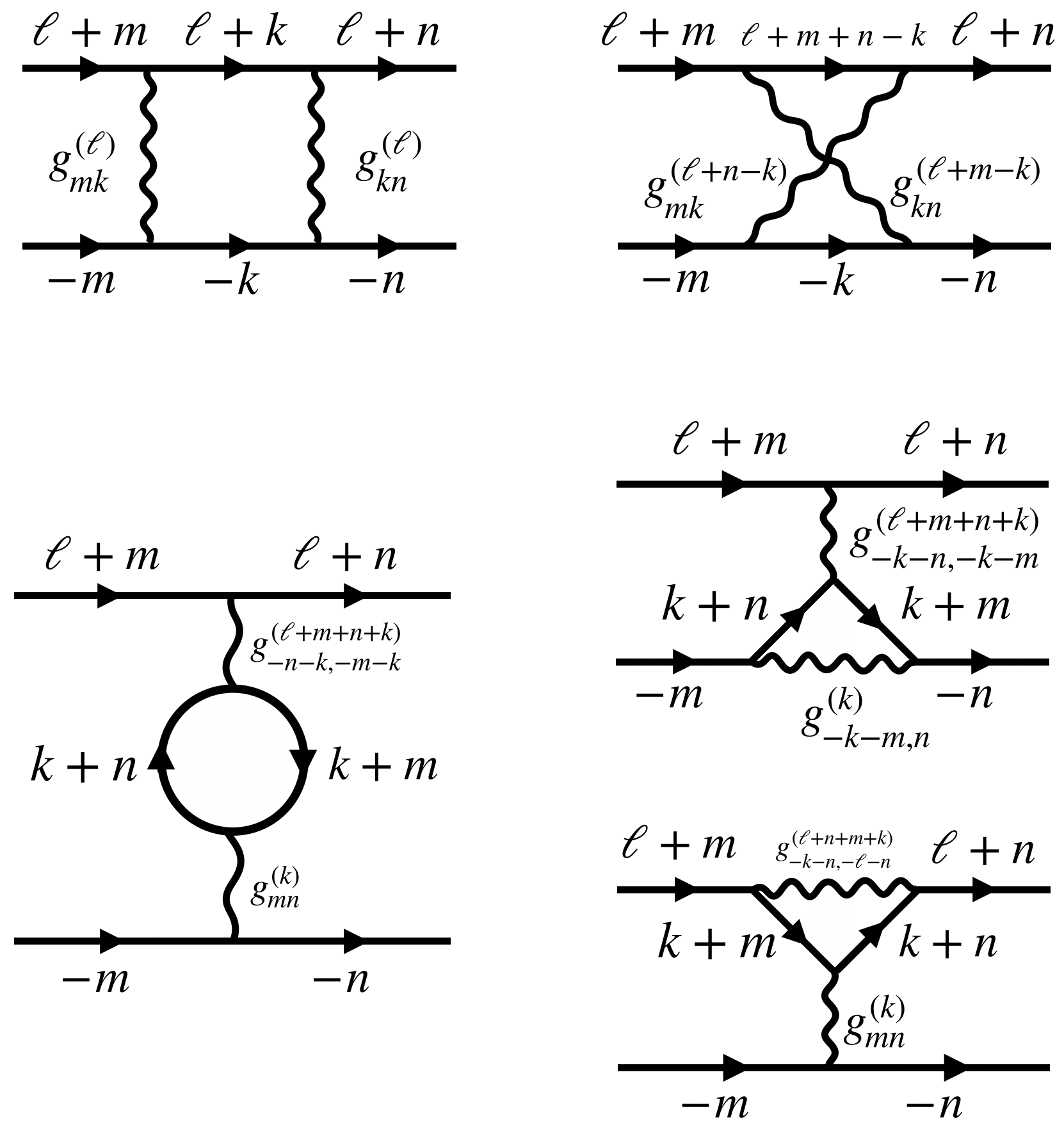}
\caption{The 1 loop Feynman diagrams renormalizing the interactions for magnetic flavor indices. Wavy lines indicate spin is conserved where they meet the fermion lines.}
\label{fig:RGflowInt}
\end{figure}

\begin{figure*}[t]
\centering
\includegraphics[width=0.95\textwidth]{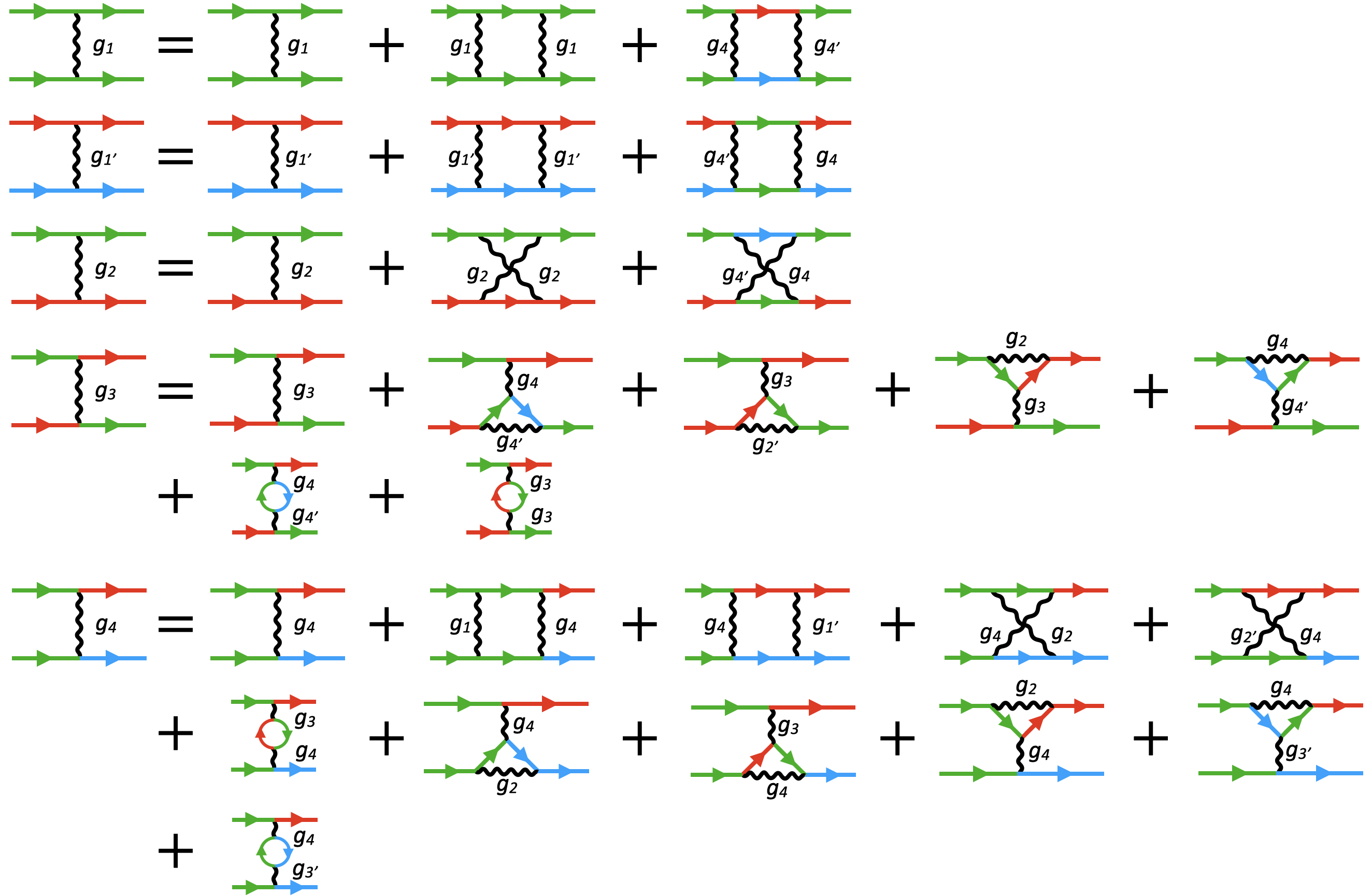}
\caption{The 1 loop Feynman diagrams RG equations for \(g_j\) channels. The magnetic flavor indices \(\ell, m, n\) are to be inserted from Fig. \ref{fig:RGflowInt}}
\label{fig:RGflowEq}
\end{figure*}

To obtain the standard 1 loop RG flow equation, we use the diagrams in Fig. \ref{fig:RGflowEq}, plugging in the magnetic flavor indices from Fig. \ref{fig:RGflowInt} and using the relations in Fig. \ref{fig:redundantGs} where necessary. This yields the RG flow equations shown in Eq. (7) in the main text.

\subsection*{Vertices}

\begin{figure*}[t]
\centering
\includegraphics[width=0.9\textwidth]{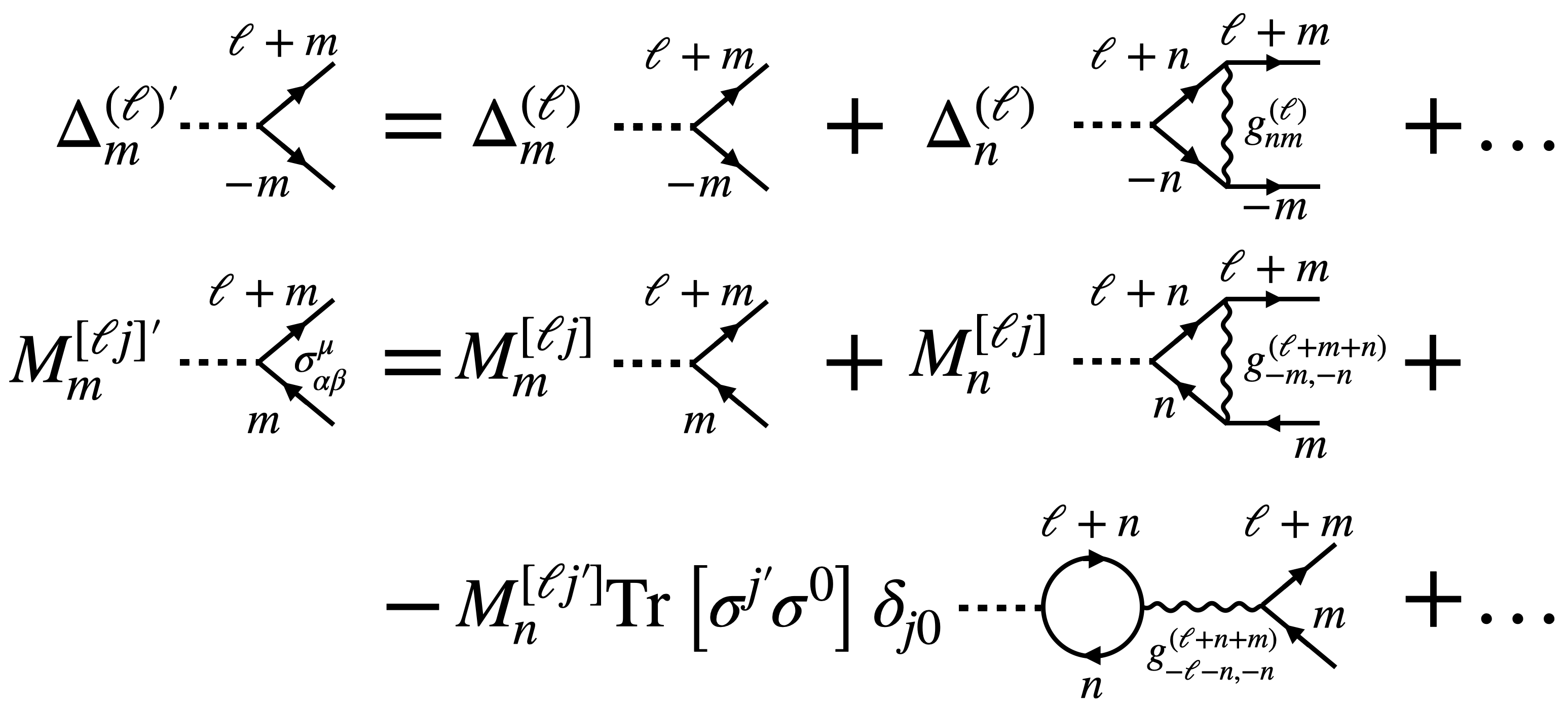}
\caption{The 1 loop Feynman diagrams renormalizing the vertices for magnetic flavor indices. The last diagram includes a trace over the spin indices and comes with an additional minus sign.}
\label{fig:RGflowVertex}
\end{figure*}

The vertices introduced in Eq. (8) also satisfy several symmetry relations. In the spin-singlet pairing channel, anti-commutation relations imply the particle-hole symmetry (PHS) relation
\[\Delta^{(\ell)}_{m;\mathrm{v}}=\Delta^{(\ell)}_{\ell-m;-\mathrm{v}}\]
In the density wave channels, hermiticity implies \(\rho^{[\ell]}_{m;0}=\rho^{[1-\ell]*}_{m+\ell;1}\), and similarly \(M^{[\ell,j]}_{m;0}=M^{[1-\ell,j]*}_{m+\ell;1}\). Here we use the notation \(M^{[\ell,j]}_{m;\mathrm{v}}\) to denote the \(j^{th}\) component of \(\mathbf{M}^{[\ell,j]}_{m;\mathrm{v}}\)
(we can include CDW as a special case with \(j=0\), \(\rho^{[\ell]}=M^{[\ell,0]}\)). The action of the MTG symmetries on the vertices is as follows:
\begin{align}
    \Delta^{(\ell)}_{m;\mathrm{v}}&\xrightarrow[]{\hat{T}_x} \Delta^{(\ell-2)}_{m+1;\mathrm{v}}\\
    \Delta^{(\ell)}_{m;\mathrm{v}}&\xrightarrow[]{\hat{T}_y}\omega^{p\ell}_q\Delta^{(\ell)}_{m;\mathrm{v}}\nonumber\\
    M^{[\ell,j]}_{m;\mathrm{v}}&\xrightarrow[]{\hat{T}_x} \omega^{-1/2}_q M^{[\ell,j]}_{m-1;\mathrm{v}}\nonumber\\
    M^{[\ell,j]}_{m;\mathrm{v}}&\xrightarrow[]{\hat{T}_y}\omega^{p\ell-1/2}_q M^{[\ell,j]}_{m;\mathrm{v}}\nonumber
\end{align}
(CDW included as \(j=0\)). Observe that SC is a charge \(2e\) order while CDW and SDW is a charge \(0e\) order (since the latter orders don't break the \(U(1)\) symmetry). The irreps for SC order were classified in \cite{ShafferWangSantos21}: there is a single \(q\) dimensional irrep for odd \(q\) and four \(q/2\) dimensional irreps for even \(q\). 
In contrast, the C/SDW orders can be seen to transform according to 1D irreps of usual order \(q\) translations forming the \(\mathbb{Z}_q^2\) group (labeled by two numbers, the eigenvalues of the order under \(\hat{T}_x\) and \(\hat{T}_y\)).
The 1 loop RG flow equations for the vertices given in Eq. (\ref{RGvertexFlowEq}) is obtained from the diagrams shown in Fig. \ref{fig:RGflowVertex} and \ref{fig:RGflowVertexEq}. This yields the RG vertex flow equations
\begin{align}\label{RGvertexFlowEq}
\dot{\Delta}^{(\ell)}_{m;0}&=-g_{nm}^{(\ell)1}\Delta^{(\ell)}_{n;0}-g_{mn}^{(\ell)4*}\Delta^{(\ell)}_{n;1}\\
\dot{\Delta}^{(\ell)}_{m;1}&=-g_{nm}^{(\ell)4}\Delta^{(\ell)}_{n;0}-g_{nm}^{(\ell)1'}\Delta^{(\ell)}_{n;1}\nonumber\\
\dot{\rho}^{[\ell]}_{m;0}&=d_{ph}\left(g_{n-m,0}^{(\ell+m-n)2}-2g_{0,-\ell}^{(\ell-m+n)3}\right)\rho^{[\ell]}_{n;0}+d_{ph}\left(g_{0,n-m-1}^{(\ell+m-n)4*}-2g_{0,-\ell}^{(\ell+m-n)4*}\right)\rho^{[\ell]}_{n;1}\nonumber\\
\dot{\rho}^{[\ell]}_{m;1}&=d_{ph}\left(g_{n-m,-1}^{(\ell+m-n)4}-2g_{0,-\ell}^{(\ell-m+n)4}\right)\rho^{[\ell]}_{n;0}+d_{ph}\left(g_{1-\ell,1-\ell+n-m}^{(\ell+m-n-1)2}-2g_{1-\ell,0}^{(\ell+m-n-1)3}\right)\rho^{[\ell]}_{n;1}\nonumber\\
\dot{M}^{[\ell]}_{m;0}&=d_{ph}\left(g_{n-m,0}^{(\ell+m-n)2}M^{[\ell]}_{n;0}+g_{0,n-m-1}^{(\ell+m-n)4*}M^{[\ell]}_{n;1}\right)\nonumber\\
\dot{M}^{[\ell]}_{m;1}&=d_{ph}\left(g_{n-m,-1}^{(\ell+m-n)4}M^{[\ell]}_{n;0}+g_{1-\ell,1-\ell+n-m}^{(\ell+m-n-1)2}M^{[\ell]}_{n;1}\right)
\,
\nonumber
\end{align}

\begin{figure*}[t]
\centering
\includegraphics[width=0.95\textwidth]{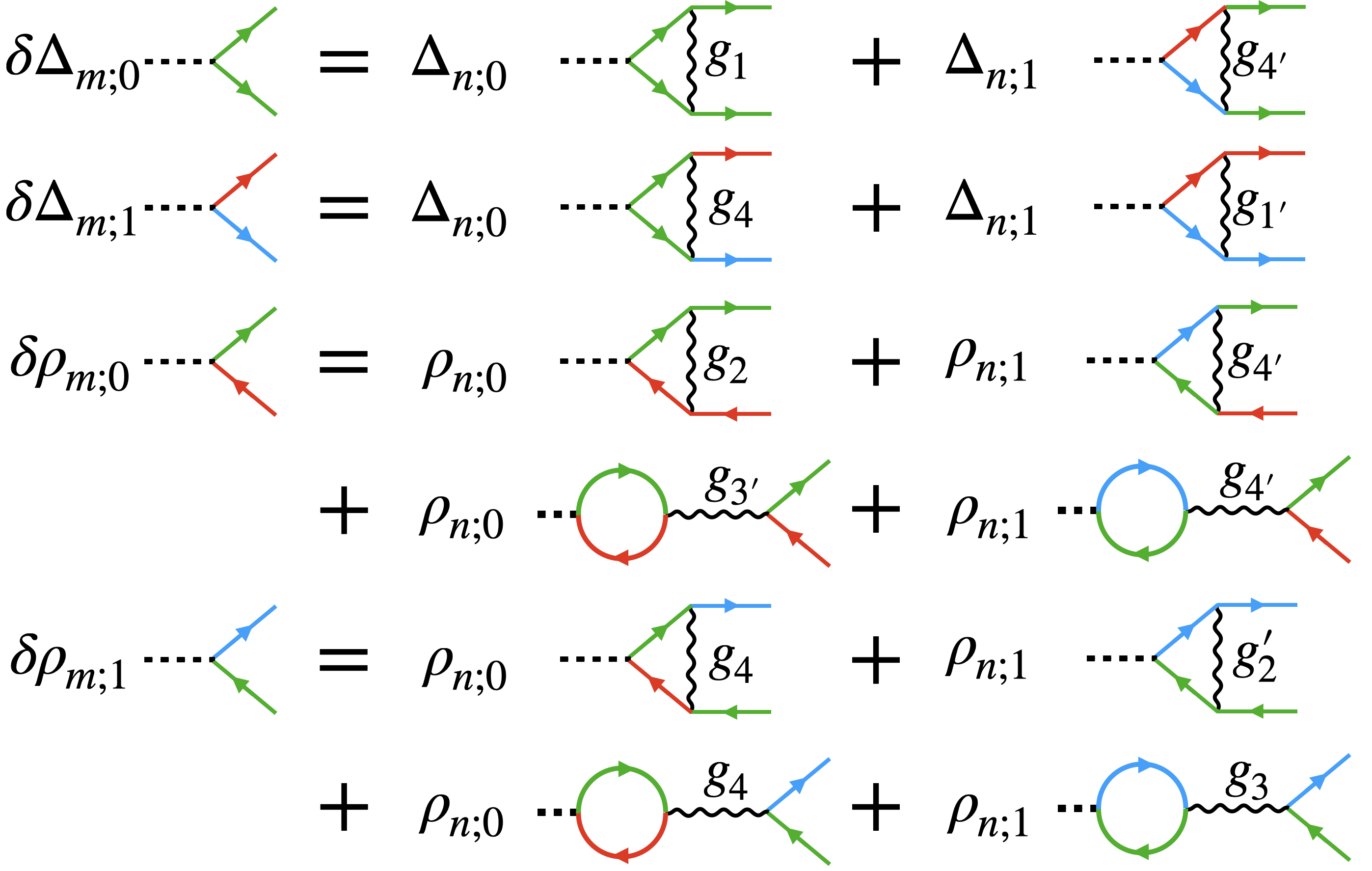}
\caption{Same as Fig. 3 in the main text. The 1 loop Feynman diagrams renormalizing the SC and CDW vertices for VHS indices (green for \(\mathrm{v}=0\), red/blue for \(\mathrm{v}=\pm1\) respectively). SDW diagrams are the same as CDW diagrams with \(M\) instead of \(\rho\). The magnetic flavor indices for each diagram can be read off from Fig. \ref{fig:RGflowVertex}.}
\label{fig:RGflowVertexEq}
\end{figure*}

Note that the MTG symmetries imply that the flow for \(\Delta^{(\ell)}_{m;\mathrm{v}}\) is the same for all \(\ell\) for odd \(q\) (i.e. each \(\ell\) is a degenerate channel, as a consequence of the irrep being \(q\) dimensional), while for even \(q\) the channels are only degenerate for \(\ell\) of the same parity and decouple into \(\Delta^{(\ell,\pm)}_{m;\mathrm{v}}=\Delta^{(\ell)}_{m;\mathrm{v}}\pm\Delta^{(\ell)}_{m+q/2;\mathrm{v}}\) that flow independently for \(+\) and \(-\)
(a total of four pairing channels, as a result of there being four \(q/2\)-dimensional irreps in this case).
The C/SDW channels, on the other hand, split into \(q\) independent equations for each \(\ell\). Moreover, note that the C/SDW equations are all of the form \(\dot{\rho}_m=G_{m-n}\rho_n\) in the magnetic flavor indices, which is a convolution. The C/SDW channels therefore further decouple into their discrete Fourier components
\[\tilde{M}^{[\ell,j]}_{k;\mathrm{v}}=\sum_{m}\omega_q^{mk}M^{[\ell,j]}_{m;\mathrm{v}}\]
which flow as
\begin{align}
\dot{\tilde{\rho}}^{[\ell]}_{k;0}&=\tilde{g}^{(\ell)\rho}_{k;00}\tilde{\rho}^{[\ell]}_{k;0}+\tilde{g}^{(\ell)\rho}_{k;01}\tilde{\rho}^{[\ell]}_{k;1}\\
\dot{\tilde{\rho}}^{[\ell]}_{k;1}&=\tilde{g}^{(\ell)\rho}_{k;10}\tilde{\rho}^{[\ell]}_{k;0}+\tilde{g}^{(\ell)\rho}_{k;11}\tilde{\rho}^{[\ell]}_{k;1}\nonumber\\
\dot{\tilde{M}}^{[\ell]}_{m;0}&=\tilde{g}^{(\ell)M}_{k;00}\tilde{M}^{[\ell]}_{k;0}+\tilde{g}^{(\ell)M}_{k;01}\tilde{M}^{[\ell]}_{k;1}\nonumber\\
\dot{\tilde{M}}^{[\ell]}_{m;1}&=\tilde{g}^{(\ell)M}_{k;10}\tilde{M}^{[\ell]}_{k;0}+\tilde{g}^{(\ell)M}_{k;11}\tilde{M}^{[\ell]}_{k;1}\nonumber
\end{align}
where
\begin{align}
    \tilde{g}^{(\ell)\rho}_{k;00}&=d_{ph}\sum_m\omega_q^{mk}\left\{g_{-m,0}^{(\ell+m)2}-2g_{0,-\ell}^{(\ell-m)3}\right\}\\
    \tilde{g}^{(\ell)\rho}_{k;01}&=d_{ph}\sum_m\omega_q^{mk}\left\{g_{-\ell-m,-\ell}^{(\ell+m)4*}-2g_{0,-\ell}^{(\ell+m)4*}\right\}\nonumber\\
    \tilde{g}^{(\ell)\rho}_{k;10}&=d_{ph}\sum_m\omega_q^{mk}\left\{g_{-\ell,-\ell-m}^{(\ell+m)4}-2g_{0,-\ell}^{(\ell-m)4}\right\}\nonumber=\tilde{g}^{(\ell)\rho*}_{k;01}\\
    \tilde{g}^{(\ell)\rho}_{k;11}&=d_{ph}\sum_m\omega_q^{mk}\left\{g_{1-\ell,1-\ell-m}^{(\ell+m-1)2}-2g_{1-\ell,0}^{(\ell+m-1),3}\right\}\nonumber\\
    \tilde{g}^{(\ell)M}_{k;00}&=d_{ph}\sum_m\omega_q^{mk}g_{-m,0}^{(\ell+m)2}\nonumber\\
    \tilde{g}^{(\ell)M}_{k;01}&=d_{ph}\sum_m\omega_q^{mk}g_{-\ell-m,-\ell}^{(\ell+m)4*}\nonumber\\
    \tilde{g}^{(\ell)M}_{k;10}&=d_{ph}\sum_m\omega_q^{mk}g_{-\ell,-\ell-m}^{(\ell+m)4}=\tilde{g}^{(\ell)M*}_{k;01}\nonumber\\
    \tilde{g}^{(\ell)M}_{k;11}&=d_{ph}\sum_m\omega_q^{mk}g_{1-\ell,1-\ell-m}^{(\ell+m-1)2}\nonumber
\end{align}
Note that C/SDW channels thus decouple into \(q^2\) \(2\times2\) systems of equations which can be easily further decoupled, all as a consequence of the MTG symmetries (which in this case act simply as regular translations). The corresponding eigen-channels flow as
\begin{align}
\dot{\tilde{\rho}}^{[\ell]}_{k;\pm}&=\gamma_{k;\pm}^{[\ell]\rho}\tilde{\rho}^{[\ell]}_{k;\pm}\\
\dot{\tilde{M}}^{[\ell]}_{k;\pm}&=\gamma_{k;\pm}^{[\ell]M}\tilde{M}^{[\ell]}_{k;\pm}\nonumber
\end{align}
where
\begin{align}
    \tilde{\rho}^{[\ell]}_{k;\pm}&=\left(\tilde{g}^{(\ell)\rho}_{k;00}-\tilde{g}^{(\ell)\rho}_{k;11}\pm\sqrt{(\tilde{g}^{(\ell)\rho}_{k;00}-\tilde{g}^{(\ell)\rho}_{k;11})^2+4\left|\tilde{g}^{(\ell)\rho}_{k;10}\right|^2}\right)\tilde{\rho}^{[\ell]}_{k;0}+2\tilde{g}^{(\ell)\rho}_{k;10}\tilde{\rho}^{[\ell]}_{k;1}\\
    \tilde{M}^{[\ell]}_{k;\pm}&=\left(\tilde{g}^{(\ell)M}_{k;00}-\tilde{g}^{(\ell)M}_{k;11}\pm\sqrt{(\tilde{g}^{(\ell)M}_{k;00}-\tilde{g}^{(\ell)M}_{k;11})^2+4\left|\tilde{g}^{(\ell)M}_{k;10}\right|^2}\right)\tilde{M}^{[\ell]}_{k;0}+2\tilde{g}^{(\ell)M}_{k;10}\tilde{M}^{[\ell]}_{k;1}\nonumber
\end{align}
and
\begin{align}
\gamma_{k;\pm}^{[\ell]\rho}&=\frac{1}{2}\left(\tilde{g}^{(\ell)\rho}_{k;00}+\tilde{g}^{(\ell)\rho}_{k;11}\pm\sqrt{(\tilde{g}^{(\ell)\rho}_{k;00}-\tilde{g}^{(\ell)\rho}_{k;11})^2+4\left|\tilde{g}^{(\ell)\rho}_{k;10}\right|^2}\right)\\
\gamma_{k;\pm}^{[\ell]M}&=\frac{1}{2}\left(\tilde{g}^{(\ell)M}_{k;00}+\tilde{g}^{(\ell)M}_{k;11}\pm\sqrt{(\tilde{g}^{(\ell)M}_{k;00}-\tilde{g}^{(\ell)M}_{k;11})^2+4\left|\tilde{g}^{(\ell)M}_{k;10}\right|^2}\right)\nonumber
\end{align}

\section{Projected Gap Functions}\label{B}

Since the RG only determines the order parameter at the VHS points, it is necessary to extend it in some way to determine the nature of the resulting phase (chiral or nodal). In principle, one needs to extend the RG calculation to the whole BZ, which is computationally prohibitive already for moderate \(q\). Even solving the self-consistent gap equation for a constant Hubbard interaction numerically is quite challenging. We therefore adopt a simpler approach and construct an ansatz gap function in real space in the \(c_{\mathbf{r}\sigma}\) basis first (e.g. standard \(s\)- or \(d\)-wave gap functions with up to nearest neighbor terms, etc.),  and then projecting onto the Hofstadter band of interest via \(d_{\mathbf{k}\alpha\sigma}=\sum_{s}\mathcal{U}^s_\alpha(\mathbf{k})c_{\mathbf{k}s\alpha}\) with the band index \(\alpha\) fixed.

The gap function at the VHSs is defined as
\[H_{SC,VHS}=\sum_{\ell,m,\mathrm{v}}\Delta^{(\ell)}_{m;\mathrm{v}}d^\dagger_{\ell+m,\mathrm{v}}d^\dagger_{-m,-\mathrm{v}}+h.c.\]
where \(\mathrm{v}=0,1\) is the VHS index and \(m=0,\dots,q-1\) is the magnetic flavor index. The full gap function defined on the reduced MBZ is
\begin{align}
H_{SC}&=\sum_{\ell,m,\mathbf{p}}\Delta^{(\ell)}_{m}(\mathbf{p})d^\dagger_{\mathbf{p},\ell+m}d^\dagger_{-\mathbf{p},-m}+h.c.=\nonumber\\
&=\sum_{m,n,\mathbf{p}}\hat{\Delta}_{mn}(\mathbf{p})d^\dagger_{\mathbf{p},m}d^\dagger_{-\mathbf{p},n}+h.c
\end{align}
where the latter is the notation of \cite{ShafferWangSantos21}. The gap function can be extended to include pairing between different bands \(\alpha\) and \(\beta\):
\begin{align}
H_{SC}&=\sum_{m,n,\alpha,\beta,\mathbf{p}}\hat{\Delta}_{m\alpha,n\beta}(\mathbf{p})d^\dagger_{\mathbf{p}\alpha m}d^\dagger_{-\mathbf{p}\beta n}+h.c=\nonumber\\
&=\sum_{m,n,s,s',\mathbf{p}}\Delta_{m,s;n,s'}(\mathbf{p})c^\dagger_{\mathbf{p}+m\mathbf{Q},s}c^\dagger_{-\mathbf{p}+n\mathbf{Q},s'}=\nonumber\\
&=\sum_{\mathbf{R},\mathbf{R}',s,s'}\Delta_{\mathbf{R},s;\mathbf{R}',s'}c^\dagger_{\mathbf{R}s}c^\dagger_{\mathbf{R}',s'}
\end{align}
\(\Delta_{m,s;n,s'}(\mathbf{p})\) is thus the gap function in the sub-lattice basis. We further define \(\Delta_{ss'}^{(\ell)}(\mathbf{k})\) with \(\mathbf{k}\) defined in the un-reduced MBZ via \(\Delta_{m,s;n,s'}(\mathbf{p})=\Delta_{ss'}^{(m+n)}(\mathbf{p}+(m-n)\mathbf{Q}/2)\). \(\Delta_{ss'}^{(\ell)}(\mathbf{k})\) is then simply the gap function in the sub-lattice basis corresponding to pairing with total momentum \(\ell\mathbf{Q}\) and defined on the original MBZ. We thus have
\begin{align}
\hat{\Delta}_{m\alpha,n\beta}(\mathbf{p})&=\sum_{ss'}\mathcal{U}_\alpha^s(\mathbf{p}+m\mathbf{Q})\mathcal{U}_\beta^{s'}(-\mathbf{p}+n\mathbf{Q})\Delta_{m,s;n,s'}(\mathbf{p})\nonumber\\
&=\sum_{ss'}\mathcal{U}_\alpha^{s+m}(\mathbf{p})\mathcal{U}_\beta^{s'+n}(-\mathbf{p})\Delta_{m,s;n,s'}(\mathbf{p})
\end{align}
The second line follows from our gauge choice \(\mathcal{U}^{s+1}_\alpha(\mathbf{k+Q})=\mathcal{U}^s_\alpha(\mathbf{k})\). We also have
\begin{align}
\Delta_{m,s;n,s'}(\mathbf{p})&=\frac{1}{N}\sum_{\mathbf{R,R}'} e^{-i\left(\mathbf{p}+m\mathbf{Q}\right)\cdot\mathbf{r}-i\left(-\mathbf{p}+n\mathbf{Q}\right)\cdot\mathbf{r}'}\Delta_{\mathbf{R},s;\mathbf{R}',s'}=\nonumber\\
&=\frac{1}{N}\sum_{\mathbf{R,R}'}e^{-i\mathbf{p}\cdot\left(\mathbf{r}-\mathbf{r}'\right)-i\mathbf{Q}\cdot\left(m\mathbf{r}+n\mathbf{r}'\right)}\Delta_{\mathbf{R},s;\mathbf{R}',s'}
\end{align}
or equivalently
\begin{align}
\Delta^{(\ell)}_{ss'}(\mathbf{k})&=\frac{1}{N}\sum_{\mathbf{R,R}'}
e^{-i\mathbf{k}\cdot\left(\mathbf{r}-\mathbf{r}'\right)-i\ell\mathbf{Q}\cdot\left(\mathbf{r}+\mathbf{r}'\right)/2}\Delta^{(\ell)}_{\mathbf{R},s;\mathbf{R}',s'}=\nonumber\\
&=\frac{1}{N}\sum_{\mathbf{R,R}'}
e^{-i\mathbf{k}\cdot\left(\mathbf{r}-\mathbf{r}'\right)-i\ell\mathbf{Q}\cdot\left(\mathbf{R}+\mathbf{R}'\right)/2}\Delta^{(\ell)}_{\mathbf{R},s;\mathbf{R}',s'}
\end{align}
Note that \(\Delta^{(\ell)}_{\mathbf{R+R}'',s;\mathbf{R}'+\mathbf{R}'',s'}=e^{i\ell\mathbf{Q\cdot R}''}\bar{\Delta}^{(\ell)}_{\mathbf{R},s;\mathbf{R}',s'}\).
The projection of \(\Delta_{m,s;n,s'}(\mathbf{p})\) (or equivalently \(\Delta_{\mathbf{R},s;\mathbf{R}',s'}\)) onto the \(\alpha\) band simply amounts to computing \(\hat{\Delta}_{m\alpha,n\alpha}(\mathbf{p})\equiv\hat{\Delta}_{mn}(\mathbf{p})\), assuming the rest of the components vanish.

We then seek \(\bar{\Delta}_{\mathbf{R},s;\mathbf{R}',s'}\) such that \(\hat{\Delta}_{mn}(\mathbf{K}_{0,v})=\Delta^{(\ell)}_{m;v}\) as found in the RG calculation. Note that thanks to the MTG symmetry, we can look at the \(\ell=m+n=0\) channel alone, the rest being obtained by simple application of \(\hat{T}_x\) symmetry (for \(q>2\) we then find the MTG symmetry of the ground state by minimizing the fourth order free energy in Supplementary Section \ref{C}). In particular, for both \(q=2\) and the lower and upper bands for \(q=3\) we found \(\Delta^{(0)}_{m;0}=-\Delta^{(0)}_{m;1}\). We also use the relations in this supplementary section to establish the action of the self-similarity symmetry \(\hat{S}\) in different bases. Recall that it imposes \(\Delta^{(0)}_{m;v}=-\Delta^{(0)}_{n;v}\) for all \(m\) and \(n\) (for \(q=2\) we find the same relation, but as a consequence of the usual \(\hat{T}_x\) symmetry). On \(\hat{\Delta}_{mn}(\mathbf{p})\), this symmetry acts as
\[\hat{\Delta}_{mn}(\mathbf{p})\xrightarrow{\hat{S}}\hat{\Delta}_{m+1,n-1}(\mathbf{p})=\left[\tau\hat{\Delta}(\mathbf{p})\tau\right]_{mn}\]
where \(\tau_{mn}=\delta_{m,n-1}\) is the shift matrix. This is in contrast to the action of \(\hat{T}_x\) itself, which acts as \(\hat{\Delta}_{mn}(\mathbf{p})\xrightarrow{\hat{T}_x}\left[\tau\hat{\Delta}(\mathbf{p})\tau^T\right]_{mn}=\hat{\Delta}_{m+1,n+1}(\mathbf{p})\). From the action of the \(\hat{S}\) in the band basis, we establish its action on the gap function components in other bases: e.g., it acts on \(\Delta_{ss'}^{(\ell)}(\mathbf{k})\) as
\[\Delta_{ss'}^{(\ell)}(\mathbf{k})\xrightarrow{\hat{S}}\Delta_{s-1,s'+1}^{(\ell)}(\mathbf{k}+\mathbf{Q})\label{DeltaL}\]
and as a convolution in real space:
\[\Delta_{\mathbf{R}s;\mathbf{R}'s'}\xrightarrow{\hat{S}}e^{-i\mathbf{Q}\cdot(\mathbf{R}-\mathbf{R}')}\sum_{X\in q\mathbb{Z}}\mathrm{sinc}\left[\frac{\pi}{q}(X+2)\right]\Delta_{\mathbf{R},s+1;\mathbf{R}'+X\hat{\mathbf{x}},s'-1}\]
where \(\mathrm{sinc}(x)=\sin x/x\) is the inverse Fourier transform of \(e^{2ik_x}\).

\section{BdG Hamiltonian, Free Energy, and Edge Modes}\label{C}

In this section we discuss the Ginzburg-Landau free energy calculation used to establish the symmetry of the HSC for \(q=3\), as well as the edge mode calculation used for generating Fig. \ref{fig:EdgeModes}. For both calculation we make use of the Bogoliubov-De Gennes (BdG) formalism.  The starting point is the mean field Hamiltonian, which in the band basis reads
\begin{align}\label{HSC}
H&=\sum_{\ell,\alpha,\mathbf{p}}\varepsilon_\alpha(\mathbf{p})d^\dagger_{\mathbf{p},\ell,\alpha}d_{\mathbf{p},\ell,\alpha}+\frac{1}{2}\sum_{m,n,\alpha,\beta,\mathbf{p}}\left[\hat{\Delta}_{\ell,\alpha;\ell'\beta}(\mathbf{p})d^\dagger_{\mathbf{p},\ell,\alpha}d^\dagger_{-\mathbf{p},\ell',\beta}+h.c.\right]+H_{\Delta^2}
\end{align}
where \(\alpha,\beta=0,\dots,q-1\) are the Hofstadter band indices, \(\ell,\ell'\) are the magnetic patch indices (we omit the spin index), and
\[H_{\Delta^2}=\sum_{\ell,n,m,\mathbf{p,p}'} \hat{\Delta}^\dagger_{\ell+m,-m}(\mathbf{p})\left[g^{-1}(\mathbf{p;p}')\right]^{(\ell)}_{mn}\hat{\Delta}_{\ell+n,-n}(\mathbf{p}')\label{HDelta2}\]
is a term quadratic in the gap function arising from the Hubbard-Stratonovich transformation and involving the inverse of the coupling tensor:
\[\sum_{o\mathbf{q}}g^{(\ell)}_{mo}(\mathbf{p;q})\left[g^{-1}(\mathbf{q;p}')\right]^{(\ell')}_{on}=\delta_{\ell,\ell'}\delta_{mn}\delta_{\mathbf{p p}'}\blue{\,.}\]
Here we omitted the band indices in the interactions for simplicity as in the weak-coupling approximation we assume only interactions within a single band play a role and inter-band interaction will not play a role below (see previous Supplementary Section \ref{B}).
In the BdG formalism we introduce the Nambu spinors \(\Psi_{\mathbf{p}\ell\alpha}=(d_{\mathbf{p}\ell\alpha},d_{-\mathbf{p}\ell\alpha}^\dagger)\), which allows us to write the Hamiltonian as
\[H=\frac{1}{2}\sum_{\substack{\ell,\ell',\mathbf{p}}{\alpha\beta}}\Psi_{\mathbf{p}\ell\alpha}^\dagger\left[\mathcal{H}_{BdG}(\mathbf{p})\right]_{\ell,\alpha;\ell',\beta}\Psi_{\mathbf{p}\ell'\beta}+H_{\Delta^2}\]
where
\[\left[\mathcal{H}_{BdG}(\mathbf{p})\right]_{\ell,\alpha;\ell',\beta}=\left(\begin{array}{cc}
     \varepsilon_\alpha(\mathbf{p})\delta_{\alpha\beta} &  \hat{\Delta}_{\ell,\alpha;\ell',\beta}(\mathbf{p})\\
     \hat{\Delta}_{\ell,\alpha;\ell',\beta}^\dagger(\mathbf{p}) & -\varepsilon_\alpha(-\mathbf{p})\delta_{\alpha\beta}
\end{array}\right)\]
is the BdG Hamiltonian. Note that when \(\hat{\Delta}_{\ell,\alpha;\ell',\beta}=0\) for \(\ell'\neq-\ell\) (i.e. when only zero total momentum pairing is present), one can remove the magnetic flavor indices \(\ell\) and work on the non-reduced MBZ instead, replacing \(\hat{\Delta}_{\ell,\alpha;-\ell,\beta}(\mathbf{p})\) with \(\hat{\Delta}_{\alpha\beta}(\mathbf{p}+\ell\mathbf{Q})\), so that the BdG Hamiltonian is a \(2q\times 2q\) matrix. In all other cases, however, the unit cell needs to be extended due to the breaking of the \(\hat{T}_y\) MTG symmetry, resulting in the \(q\)-fold folding of the MBZ into the rMBZ, in which case we have to work with a \(2q^2\times 2q^2\) BdG Hamiltonian.

\subsection{Ginzburg-Landau Free Energy}

To obtain the gap function in the mean field approach we need to minimize the free energy, which yields the self-consistent gap equation. The free energy is in turn obtained from Eq. (\ref{HSC}) by integrating out the \(\Psi_{\mathbf{p}\ell\alpha}\) fields from the partition function. For this part of the calculation we assume that the pairing happens only in one band \(\alpha\) and so drop the band index. Using the Matsubara formalism we then find
\[\mathcal{F}=-T\sum_{\omega,\mathbf{p}}\text{Tr}\left[\log\beta\mathcal{G}^{-1}(i\omega,\mathbf{p})\right]+H_{\Delta^2}\label{Fmicro}\]
where \(\omega=(2\pi j+1) T\) with integer \(j\) are the Matsubara frequencies and we defined the Gor'kov Green's function
\begin{align}
\mathcal{G}(i\omega,\mathbf{p})&=\left(i\omega-\mathcal{H}_{BdG}(\mathbf{p})\right)^{-1}=\nonumber\\
&=\left(\begin{array}{cc}
    \hat{G}(i\omega,\mathbf{p}) & \hat{F}(i\omega,\mathbf{p}) \\
    \hat{F}^\dagger(i\omega,\mathbf{p}) & -\hat{G}^T(-i\omega,-\mathbf{p})
\end{array}\right)\blue{\,.}
\end{align}
Minimizing \(\mathcal{F}\) with respect to \(\hat{\Delta}^\dagger\) gives the gap equation
\[\hat{\Delta}_{\ell+n,-n}(\mathbf{p})=T\sum_{\omega\mathbf{p}'m}g^{(\ell)}_{nm}(\mathbf{p;p}')\hat{F}_{\ell+m,-m}(i\omega,\mathbf{p}')\label{GapEq}\\\blue{\,.}\]
Close below \(T_c\) we can expand the free energy and the Green's functions in powers of the gap function and obtain the linearized gap equation (see the appendix in \cite{ShafferWangSantos21} for details):
\[\hat{\Delta}_{\ell+n,-n}(\mathbf{p})=-\log\frac{1.13\Lambda}{T}\sum_{\mathbf{p}'m}\nu(\mathbf{p}')g^{(\ell)}_{nm}(\mathbf{p;p}')\hat{\Delta}_{\ell+m,-m}(\mathbf{p}')\]
where \(\nu(\mathbf{p})\) is the momentum resolved density of states at the Fermi level and \(\Lambda\) is the high energy cutoff. We note that the linearized gap equation is equivalent to the 1 loop RG flow equation for the SC vertex and both give the same form of the gap function.

As shown in \cite{ShafferWangSantos21} and discussed above, for odd \(q\) the gap function belongs to a \(q\)-dimensional irrep which means that just as in the 1 loop RG flow there are \(q\) degenerate solutions of the linearized gap equation that we can label \(\hat{\Delta}^{(L)}\) with \(L=0,\dots,q-1\) labeling the total momentum \(L\mathbf{Q}\) of the corresponding Cooper pairs. The solutions are picked such that
\begin{align}
    \hat{\Delta}^{(L)}&\xrightarrow[]{\hat{T}_x}\hat{\Delta}^{(L-2)}\label{DeltaLT1}\blue{\,,}\\
    \hat{\Delta}^{(L)}&\xrightarrow[]{\hat{T}_y}\omega^{pL}_q\hat{\Delta}^{(L)}\label{DeltaLT2}\,.
\end{align}
\(\hat{\Delta}^{(L)}\) then form \(q\) components of the irrep that are eigenmodes of the \(\hat{T}_y\) symmetry and are generated by the \(\hat{T}_x\) symmetry \cite{ShafferWangSantos21}. Because the linearized gap equation is linear (as is the 1 loop RG vertex flow equation), any linear combination
\[\hat{\Delta}(\mathbf{p})=\sum_L\eta_L\hat{\Delta}^{(L)}(\mathbf{p})
\label{eq: irrep expansion}\]
is a solution for any choice of the complex parameters \(\eta_L\). The vector \(\boldsymbol{\eta}=(\eta_0,\dots,\eta_{q-1})\) constitutes the order parameter that is selected spontaneously once non-linear terms are included.

In order to obtain the higher order terms we take Eq. (\ref{eq: irrep expansion}) as the ansatz, plug it into the free energy in Eq. (\ref{Fmicro}) and sum over the momentum \(\mathbf{p}\), which yields the macroscopic Ginzburg-Landau free energy
\[\mathcal{F}=a\left|\boldsymbol{\eta}\right|^2+\sum_{MN}b_{MN}\sum_L\eta_{L+M}^*\eta_{L-M}^*\eta_{L+N}\eta_{L-N}\,
\label{F}\]
where the parameters \(a\) and \(b_{MN}\) are
\begin{align}
    a&=H_{\Delta^2}-\log\frac{1.13\Lambda}{T}\sum_\mathbf{p}\nu(\mathbf{p})\text{Tr}\left[\hat{\Delta}^{(L)\dagger}\hat{\Delta}^{(L)}\right]\blue{\,,}\qquad
    b_{MN}&=\frac{7\zeta(3)}{32\pi^2T^2}\sum_\mathbf{p}\nu(\mathbf{p})\text{Tr}\left[\hat{\Delta}^{(L+M)\dagger}\hat{\Delta}^{(L+N)}\hat{\Delta}^{(L-M)\dagger}\hat{\Delta}^{(L-N)}\right]\blue{\,.}
\end{align}
In order to compute the fourth order \(b_{MN}\) coefficients, we note that the point with high DOS \(\nu(\mathbf{p})\) dominate in the sum, so we can get an approximate expression by restricting the sum only to the VHS points \(\mathbf{K}_{\ell,\mathrm{v}}\). Minimizing the resulting Ginzburg-Landau free energy Eq. (\ref{F}) for the solutions we find in RG for \(q=3\), we find that the solutions are of the form \(|\eta_L|=\eta\), \(\arg[\eta_L]=\pm 4\pi/3 \delta_{LM}\) for some fixed \(M=0,1,2\), for a total of six solutions. As shown in \cite{ShafferWangSantos21}, these solutions are precisely the solutions symmetric under \(\omega^{pM}_3\hat{T}_x\hat{T}_y^{\pm 1}\) for the \(+\) and \(-\) cases respectively (equivalently, the gap functions pick up a phase of \(\omega^{-2pM}_3\) under \(\hat{T}_x\hat{T}_y^{\pm 1}\)). The fact that there are six solutions instead of three as one would expect from the breaking of the \(\hat{T}_x\) symmetry is due to the fact that the \(\hat{C}_4\) symmetry is also broken and maps \(\hat{T}_x\hat{T}_y\) to \(\hat{T}_y\hat{T}_x^{-1}\). In all cases the ground states have a \(\mathbb{Z}_3\) symmetry.

\subsection{Edge Mode Calculation}

In order to compute the edge modes, we considered the BdG Hamiltonian for the gap function \(\Delta_{ss'}(\mathbf{k})\) of the form in Eq. (12) put on a vertical cylinder, i.e. assuming a system periodic in the \(x\) direction but with open boundary conditions in the \(y\) direction. More precisely, we consider the \(\hat{T}_x\hat{T}_y\) symmetric gap functions that are linear combinations \(\sum_L\eta_L\Delta_{ss'}^{(L)}(\mathbf{k})\) satisfying \(\Delta_{ss'}^{(L)}(\mathbf{k})=\Delta_{s+1,s'+1}^{(L+2)}(\mathbf{k})\) with \(\eta_L\) found in the previous paragraph and with \(\Delta_{ss'}^{(L)}(\mathbf{p}+\ell\mathbf{Q})=\Delta_{L/2-\ell,s;L/2+\ell,s'}(\mathbf{p})\) as defined in Supplementary Section \ref{B}. We then perform an inverse Fourier transform on \(k_y\) to obtain the total hybrid gap function
\[\Delta_{m,y,s;n,y',s'}(p_x)=\eta_{m+n}\Delta_0\left[\delta_{yy'}(1-\cos p_x)-\delta_{y,y'-1}\omega_3^{-p(s-s'+m-n)}+\delta_{y,y'+1}\omega_3^{p(s-s'+m-n)}\right]\]
where for concreteness we took \(\eta_L=\omega_3^{2\delta_{L,0}}\). The BdG Hamiltonian in this hybrid basis is then
\[\left[\mathcal{H}_{BdG}(p_x)\right]_{m,y,s;n,y',s'}=\left(\begin{array}{cc}
     \mathcal{H}^{(0)}_{y,s;y',s'}(p_x)\delta_{mn} &  \Delta_{m,y,s;n,y',s'}(p_x)\\
     \Delta^*_{n,y',s';m,y,s}(p_x) & -\mathcal{H}^{(0)*}_{y,s;y',s'}(-p_x)\delta_{mn} 
\end{array}\right)\]
where
\[\mathcal{H}^{(0)}_{y,s;y',s'}(p_x)=\delta_{ss'}\left(-\mu\delta_{yy'}-2t\delta_{y,y'-1}\omega_3^{ps}-2t\delta_{y,y'+1}\omega_3^{-ps}\right)-t\delta_{yy'}\left(\delta_{s,s'+1}e^{ip_x}+\delta_{s,s'-1}e^{-ip_x}\right)\]
is the normal state Hofstadter Hamiltonian in the hybrid basis. Taking the number of lattice sites along the \(y\) direction to be \(N_y\), the hybrid BdG Hamiltonian is a \(2q^2N_y\times 2q^2N_y\) Hamiltonian. In Fig. \ref{fig:EdgeModes} we used \(N_y=100\) with \(\mu=\pm2.44\) and \(\Delta_0=0.02\) and \(\Delta_0=0.2\) respectively (setting \(t=1\)). We note that one could compute the Chern number directly from the bulk spectrum; however, interestingly we found that this numerical computation is more challenging than the edge mode computation.

 
\end{document}